\documentclass[a4paper,11pt]{article}
\pdfoutput=1
\usepackage[utf8x]{inputenc}
\usepackage{jheppub}
\usepackage{epsf}
\usepackage{graphicx}	% needed for figures
\usepackage{dcolumn}	% needed for some tables
\usepackage{bm}			% for math
\usepackage{amssymb, amsmath}
\usepackage{wasysym}
\usepackage{lipsum, color}
\usepackage{physics}
\usepackage{bbm}		% \mathbbm{1} - the unit matrix.
\usepackage{multirow}

\usepackage{slashed}
\usepackage{hyperref}
\hypersetup{colorlinks, citecolor=bluscuro, linkcolor=black, urlcolor=bluscuro}
\definecolor{rossos}{cmyk}{0,1,1,0.55}
\definecolor{bluscuro}{rgb}{0.15, 0.2, .85}
\definecolor{bluchiaro}{cmyk}{1,.3,0.,0.1}

\let\oldquote\quote
\renewcommand\quote{\scriptsize\oldquote}
\let\oldquotation\quotation
\renewcommand\quotation{\scriptsize\oldquotation}

\newcommand{\gsim}{\lower.7ex\hbox{$\;\stackrel{\textstyle>}{\sim}\;$}}
\newcommand{\lsim}{\lower.7ex\hbox{$\;\stackrel{\textstyle<}{\sim}\;$}}

\def\beq{\begin{equation}}
\def\eeq{\end{equation}}
\def\be{\begin{equation}}
\def\ee{\end{equation}}
\def\bea{\begin{eqnarray}}
\def\eea{\end{eqnarray}}
\def\bmat{\begin{pmatrix}}
	\def\emat{\end{pmatrix}}
\def\bei{\begin{itemize}}
	\def\eei{\end{itemize}}

\usepackage[usenames,dvipsnames,svgnames]{xcolor}
\usepackage[normalem]{ulem}

\makeatletter
\def\section{\@startsection {section}{1}{\z@}{-3.5ex plus -1ex minus
		-.2ex}{2.3ex plus .2ex}{\large\bf}}
\def\subsection{\@startsection{subsection}{2}{\z@}{-3.25ex plus -1ex
		minus -.2ex}{1.5ex plus .2ex}{\normalsize\bf}}
\makeatother
\makeatletter

\@addtoreset{equation}{section}

\makeatother

\usepackage{graphicx}  % needed for figures
\usepackage{dcolumn}   % needed for some tables
\usepackage{bm,relsize}        % for math
\usepackage{amssymb, amsmath}
\usepackage{wasysym}
\usepackage{slashed}
\usepackage{bbold} % For Aviv
\usepackage{jheppub}
\usepackage{mathtools}
\usepackage{comment}
\usepackage{cancel}
\usepackage{ulem}

\usepackage{feyn}
\def\beq{\begin{equation}}
\def\eeq{\end{equation}}

\begin{document}

\global\long\def\com#1#2{\underset{{\scriptstyle #2}}{\underbrace{#1}}}

\global\long\def\comtop#1#2{\overset{{\scriptstyle #2}}{\overbrace{#1}}}

\global\long\def\ket#1{\left|#1\right\rangle }

\global\long\def\bra#1{\left\langle #1\right|}

\global\long\def\braket#1#2{\left\langle #1|#2\right\rangle }

\global\long\def\op#1#2{\left|#1\right\rangle \left\langle #2\right|}

\global\long\def\opk#1#2#3{\left\langle #1|#2|#3\right\rangle }

\global\long\def\L{\mathcal{L}}

\title{Unveiling dark fifth forces with linear cosmology}

\abstract{We initiate the exploration of the cosmology of dark fifth forces: new forces acting solely on Dark Matter. We focus on long range interactions which lead to an effective violation of the Equivalence Principle on cosmological scales today. At the microscopic level, the dark fifth force can be realized by a light scalar with mass smaller than the Hubble constant today ($\lesssim 10^{-33}\,\text{eV}$) coupled to Dark Matter. We study the behavior of the background cosmology and linear perturbations in such a Universe. At the background level, the new force modifies the evolution of the Dark Matter energy density and thus the Hubble flow. At linear order, it modifies the growth of matter perturbations and generates relative density and velocity perturbations between Dark Matter and baryons that grow over time. We derive constraints from current CMB and BAO data, bounding the strength of the dark fifth force to be less than a percent of gravity. These are the strongest constraints to date. We present potential implications of this scenario for the Hubble tension and discuss how our results are modified if the light scalar mediator accounts for the observed density of the Dark Energy. Finally, we comment on the interplay between our constraints and searches for violations of the Equivalence Principle in the visible sector.}

\author[a]{Maria Archidiacono,}
\author[a,c]{Emanuele Castorina,}
\author[b,c]{Diego Redigolo,}
\author[c,d]{Ennio Salvioni}
\affiliation[a]{Dipartimento di Fisica ``Aldo Pontremoli'', Universit\`a degli Studi di Milano,\\ Via Celoria 16, 20133 Milan, Italy}
\affiliation[b]{INFN Sezione di Firenze,\\ Via Sansone 1, 50019 Sesto Fiorentino, Italy}
\affiliation[c]{Theoretical Physics Department, CERN,\\
Esplanade des Particules 1, 1211 Geneva, Switzerland}
\affiliation[d]{Dipartimento di Fisica e Astronomia ``Galileo Galilei'', Universit\`a di Padova\\and INFN Sezione di Padova,\\
Via Marzolo 8, 35131 Padua, Italy}
\date{\today}

\preprint{CERN-TH-2022-066}

\emailAdd{\hbox{maria.archidiacono@unimi.it, emanuele.castorina@unimi.it,}\\diego.redigolo@fi.infn.it, ennio.salvioni@cern.ch}

\maketitle

\section{Introduction}\label{sec:intro}

The existence of Dark Matter (DM) is supported by a large variety of observations. This strong evidence, however, concerns only its gravitational interactions, providing no insights on its microscopic properties. Current data leave wide open the landscape of possible DM masses and interactions, and besides the development of a large number of experimental strategies, we are just scratching the surface of the theoretically plausible microscopic theories of DM. 

Facing this plenitude of theoretical possibilities, it is interesting to take a more data-driven methodology and focus on what we can learn about the nature of DM using present and forthcoming observations. Following this approach, we initiate the exploration of DM scenarios that can be uniquely tested with cosmological measurements at an exquisite level of precision. These are dark sectors where DM is self interacting through a dark fifth force which has a range comparable to the size of the observable Universe. 

Cosmology offers the unique possibility of studying the very long range dynamics of DM, where the presence of any force other than gravity results in an effective violation of the weak Equivalence Principle (EP). The idea is to use the Universe as a ``scale'' and try to measure by how much the inertial and gravitational masses of DM can differ in light of the available data. In this sense, cosmological observations can be viewed as dark fifth force experiments, in analogy with the many tests of long range forces in the visible sector~\cite{Adelberger:2003zx,Will:2014kxa,Tino:2020nla}. 

In this paper we begin a systematic study of the cosmological signatures of dark fifth forces. As the first step we focus on the constraints from linear cosmology, looking at the current Cosmic Microwave Background (CMB) data from Planck~\cite{Planck:2018vyg} and Baryon Acoustic Oscillations (BAO) data~\cite{Kazin:2014qga,Beutler:2011hx,BOSS:2016wmc,Ross:2014qpa}. For concreteness, we consider the case where the dark fifth force is mediated by trilinear interactions of the DM with a light scalar mediator, whose mass we take to be lighter than or equal to the size of the Universe today ($m_\varphi \lesssim H_0 \simeq 10^{-33}$~eV). We leave the exploration of constraints from non-linear cosmology, as well as different mass regimes and non-minimal realizations of a dark fifth force, for future works. 

The presence of a long range force between DM particles severely affects both the cosmological background and the linear regime of perturbations, resulting in a strong upper bound on its strength. We will show this is constrained to be roughly two orders of magnitude weaker than gravity, as long as the mass of the scalar and its initial displacement from the origin are small enough to make its contribution to the vacuum energy negligible. This scenario, which we call a ``pure fifth force'' or 5F for short, was first studied in Refs.~\cite{Damour:1990tw,Frieman:1991zxc,Casas:1991ky,Gradwohl:1992ue} (see also later discussions of possible observational imprints~\cite{Anderson:1997un,Damour:2002nv,Keselman:2009nx,Audren:2013dwa,Audren:2014hza}, including on galactic scales~\cite{Kesden:2006vz,Desmond:2020gzn}).

In the opposite regime, the light scalar vacuum energy can account for the observed amount of Dark Energy (DE) today. In this scenario, which we call Coupled Dark Energy (CDE), the scalar is a quintessence field coupled to the DM through trilinear interactions and minimally coupled to gravity. Such interaction between DM and DE was first studied in Refs.~\cite{Wetterich:1994bg,Amendola:1999er,Farrar:2003uw} and has been subject of intense activity in subsequent literature~\cite{Amendola:2003wa,Pettorino:2008ez,Valiviita:2008iv,Gavela:2009cy,Saracco:2009df,Pettorino:2012ts,Pettorino:2013oxa,Pourtsidou:2013nha,Costa:2014pba,Gleyzes:2015pma,Gleyzes:2015rua,DAmico:2016jbm,Agrawal:2019dlm,Gomez-Valent:2020mqn}. In our simple setup and with natural values for the scalar potential, the tuning of the initial displacement of the scalar field required to match the value of the Cosmological Constant (CC) does not allow us to take the mass of the scalar arbitrarily small and at the same time retain an observable coupling with DM. Saturating this condition (i.e.~taking the highest possible mass $m_\varphi = H_0$) yields constraints on the strength of the coupling similar to the ones for a pure fifth force.

Finally, we take a look at the generic implications of the presence of a dark fifth force on a consistent relativistic theory of DM. Requiring naturalness of the scalar potential forces the DM mass to be below about $10^{-2}$~eV. This bound could be overcome in non-minimal models (for instance imposing supersymmetry in the dark sector), but it indicates a generic correlation between a dark sector endowed with a fifth force and ultralight DM produced in the early Universe through the misalignment mechanism~\cite{Preskill:1982cy,Abbott:1982af,Dine:1982ah}. We explore the parameter space of simple axion DM scenarios and discuss the interplay between the dark fifth force constraints derived here and fifth force constraints in the visible sector.

This paper is organized as follows. In Sec.~\ref{sec:setup} we illustrate our framework, deriving the equations that govern the dynamics of DM in the presence of a fifth force and of the fifth force field itself. We then study the behavior of the background in Sec.~\ref{sec:bkg} and move to the linear perturbations in Sec.~\ref{sec:fluctuations}. The constraints derived from CMB and BAO data are discussed in Sec.~\ref{sec:constraints}. In Sec.~\ref{sec:implications_particle} we comment on the interplay between our constraints and searches for EP violation in the visible sector. Section~\ref{sec:conclusions} contains our conclusions. In a triptych of appendices (\ref{sec:appendixFields},~\ref{app:pert_SG},~\ref{app:fluid_eqs}) we give further technical details about the equations governing the cosmological dynamics. %Appendix~\ref{app:lensing} contains our constraints including CMB Lensing.   

\section{Framework}\label{sec:setup}

In this section we show how to describe the cosmological imprints of a dark fifth force. We start in Sec.~\ref{sec:particles} by studying the behavior of a DM particle in the presence of an external gravitational field plus a scalar fifth force. We then derive the equations governing a fluid of DM particles in the non-relativistic limit. This limit is the one relevant for DM, although in principle our derivation could apply to relativistic fluids like neutrinos~\cite{Esteban:2021ozz,Esteban:2022rjk}. The relativistic expressions are provided in Appendix~\ref{sec:appendixFields}. In Sec.~\ref{sec:fields} we give explicit examples of full relativistic quantum field theories realizing this fluid behavior and outline their properties.  

We assume the Universe is described by a flat FLRW metric and consider only scalar perturbations around it, which can be written in Newtonian gauge as
\begin{align}
    ds^2 = -\big(1+2 \Psi(\vb{x},t)\big)\dd t^2 + a^2(t) \big(1+2 \Phi(\vb{x},t) \big) \delta_{ij} \dd x^i \dd x^j\ ,
\end{align}
where $\Psi$ and $\Phi$ coincide with the two gauge invariant Bardeen potentials~\cite{Bardeen:1980kt}.
For some applications we will make extensive use of other gauges, in particular the synchronous gauge, to which coordinate transformations are well known~\cite{Ma:1995ey}.\footnote{The synchronous gauge is a full gauge fixing of the metric only if the perturbations are defined in a frame comoving with one (or more) pressureless species that follows the standard General Relativity geodesics. This is not the case for DM interacting with new light degrees of freedom, hence strictly speaking the synchronous gauge cannot be defined in our setup. We then defined the synchronous gauge with respect to a small fraction of regular cold Dark Matter (CDM) $\rho_c$, not subject to the new long range force. In particular, we set $\Omega_{c} = 10^{-5}$ today and we checked that this value has no observational impact given the current uncertainties of cosmological data.} The equations governing the perturbations in synchronous gauge are provided in Appendix~\ref{app:pert_SG}.

\subsection{Particles and fluids}~\label{sec:particles}
In the classical limit, we can derive the worldline of a DM particle (independently of its spin) in an external background with a non trivial metric $g_{\mu\nu}$ and a scalar fifth force $s$~\cite{Farrar:2003uw} as
\begin{align}
    S_\chi = - \int \dd \lambda~ m_\chi(s) \sqrt{- g_{\mu \nu} \frac{\dd x^\mu}{\dd \lambda}\frac{\dd x^\nu}{\dd \lambda}}\, ,\label{eq:wordline}
\end{align}
for some affine parameter $\lambda$. We take $m_\chi(s)$ to be a function of $s$ only, whose explicit form depends on the microscopic interactions between the DM and the scalar fifth force. By varying the action we find the geodesic equation
\begin{align}
\label{eq:geo1}
    \frac{\dd^2 x^\mu}{\dd \sigma^2}  + \Gamma^\mu_{\;\nu \rho} \frac{\dd x^\nu}{\dd \sigma}\frac{\dd x^\rho}{\dd \sigma} + \frac{\partial \log m_\chi(s)}{\partial s} \frac{\partial s}{\partial x^\nu} \left( g^{\mu \nu}+\frac{\dd x^\mu}{\dd \sigma}\frac{\dd x^\nu}{\dd \sigma}\right) = 0\,,
\end{align}
where $\dd \sigma = \sqrt{- g_{\mu \nu} \frac{\dd x^\mu}{\dd \lambda}\frac{\dd x^\nu}{\dd \lambda}}\, \dd \lambda$ and the $\Gamma$'s are the usual Christoffel symbols. The fact that the DM mass does not drop out from the integral in Eq.~\eqref{eq:wordline} indicates a violation of the EP in the dark sector, which results in a modification of the DM geodesic equation controlled by the space-time profile of the scalar field $s$. 

We define the 4-momentum of the $\chi$ particles as 
\begin{equation}
P^\mu \equiv \frac{\dd x^\mu}{\dd \lambda}=\big( (1-\Psi)E,(1-\Phi)p^i/a \big)\ ,
\end{equation}
with the mass shell condition $g_{\mu \nu} P^\mu P^\nu = - m_\chi^2(s)$ providing the dispersion relation $E=\sqrt{m_\chi^2(s)+p^2}$. This allows us to rewrite the geodesic equation as 
\begin{align}
\label{eq:geo2}
    \frac{\dd P^\mu}{\dd \lambda} +  \Gamma^\mu_{\;\nu \rho}  P^\nu P^\rho + \frac{1}{2}\frac{\partial\, m^2_\chi(s)}{\partial s} \frac{\partial s}{\partial x^\nu}g^{\mu \nu} = 0\, ,
\end{align}
which in the non-relativistic limit and in flat space-time reduces to the form $\Ddot{x}^i = -
\nabla^i \Psi - (\partial \log m_\chi(s)/\partial s) \nabla^i s$. We clearly see that the particle acceleration feels both the gravitational force (the gradient of the gravitational potential) and the scalar fifth force.

The DM single particle phase space density $f_\chi(\vb{x},t,\vb{p})$ satisfies the Vlasov equation~\cite{Bertschinger:1993xt} 
\begin{align}\label{eq:BE}
    \frac{\partial f_\chi}{\partial t} + 
    \frac{\dd x^i}{\dd t}\frac{\partial f_\chi}{\partial x^i} + \frac{\dd p^i}{\dd t}\frac{\partial f_\chi}{\partial p^i} = 0\,,
\end{align}
where we treat the DM as a collisionless fluid and encode its interaction with the fifth force in the geodesic equation, very much as we do with gravity. This treatment is consistent as long as the fifth force mediator $s$ has a large occupation number and can be viewed as a coherent field acting on test particles.

We then proceed and compute the first two moments of the Vlasov equation, assuming $\chi$ to be non-relativistic and expanding to the linear order in the metric perturbations. The resulting continuity and Euler equations read
\begin{align}
   \dot{\rho}_\chi + 3 (H + \dot{\Phi}) \rho_\chi + \frac{1}{a}\nabla_i( \rho_\chi v_\chi^i) - \dot{s}\, \frac{\partial \log m_\chi(s)}{\partial s} \rho_\chi =\;& 0\, ,\qquad\label{eq:continuity}
 \\
 (\rho_\chi v_\chi^i)^{\boldsymbol{\cdot}} + \frac{1}{a}\nabla_j \Sigma^{ij}_{\chi}  + 4 H \rho_\chi v_\chi^i +\frac{\rho_\chi}{a}\nabla^i \Psi + \frac{\rho_\chi}{a}\frac{\partial \log m_\chi(s)}{\partial s} \nabla^i s =\;& 0\, ,\qquad\label{eq:euler}
\end{align}
where $\rho_\chi = m_\chi \int \dd^3 p  f_\chi/(2\pi)^3$ is the energy density, $v^i_\chi = \rho_\chi^{-1}\int \dd^3 p\hspace{0.2mm} p^i f_\chi/(2\pi)^3$ is the fluid velocity, $H$ is the Hubble parameter and $()^{\boldsymbol{\cdot}}$ indicates derivatives with respect to the coordinate time $t$. The equations are fully non-linear in the $\chi$ density and velocity perturbations. The piece of Eq.~\eqref{eq:euler} containing the second moment of the phase space distribution, $\Sigma^{ij}_{\chi} = m_\chi^{-1} \int \dd^3 p\hspace{0.2mm} p^i p^j f_\chi /(2\pi)^3$, needs to be included for a perturbative treatment beyond linear theory~\cite{Baumann:2010tm,Carrasco:2012cv}. 

The presence of the new long range force modifies the Euler equation in Eq.~\eqref{eq:euler}, introducing a new acceleration term similar to the one induced by the gravitational potential, but controlled by the field dependence of the DM mass. This was expected, since the classical non-relativistic $1/r$ potential generated by the exchange of the fifth force adds to the gravitational potential in flat space-time. Interestingly, the effect of the new force can be rewritten as the gradient of the inertial mass, $\rho_\chi (\partial \log m_\chi(s)/\partial s)\nabla^i s=n_\chi\nabla^i m_\chi(s)$, which shows once again how the presence of a scalar force induces a violation of the EP. 

The continuity equation in Eq.~\eqref{eq:continuity} is also modified, reflecting the fact that only the total energy density of $\chi$ and $s$ is conserved. This introduces an explicit dependence of the DM density on the time evolution of the fifth force field $s\,$, which will impact the behavior of the cosmological background (discussed in Sec.~\ref{sec:bkg}) and the growth of density and velocity perturbations (discussed in Sec. \ref{sec:fluctuations}). These effects cannot be easily decoupled parametrically from the modification of the Euler equation and need to be included in a complete description of the fifth force dynamics. As a result, the imprints of a fifth force in the cosmological observables cannot be fully captured by looking purely at the non-relativistic limit of the Vlasov equation, as done for example in Refs.~\cite{Gradwohl:1992ue,Creminelli:2013nua}.  

A completely equivalent way of deriving the fluid equations above would have been to start from the energy-momentum tensor for the microscopic degrees of freedom, and then rewrite it in terms of the macroscopic fluid variables. We explicitly show this derivation in Appendix~\ref{sec:appendixFields}.

\subsection{Fields}~\label{sec:fields}
The continuity and Euler equations for DM should be complemented with the equations describing the dynamical evolution of the scalar fifth force and gravity. In general, the microscopic Lagrangian controlling the fifth force and DM dynamics can be parametrized as 
\begin{equation}
\mathcal{L} = K_\chi - V_\chi + K_s - V_s - V_{\text{int}}\ ,
\end{equation}
where $K_{\chi}$ is a canonical kinetic term for the DM and $V_\chi$ is its potential energy.\footnote{We stress that, despite the notation, the DM is not assumed here to be a scalar.} On the other hand, $K_{s}$ and $V_{s}$ are the kinetic and potential energy of the fifth force field, and $V_{\text{int}}$ parametrizes the interactions between the two species. To strengthen the analogy with the gravitational potential we define the 5th force kinetic term as $K_s \equiv -\, \partial^\mu s\partial_\mu s /(2G_s)$, where $s$ is dimensionless, like metric perturbations are, and $G_s$ is a dimensionful constant analogous to $G_N$. We thus define the dimensionless ratio
\begin{align}
 \beta \equiv \frac{G_s}{4 \pi G_N}  \,,\label{eq:betadef}
\end{align}
which determines the strength of the 5th force with respect to gravity. The Einstein and Klein-Gordon (KG) equations are written as 
\begin{align}
G^{\mu\nu}=&\;8\pi G_N \sum_{x\, =\, \chi,\, s,\, \ldots}  T^{\mu\nu}_{x}  \ , \label{eq:EE_general}  \\ 
\Box s - G_s \frac{\partial V_s}{\partial s} =&\; G_s \frac{\partial V_{\rm int}}{\partial s} \ ,  \label{eq:KG_general} 
\end{align}
where $\Box s \equiv \frac{1}{\sqrt{-g}}\, \partial_\mu \left( \sqrt{-g}\, g^{\mu\nu} \partial_\nu s \right)$ is the KG operator. Energy-momentum conservation for the sum of the DM and fifth force fluids requires $\nabla_\mu (T^{\mu\nu}_\chi + T^{\mu\nu}_s) = 0\,$, but the individual stress tensors are not conserved. We thus have the freedom to choose which stress tensor the interaction term $V_{\rm int}$ is contributing to. We find it convenient to package the interaction in the DM stress tensor, which makes transparent the appearance of the field dependent mass $m_\chi(s)$ and its relation with the interaction term $V_{\text{int}}$. With this choice we can write $\nabla_\mu T_\chi^{\mu\nu} = - \nabla_\mu T_s^{\mu\nu} = -  (\partial V_{\rm int}/\partial s) \partial^\nu s$, where the DM energy density is given by $\rho_\chi = 2(V_\chi+V_{\rm int})$ assuming the DM to be pressureless (see Appendix~\ref{sec:appendixFields} for a detailed discussion).   

At this point, we pause briefly to discuss the decoupling limit of the 5th force, which is not entirely trivial due to the use of the dimensionless field variable $s$. The rewriting $K_s = -\, \partial^\mu s \partial_\mu s /(8\pi G_N \beta)$ makes it apparent that in the decoupling limit $\beta\to 0$, $s \sim \beta^{1/2}$ is required in order to keep fixed the kinetic term of the 5th force field. Therefore the last piece on the left-hand side of the continuity and Euler equations,~\eqref{eq:continuity} and~\eqref{eq:euler}, vanishes. The decoupling is also manifest in the KG equation~\eqref{eq:KG_general}, where the right-hand side vanishes because $V_{\rm int}$ is at least linear in $s$. To inspect the decoupling in the Einstein equation~\eqref{eq:EE_general} it is useful to recall the structure of the DM stress tensor, 
\begin{equation}
(T_{\chi})_{\mu\nu} = (\mathrm{derivative}\;\mathrm{terms}) - g_{\mu\nu} V_\chi - g_{\mu\nu} V_{\rm int}\,,
\end{equation}
whose last term vanishes for $\beta \to 0$. In summary, in the decoupling limit we recover two non-interacting species (and fluids) with separately conserved stress tensors, as expected.

So far our discussion has been general, but we now focus on the minimal scenarios of interest in this work, where both $V_{\chi}$ and $V_{\rm int}$ are quadratic in $\chi$. We assume $V_\chi$ to contain only a mass term for DM while the DM interaction with the fifth force takes the schematic form $V_{\rm int} \sim \chi^2 F(s)$, with $F(s)$ some function of the 5th force field. As discussed in Appendix~\ref{sec:appendixFields}, in this case the relation $\rho_\chi = 2(V_\chi+V_{\rm int})$ can be written as
\begin{equation} \label{eq:micro_macro}
\frac{\partial V_{\rm int}}{\partial s}  = \rho_\chi \frac{\partial \log m_\chi (s)}{\partial s}\;,
\end{equation}
which provides an explicit map between the microscopic interactions in the field theory picture (the left hand side of the equation) and the fluid description (the right hand side of the equation).

We now write two explicit particle physics models as examples of the microscopic interpretation to the parameters $G_s$ and $m_\chi (s)$. The simplest way to realize a dark fifth force scenario is through a trilinear interaction between the Dark Matter particle $\chi$ and the scalar mediator $\varphi$ (taken to have canonical field dimension). While in this paper we focus on this simple case, we emphasize that our formalism applies to any interaction of the form $V_{\rm int} \sim \chi^2 F(s)$, provided $F(s)$ can be expanded as a power series $\sum_{n\, \geq\, 1} c_n s^n$ where $c_n$ are dimensionless coefficients.

Explicitly we can write 
\begin{align}
   \mathcal{L}_{\rm fermion\;DM} \,=&\, -\frac{1}{2}\overline{\chi} ( \stackrel{\rightarrow}{\slashed{\nabla}} - \stackrel{\leftarrow}{\slashed{\nabla}}) \chi - m_\chi \overline{\chi}\chi -\frac{1}{2} \partial_\mu\varphi \partial^\mu \varphi - V_s(\varphi)- g_D \varphi \overline{\chi}\chi \, ,\label{eq:Lag_fermion}\\
\mathcal{L}_{\rm scalar\;DM} \,=&\, -\frac{1}{2} \partial_\mu\chi \partial^\mu \chi - \frac{1}{2}m_\chi^2\chi^2 -\frac{1}{2} \partial_\mu\varphi \partial^\mu \varphi -V_s(\varphi)- g_D m_\chi \varphi \chi^2 \, , \label{eq:Lag_scalar}
\end{align}
where the DM $\chi$ is a Dirac fermion in Eq.~\eqref{eq:Lag_fermion} or a real scalar in Eq.~\eqref{eq:Lag_scalar}. The non relativistic potential generated in Minkowski space-time by the exchange of the scalar fifth force with the DM particles as external sources can be written as $V(r) = -\, g_D^2 e^{-m_\varphi r}/(4\pi r)$ independently of the DM spin. As a consequence, at distances $r \ll m_\varphi^{-1}$ the long range force is equivalent to a shift in Newton's constant, $G_N\to G_N(1+\beta)$, where $\beta$ was given in Eq.~\eqref{eq:betadef} and we defined 
\begin{equation} \label{eq:param_id}
s = G_s^{1/2}\varphi\,, \qquad G_s = \frac{g_D^2}{m_\chi^2}\,.
\end{equation}
The field dependent DM mass is $m_\chi (s) = m_\chi (1 + s)$ for fermionic DM and $m_\chi(s) = m_\chi \sqrt{1+2 s}\,$ for real scalar DM. 

A non-polynomial dependence of the DM mass on the fifth force field can in principle be realized if $\varphi$ is a non-minimally coupled scalar in Brans-Dicke theories~\cite{Brans:1961sx}, a string theory dilaton~\cite{Damour:1994zq}, or the dilaton of spontaneously broken conformal symmetry~\cite{Komargodski:2011vj}. The cosmological implications of $m_\chi (\varphi) = m_\chi e^{-\beta \varphi/M_{\text{Pl}}}$ for an ultralight scalar were considered already in the seminal papers on coupled quintessence~\cite{Wetterich:1994bg,Amendola:1999er}. In this work we do not consider these scenarios, leaving a detailed analysis for future study.

\subsection{Fifth force potential and naturalness}~\label{sec:naturalness}
As our goal is to study the phenomenology of a new long range attractive force operating in the dark sector, the force mediator $\varphi$ is assumed to have physical mass smaller than or equal to the Hubble scale today and to interact with DM through trilinear interactions like the ones in Eq.~\eqref{eq:Lag_fermion} or Eq.~\eqref{eq:Lag_scalar}. In general, we can consider the renormalizable potential for the scalar fifth force mediator
\begin{equation}
V_s (\varphi) = \frac{1}{2} m^2_{\varphi} \varphi^2 + \frac{c_\varphi}{3}\varphi^3 + \frac{\lambda_\varphi}{4}\varphi^4\,, 
\end{equation}
where naturalness requires its parameters to be as large as the size of the radiative corrections controlled by the interaction strength with DM. Explicitly we find for both scalar and fermionic DM  
\begin{equation}
m^2_{\varphi} \gtrsim \frac{\beta}{(4\pi)^2}\frac{m_\chi^4}{M_{\rm Pl}^2}\,,\qquad  c_\varphi \gtrsim  \frac{\beta^{3/2}}{(4\pi)^2}\frac{m_\chi^4}{M_{\rm Pl}^3}\,,\qquad \lambda_{\varphi} \gtrsim  \frac{\beta^2}{(4\pi)^2}\frac{m_\chi^4}{M_{\rm Pl}^4}\;,\label{eq:naturalness}
\end{equation}
where $M_{\rm Pl} = (8\pi G_N)^{-1/2} = 2.4\times 10^{18}$~GeV is the reduced Planck mass and we used the definitions in Eq.~\eqref{eq:betadef} and Eq.~\eqref{eq:param_id} in the estimates above. For simplicity, in this paper we take $V_s$ to contain just a mass term $m_\varphi^2 \varphi^2/2$. Nonetheless, we have checked that a very similar cosmology is obtained by considering for instance a purely quartic potential with effective coupling $\lambda_\varphi = \beta\hspace{0.2mm} m_\varphi^2 / (\bar{s}^2 M_{\rm Pl}^2)$, where $\bar{s}$ is the typical field value during matter domination (see for example Fig.~\ref{fig:s_bkg_beta0p005}). Taking a linear combination of mass, cubic and quartic would also leave our results qualitatively unchanged. 

We recall here that for a scalar field with physical mass much larger than $H_0$, the purely quadratic and purely quartic potentials result in very different late-time evolutions, in particular for the cycle-averaged background energy density of the scalar, as thoroughly studied in the decoupled case~\cite{Turner:1983he,Poulin:2018cxd,Agrawal:2019lmo}. An extension to include a coupling to DM has been made in Ref.~\cite{Karwal:2021vpk}, where a scalar field with purely quartic potential playing the role of Early DE was considered.\footnote{The formal correspondence between our description and the one in Ref.~\cite{Karwal:2021vpk} is $\{m_\chi (\varphi), \rho_\chi, V_s(\varphi) \} \leftrightarrow \{ A(\phi), \tilde{\rho}_{\rm dm}A(\phi)^4, V(\phi)\}$, so that their perturbation equations in synchronous gauge match ours (reported in Appendix~\ref{app:pert_SG}) after the identifications $\delta_\chi \leftrightarrow \tilde{\delta}_{\rm dm} + 4 (\partial \log A(\phi)/\partial \phi) \delta \phi$ and $\theta_\chi \leftrightarrow \tilde{\theta}_{\rm dm}$.} 

Requiring $m_\varphi \lesssim H_0$ and the naturalness of the fifth force potential in Eq.~\eqref{eq:naturalness} can be read as an upper bound on the DM mass (see also Ref.~\cite{DAmico:2016jbm}),
\begin{equation}\label{eq:naturalness_bound}
m_\chi  \lesssim \beta^{-1/4} \left(4\pi\, m_\varphi M_{\rm Pl}\right)^{1/2} \simeq 0.02\;\mathrm{eV} \left( \frac{0.01}{\beta}\right)^{1/4} \left( \frac{m_\varphi}{H_0} \right)^{1/2}\,,
\end{equation}
where we set as reference for $\beta$ the ballpark value of our cosmological constraints, as derived later in Sec.~\ref{sec:constraints}. Therefore, a relativistic field-theoretical description of the dark fifth force mediator without fine-tuning requires the DM to be an ultralight boson. In this mass region the DM may for instance be an axion, produced via misalignment or other non-thermal mechanisms. While this observation does not affect our analysis of the cosmology, we will return to its implications for particle physics in Sec.~\ref{sec:implications_particle}. One should also keep in mind that the naturalness bound in Eq.~\eqref{eq:naturalness} can in principle be circumvented in non minimal DM models, for instance enforcing supersymmetry in the dark sector. In such non minimal scenarios the DM could be heavier and possibly fermionic.  

Having introduced our theoretical framework, we are now ready to present the background and first order cosmological dynamics in the next two sections. For definiteness, when providing numerical results we always refer to a representative model with scalar DM and purely quadratic fifth force potential, namely $-\mathcal{L} = (\partial_\mu \chi \partial^\mu \chi + m_\chi^2 (s) \chi^2)/2 + M_{\rm Pl}^2(\partial_\mu s \partial^\mu s + m_\varphi^2 s^2)/\beta$ with $m_\chi (s) = m_\chi \sqrt{1+2s\,}$. However, we emphasize again that our findings have little dependence on the DM spin and the precise form of the $\varphi$ potential.

\section{Cosmological Background}\label{sec:bkg}

At the background level, the momentum of the $\chi$ particles redshifts with the expansion of the Universe like the one of all other particles, $\dot{p}^i = -H p^i$. This guarantees the existence of a frame where the spatial part of the fluid 4-velocity of each species is zero, hence homogeneity and isotropy are preserved. We thus split each variable in a background component plus a perturbation around it, with the former depending only on time. For example $\rho_\chi = \bar{\rho}_\chi(t) + \delta \rho_\chi (\vb{x},t)$ and $s = \bar{s}(t) + \delta s(\vb{x},t)$, and so on for the other species.

The equation of motion for the background DM density is directly obtained from the continuity equation~\eqref{eq:continuity}. Switching to conformal time $\tau$ we have
\begin{align}\label{eq:chi_bkg}
    \bar{\rho}^\prime_\chi + 3 \mathcal{H}\bar{\rho}_\chi = \bar{\rho}_\chi \frac{\partial \log m_\chi(s)}{\partial s} \bar{s}^\prime\,,
\end{align}
where $()'$ denotes derivatives with respect to $\tau$ and $\mathcal{H} = a'/a$ is the conformal Hubble parameter. The field dependent mass and its derivatives are, here and throughout the text, evaluated in the background. Eq.~\eqref{eq:chi_bkg} can also be derived by writing in the non-relativistic limit $\bar{\rho}_\chi = m_\chi (s) \bar{n}_\chi$ and imposing conservation of the comoving number density, $\bar{n}^\prime_\chi + 3 \mathcal{H} \bar{n}_\chi = 0$. We see that while the number density of DM scales proportionally to the volume (i.e. $\bar{n}_\chi\sim a^{-3}$), the energy density does not, unless $\bar{s}$ is constant. As we show below, the KG equation forces $\bar{s}' < 0$ throughout the cosmological history so that the right-hand side of Eq.~\eqref{eq:chi_bkg} is negative and non-zero, corresponding to a net energy transfer from the DM to the 5th force field which modifies the redshift behavior of the DM energy density. 

The KG equation for the scalar field is 
\begin{align}
    \bar{s}^{\prime\prime} + 2 \mathcal{H} \bar{s}^\prime + G_s a^2  V_{s,s} + G_s  a^2  \bar{\rho}_\chi \frac{\partial \log m_\chi(s)}{\partial s} = 0\,,
\end{align}
where $V_{s,s} \equiv \partial V_s/\partial s$, and for the remainder of this paper by $V_s$ we always indicate the self interaction of $s$ in the background, and similarly for its derivatives. The above equation is derived starting from the Lagrangian for the fields, Eq.~\eqref{eq:KG_general}, and rewriting the source term as a function of the $\chi$ fluid variables according to Eq.~\eqref{eq:micro_macro}. At the background level the energy density and pressure of the 5th force mediator are
\begin{align}\label{eq:rho_p_s}
    \bar{\rho}_s = \frac{(\bar{s}^\prime)^2}{2G_s a^2} + V_s\,,\qquad \overline{\mathcal{P}}_s = \frac{(\bar{s}^\prime)^2}{2G_s a^2} - V_s\,,
\end{align}
satisfying the equation of motion
\begin{align}\label{eq:fluid_s}
    \bar{\rho}^\prime_s + 3 \mathcal{H}\bar{\rho}_s(1 + w_s) = -\, \bar{\rho}_\chi \frac{\partial \log m_\chi(s)}{\partial s} \bar{s}^\prime\,,
\end{align}
where $w_s \equiv \overline{\mathcal{P}}_s/\bar{\rho}_s\,$ is the equation of state parameter of the 5th force fluid. The sum of this equation with~\eqref{eq:chi_bkg} is identically zero on the right hand side, reflecting the energy conservation for the total fluid.

\subsection{New cosmological parameters}
The background evolution depends on three new parameters beyond the ones of $\Lambda$CDM: i) the dimensionless quantity $\beta$ as defined in Eq.~\eqref{eq:betadef} determines the coupling strength of the 5th force in units of $G_N\,$; ii) the mass of the 5th force mediator $m_\varphi$ is assumed to be smaller than $H_0\,$ in order for the interaction to be long range till today; iii) the dimensionless initial background value of the 5th force field $\bar{s}_{\rm ini}$ controls the size of its contribution to the vacuum energy, together with $m_\varphi$ and $\beta$.\footnote{The initial value of the derivative $\bar{s}^\prime_{\rm ini}$ is also required to solve the KG equation, but this is determined as a function of the other input parameters as discussed in Sec.~\ref{sec:bkg_behavior}.}  

We consider two physically distinct scenarios where the scalar field $s$ plays different roles in the cosmological history:
\begin{itemize}
\item {\bf Pure Fifth Force (5F):} the scalar field $s$ leaves observable imprints only by mediating a long range force between DM particles, whilst its energy density $\bar{\rho}_s$ remains negligible throughout the cosmological history. In this case we set $\bar{s}_{\rm ini}$ to a very small value (for definiteness we choose $\bar{s}_{\rm ini} = 10^{-4}$) and determine the value of the CC density parameter $\omega_{\Lambda}$ that satisfies the closure condition today, $\sum_i \omega_i^0 = h^2$. In the 5F scenario the evolution of the $s$ background undergoes the phases {\bf A} and {\bf B} described in Sec.~\ref{sec:bkg_behavior}.

\item {\bf Coupled Dark Energy (CDE):} in addition to mediating a 5th force, $s$ also accounts for the DE. In this case we set the CC to zero, $\omega_\Lambda = 0$, and fix the initial field value $\bar{s}_{\rm ini}$ by imposing the closure condition $\sum_i \omega_i^0 = h^2$. The fine-tuning of the CC is thus replaced by the fine-tuning of the initial conditions for $s$. The required initial field value is always much larger than in the case of a pure 5th force: assuming a quadratic potential $V_s$, for a scalar mass $m_\varphi\sim H_0$ one finds $\bar{s}_{\rm ini} \sim \mathcal{O}(1)$ whereas for smaller masses the initial condition scales as $\bar{s}_{\rm ini}\sim H_0/m_\varphi$.\footnote{Since $\bar{s}$ undergoes only a modest relative change across cosmic time, the vacuum energy can be estimated directly from its initial value. Requiring the scalar to account for the DE today imposes $(V_s)_{\rm ini} = m_\varphi^2 M_{\rm Pl}^2 \bar{s}_{\rm ini}^2/\beta \sim 3 H_0^2 M_{\rm Pl}^2\,$, from which the scaling $\bar{s}_{\rm ini}\sim H_0/m_\varphi$ follows.} In the CDE scenario the $s$ background evolution consists of phases {\bf A}, {\bf B}, and {\bf C} as detailed in Sec.~\ref{sec:bkg_behavior}.
\end{itemize}
Finally, the energy density of the $\chi$ particles, $\omega_\chi$, replaces the one of standard CDM. For practical reasons, Boltzmann codes like CLASS~\cite{Lesgourgues:2011re,Blas:2011rf} take as input the present day value of the energy density of each species, $\omega_i^0$, and evolve it back to the desired initial redshift using the well known behavior with the scale factor of CDM, baryon, photons, etc. This is not possible in our case, because we do not know a priori the exact time dependence of $\omega_\chi$ and $\omega_s$. For consistency with the other species, we use as input parameter the energy density that the $\chi$ particles would have had they evolved with $a^{-3}$ like CDM. We dub this parameter $\widetilde{\omega}_d$ and set the initial density of $\chi$ as 
\begin{equation}
\omega_\chi(a_{\rm ini}) \equiv \omega_\chi^{\rm ini} = \widetilde{\omega}_d\, a_{\rm ini}^{-3}\,.
\end{equation}
Clearly, $\omega_{\chi}^0 \neq \widetilde{\omega}_d$ and the procedures described above enforce flatness.

\subsection{Background behavior}\label{sec:bkg_behavior}
The background field $\bar{s}$ exhibits a remarkably rich cosmological history, consisting of several successive epochs characterized by distinct dynamics. Its evolution directly impacts the one of the DM density, as can be seen by solving Eq.~\eqref{eq:chi_bkg}:
\begin{equation} \label{eq:om_chi_formal}
\frac{\omega_\chi (a)}{\widetilde{\omega}_d} = \frac{m_{\chi} (\bar{s})}{m_{\chi}(\bar{s}_{\rm ini})}\, a^{-3}\,.
\end{equation}
Since $\bar{s}$ is a monotonically {\it decreasing} function of time,\footnote{ This applies generally during the radiation and matter dominated eras, when the evolution is driven by the interaction with DM. That it also holds during the DE dominated epoch depends on our choice of quadratic potential for $s$.} $\chi$ redshifts {\it faster} than CDM. We now present an analytical understanding of the solutions across cosmic time. The analytical approximations we provide are in very good agreement with numerical results, which are presented in Sec.~\ref{eq:numerics_background} below.

\subsubsection*{A) Radiation dominated era: $\bar{\rho}_s \propto a^{-2}$}
\begin{figure}
    \centering
    \includegraphics[width = 0.49\textwidth]{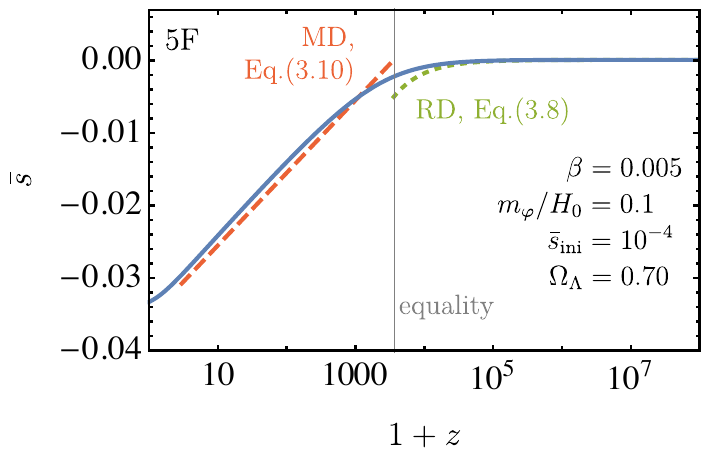}\hspace{1mm}
     \includegraphics[width = 0.48\textwidth]{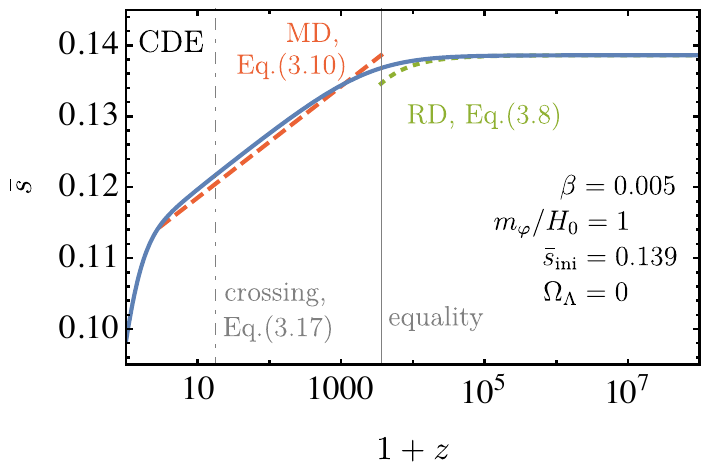}
    \caption{\emph{(Left)} The time evolution of $\bar{s}$ in a pure 5th force scenario, assuming $\beta = 0.005$. In the analytical solution for the MD epoch, Eq.~\eqref{eq:s_MD}, we have set $\bar{s}_{\rm eq} = \bar{s}_{\rm ini}\,$. \emph{(Right)} The same as in the left panel, but now assuming $s$ is a Coupled Dark Energy field. The vertical dot-dashed line corresponds to the redshift at which the $s$ equation of state parameter crosses zero as it transitions from $w_s = +1$ to $w_s = -1$, derived analytically in Eq.~\eqref{eq:crossing}.}
    \label{fig:s_bkg_beta0p005}
\end{figure}

\begin{figure}
    \centering
    \includegraphics[width = 0.49\textwidth]{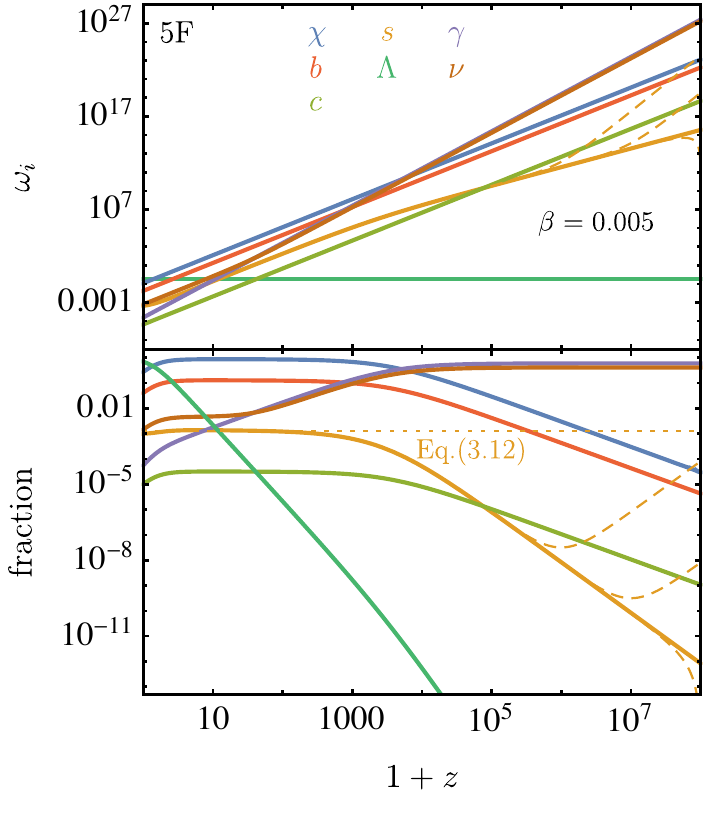}\hspace{1mm}
     \includegraphics[width = 0.49\textwidth]{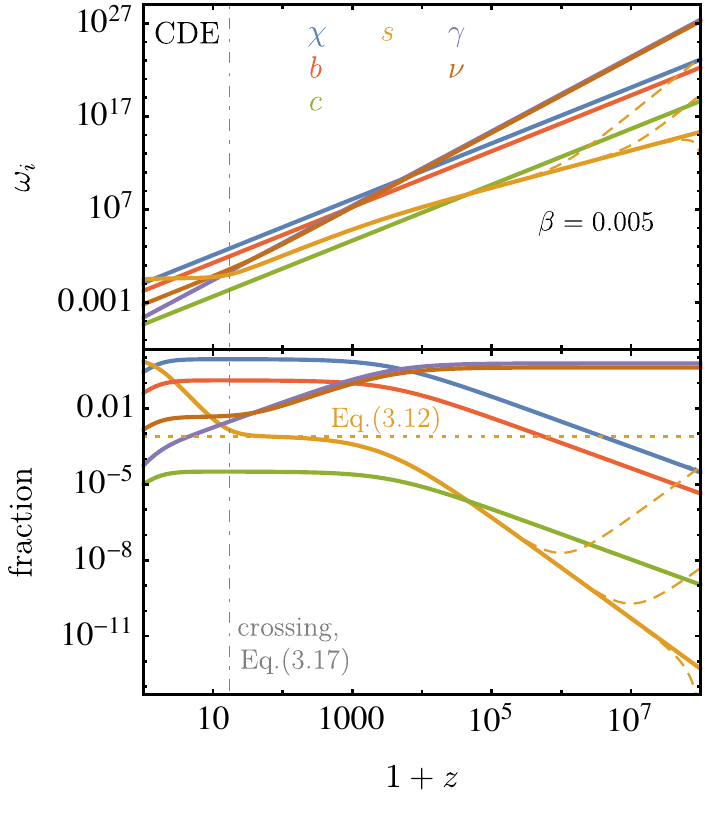}
    \caption{\emph{(Left)} In the top panel, the time evolution of the energy density of the different species in a pure 5th force scenario, assuming $
    \beta = 0.005$. The bottom panel shows the fractional contributions. For the scalar field $s$, the light orange dotted line shows the analytical estimate of $\bar{\rho}_s/\bar{\rho}_{\rm tot}$ during early MD, as found in Eq.~\eqref{eq:rhos_MD}. In both panels, light orange dashed curves show the scalar field evolution starting from initial values of $\bar{s}'$ different from the natural expectation determined by Eq.~\eqref{eq:s_RD}: from bottom to top, $\bar{s}^\prime_{\rm ini} = \{0, 10^2, 10^4\} (\bar{s}_{\rm ini}^\prime)_{\rm natural}\,$. \emph{(Right)} The same as in the left panels, but now assuming $s$ is a Coupled Dark Energy field.}
    \label{fig:rho_bkg_beta0p005}
\end{figure}

\begin{figure}
    \centering
    \includegraphics[width = 0.49\textwidth]{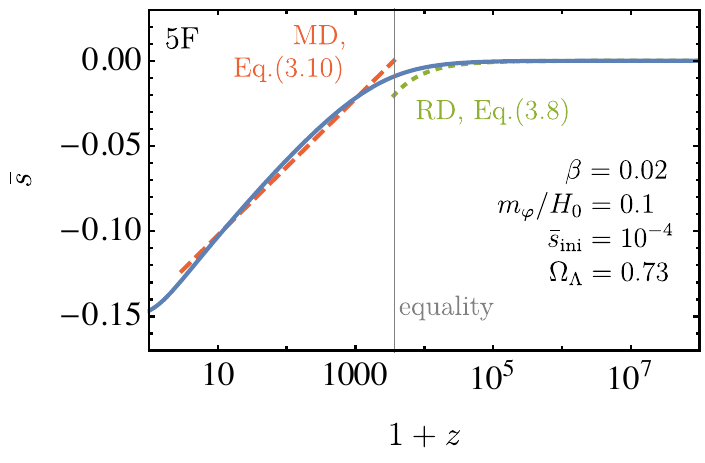}\hspace{1mm}
     \includegraphics[width = 0.48\textwidth]{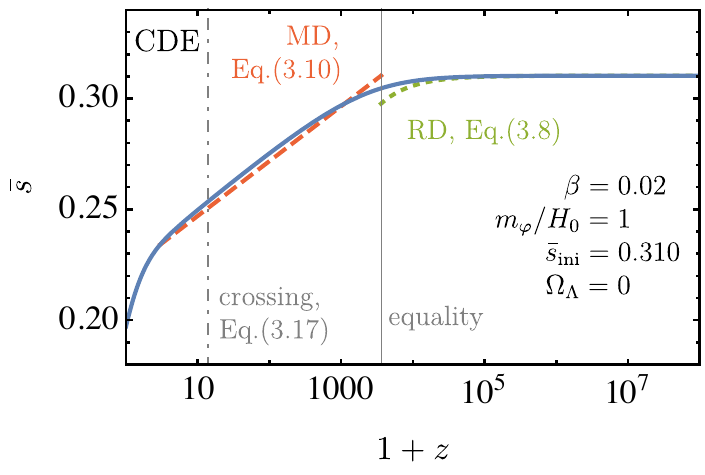}
    \caption{Same as Fig.~\ref{fig:s_bkg_beta0p005}, but for $\beta = 0.02$.}
    \label{fig:s_bkg_beta0p02}
\end{figure}

During radiation domination (RD, where $\mathcal{H}=1/\tau$ at zeroth order in $\beta$) the $s$ self-interaction can be neglected in the KG equation, $V_{s,s} \ll \bar{\rho}_\chi (\partial \log m_\chi (s)/\partial s)$. We proceed perturbatively in $\beta$, making the approximations
\begin{equation}\label{eq:anal_approx}
\frac{\partial \log m_\chi (s)}{\partial s} = \mathrm{constant}\,,\qquad \Omega_\chi (a) = \widetilde{\Omega}_d\hspace{0.2mm} a^{-3}\,, \end{equation}
where $\widetilde{\Omega}_d = \widetilde{\omega}_d h^{-2}$. The KG equation is then solved by
\begin{equation}\label{eq:s_RD}
\bar{s} \simeq \bar{s}_{\rm ini} - \frac{3\beta}{4} \frac{\partial \log m_\chi (s)}{\partial s} \frac{\widetilde{\Omega}_d}{(\Omega_r^{0})^{1/2}} H_0 (\tau - \tau_{\rm ini})\,, \qquad (\mathrm{RD})
\end{equation}
where a solution scaling as $\tau^{-1}$ has been discarded, and we have used the leading-order expression $a = (\Omega_r^{0})^{1/2} H_0 \tau$, with $\Omega_r^0 = \Omega_\gamma^0 + \Omega_\nu^0$ if neutrino masses are neglected. Quantitatively, the linear dependence on $a$ implies that $\bar{s}$ remains approximately constant until matter-radiation equality, as seen in Figs.~\ref{fig:s_bkg_beta0p005} and~\ref{fig:s_bkg_beta0p02}. However, several insights can be extracted from Eq.~\eqref{eq:s_RD}. First, it determines the natural value of $\bar{s}'_{\rm ini}$ as a function of the other input parameters. Second, via Eq.~\eqref{eq:rho_p_s} it establishes that for the very small potentials $V_s$ relevant to this work, $w_s = +1$ throughout RD. However, the energy density does not scale as $a^{-6}$, as it would for a decoupled field with $w = +1$, but redshifts as $\bar{\rho}_s \propto a^{-2}$ instead. Consistency with the fluid equation~\eqref{eq:fluid_s} then requires the following quantity to be a constant of motion,
\begin{equation}\label{eq:doom}
    \frac{\bar{\rho}_\chi (\partial \log m_\chi (s)/\partial s) \bar{s}'}{3\mathcal{H} \bar{\rho}_s(1 + w_s)} = -\, \frac{2}{3}\;,\qquad (\mathrm{RD})
    \end{equation}
%\end{equation}
which can be easily checked using the approximate analytical solution and is verified numerically to high accuracy. The fraction of the total energy density stored in $s$ grows rapidly as $\bar{\rho}_s/\bar{\rho}_{\rm tot} \propto a^2$ until matter-radiation equality. This solution is in fact an attractor, to which trajectories with initial velocity $\bar{s}'_{\rm ini}$ larger or smaller than the one fixed by Eq.~\eqref{eq:s_RD} rapidly converge. Such trajectories correspond to a nonzero coefficient for the $\tau^{-1}$ solution that was discarded in Eq.~\eqref{eq:s_RD}. If the coefficient is negative, $\bar{s}'_{\rm ini}$ is larger than the natural value and $\bar{\rho}_s$ undergoes an initial period of kination with $a^{-6}$ scaling before converging to the attractor. If the coefficient is positive, there is an initial phase of kination that ends when $\bar{\rho}_s$ reaches zero, after which the trajectory very quickly converges to the attractor. These features are clearly visible in Figs.~\ref{fig:rho_bkg_beta0p005} and~\ref{fig:rho_bkg_beta0p02}. 

\subsubsection*{B) Early matter dominated era: $\bar{\rho}_s \propto a^{-3}$}
At least for the early part of matter domination (MD, where $\mathcal{H}=2/\tau$ at leading order in $\beta$), the self interaction potential $V_s$ remains negligible both in the KG equation and in the $s$ equation of state. Making again the approximations in Eq.~\eqref{eq:anal_approx}, the KG equation is solved by
\begin{equation}\label{eq:s_MD}
\bar{s} \simeq \bar{s}_{\rm eq} - 2\beta  \frac{\partial \log m_\chi (s)}{\partial s} f_\chi  \log\frac{\tau}{\tau_{\rm eq}}\,, \qquad (\mathrm{MD})
\end{equation}
where we defined 
\begin{equation}
f_\chi \equiv \frac{\bar{\rho}_\chi}{\bar{\rho}_m} \simeq \frac{\widetilde{\Omega}_d}{\widetilde{\Omega}_d + \Omega_b^0}\ ,\label{eq:deffchi}
\end{equation}
with the last equality holding at zeroth order in $\beta$. A solution scaling as $\tau^{-3}$ has been discarded in order to match to the evolution in RD and the leading-order expression of the scale factor $a = (\widetilde{\Omega}_d + \Omega_b^0) H_0^2 \tau^2/4$ has been used. 

Since $w_s = +1$, one immediately obtains $\bar{\rho}_s \propto a^{-3}$ and a simple estimate for the (constant) fraction of energy density in $s$ (see also Ref.~\cite{Amendola:1999er}),
\begin{equation}\label{eq:rhos_MD}
\frac{\bar{\rho}_s}{\bar{\rho}_{\rm tot}} \simeq \frac{\beta}{3}\left( \frac{\partial \log m_\chi (s)}{\partial s} f_\chi \right)^2\,. \qquad (\mathrm{MD})
\end{equation}
In addition, we derive perturbative solutions up to $\mathcal{O}(\beta)$ for Hubble and the DM density fraction, focusing on the 5F scenario where $\bar{s}_{\rm eq} \approx \bar{s}_{\rm ini} \approx 0$ in Eq.~\eqref{eq:s_MD}. The second Friedmann equation is written as
\begin{align}
a'' = \frac{4\pi G_N }{3} a^3 (\bar{\rho}_{\rm tot} - 3 \overline{\mathcal{P}}_{\rm tot}) \simeq  (\widetilde{\Omega}_d + \Omega_b^0) \frac{H_0^2}{2} \left[1 - 2 \beta f_\chi^2 \bigg( \frac{\partial \log m_\chi (s)}{\partial s} \bigg)^2 \bigg( \log \frac{\tau}{\tau_{\rm eq}} + \frac{1}{3} \bigg)  \right]\,, 
\end{align}
and it is solved by
\begin{align}
    a =  \frac{1}{4} (\widetilde{\Omega}_d + \Omega_b^0) H_0^2 \tau^2 \bigg(1 - 2 \beta f_\chi^2  \bigg( \cfrac{\partial \log m_\chi (s)}{\partial s} \bigg)^2  \log \frac{\tau}{\tau_{\rm eq}} \bigg)\,,\qquad (\mathrm{MD, 5F}) \label{eq:a_MD_5F}
\end{align}
where non-log-enhanced terms have been dropped, since they are numerically negligible. Thus,
\begin{equation} \label{eq:Hubble_MD}
H(a) = \frac{a'}{a^2} = H_0 (\widetilde{\Omega}_d + \Omega_b^0)^{1/2} a^{-3/2}\bigg( 1 - \frac{\beta}{2}f_\chi^2 \bigg( \frac{\partial \log m_\chi (s)}{\partial s} \bigg)^2 \log \frac{a}{a_{\rm eq}} \bigg)\,.\quad (\mathrm{MD, 5F}) 
\end{equation}
We also exploit Eq.~\eqref{eq:om_chi_formal} to write a perturbative expression for the DM density fraction,
\begin{equation}\label{eq:omchi_MD}
\frac{\omega_\chi (a)}{\widetilde{\omega}_d\, a^{-3}} = 1 - \beta f_\chi  \bigg( \frac{\partial \log m_\chi (s)}{\partial s} \bigg)^2  \log \frac{a}{a_{\rm eq}}\,.  \qquad  (\mathrm{MD, 5F})  
\end{equation}
The appearance of the logarithms in Eqs.~\eqref{eq:Hubble_MD} and~\eqref{eq:omchi_MD} implies that the deviation from $\Lambda$CDM of the background evolution is parametrically enhanced  by a factor $\lesssim \log\, (1 + z_{\rm eq}) \approx 8$ with respect to the naively expected size of $\mathcal{O}(\beta)$. The log-enhanced effects encoded by these formulae match well the numerical results, as shown in Fig.~\ref{fig:hubble}. 

\begin{figure}
    \centering
    \includegraphics[width = 0.49\textwidth]{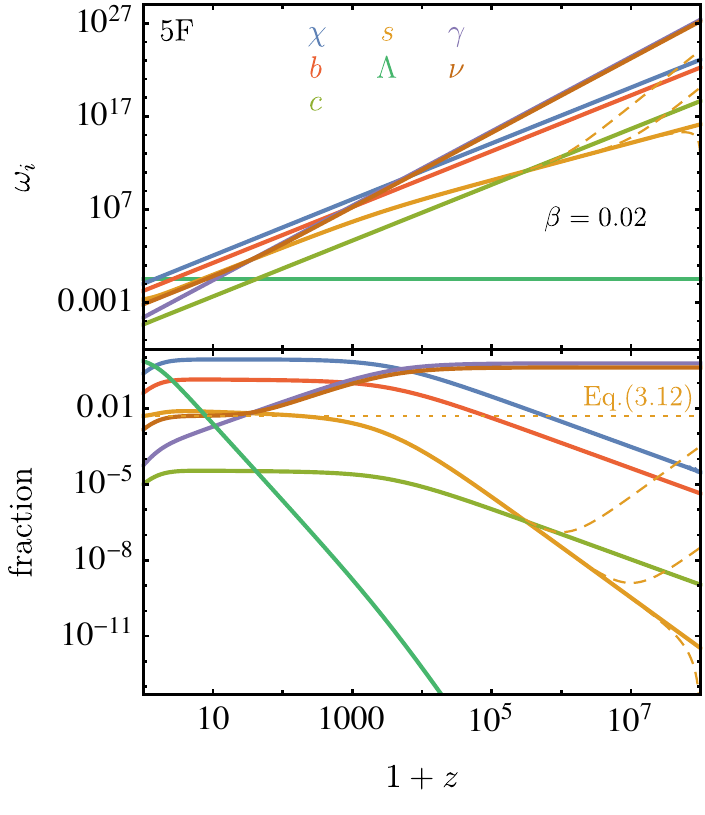}\hspace{1mm}
     \includegraphics[width = 0.49\textwidth]{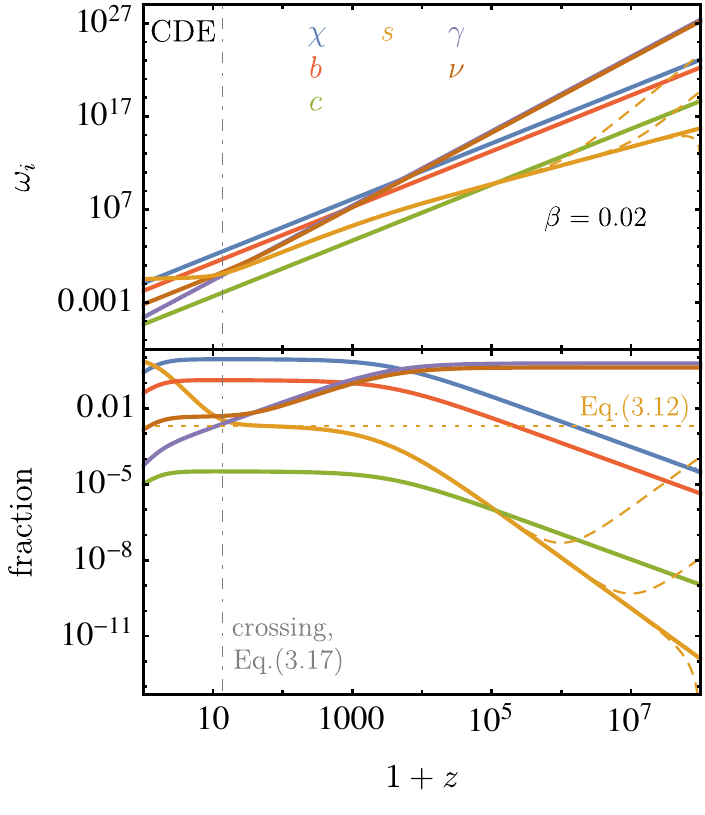}
    \caption{Same as Fig.~\ref{fig:rho_bkg_beta0p005}, but for $\beta = 0.02$.}
    \label{fig:rho_bkg_beta0p02}
\end{figure}

If $s$ is a pure 5th force field, the above solutions persist until the CC comes to dominate the total energy density of the Universe. In this final phase of CC domination the dynamics is modified, but Eq.~\eqref{eq:rhos_MD} still provides a good estimate of the $s$ energy density fraction today, as can be seen in the left panels of Figs.~\ref{fig:rho_bkg_beta0p005} and~\ref{fig:rho_bkg_beta0p02}.

\subsubsection*{C) Late matter dominated era: $\bar{\rho}_s \simeq \mathrm{constant}\,$}
\begin{figure}
    \centering
    \includegraphics[width = 0.45\textwidth]{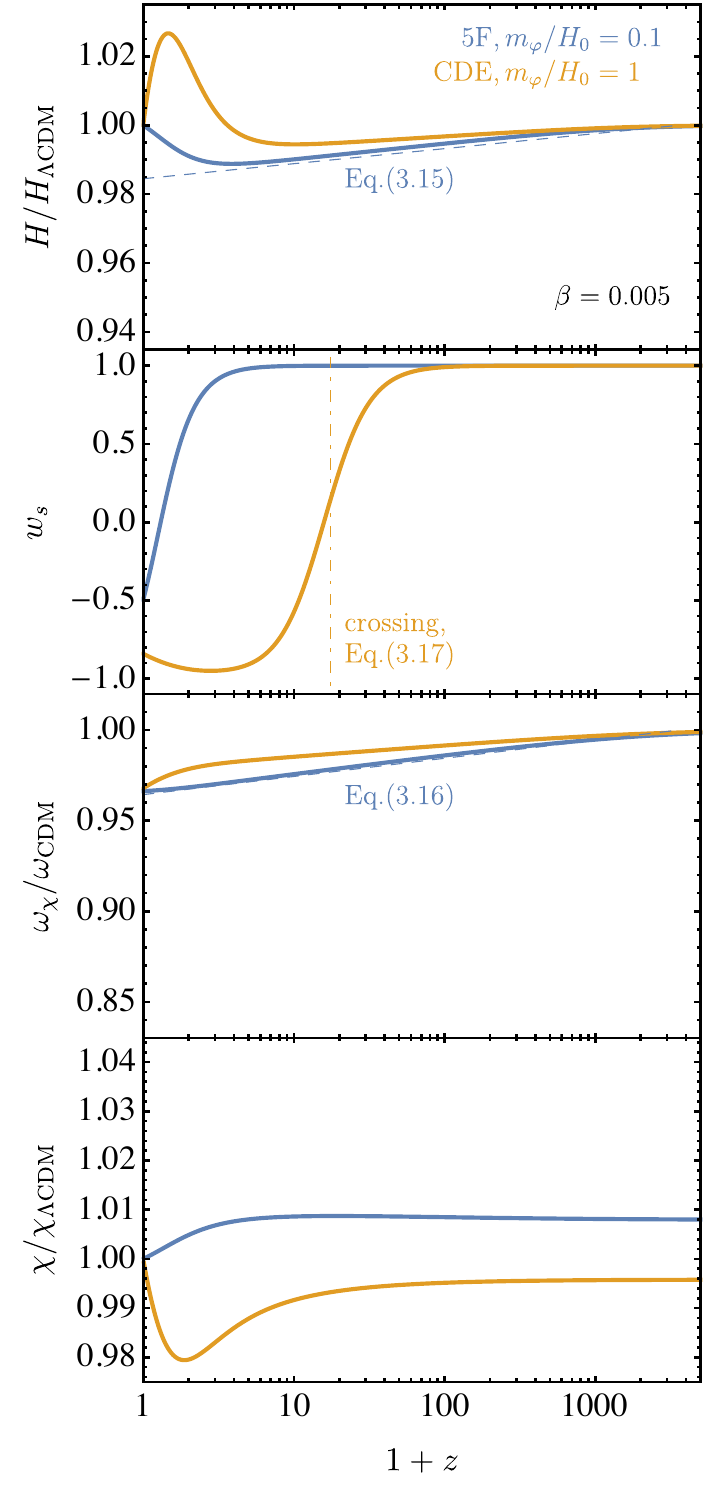}\hspace{2mm}
     \includegraphics[width = 0.45\textwidth]{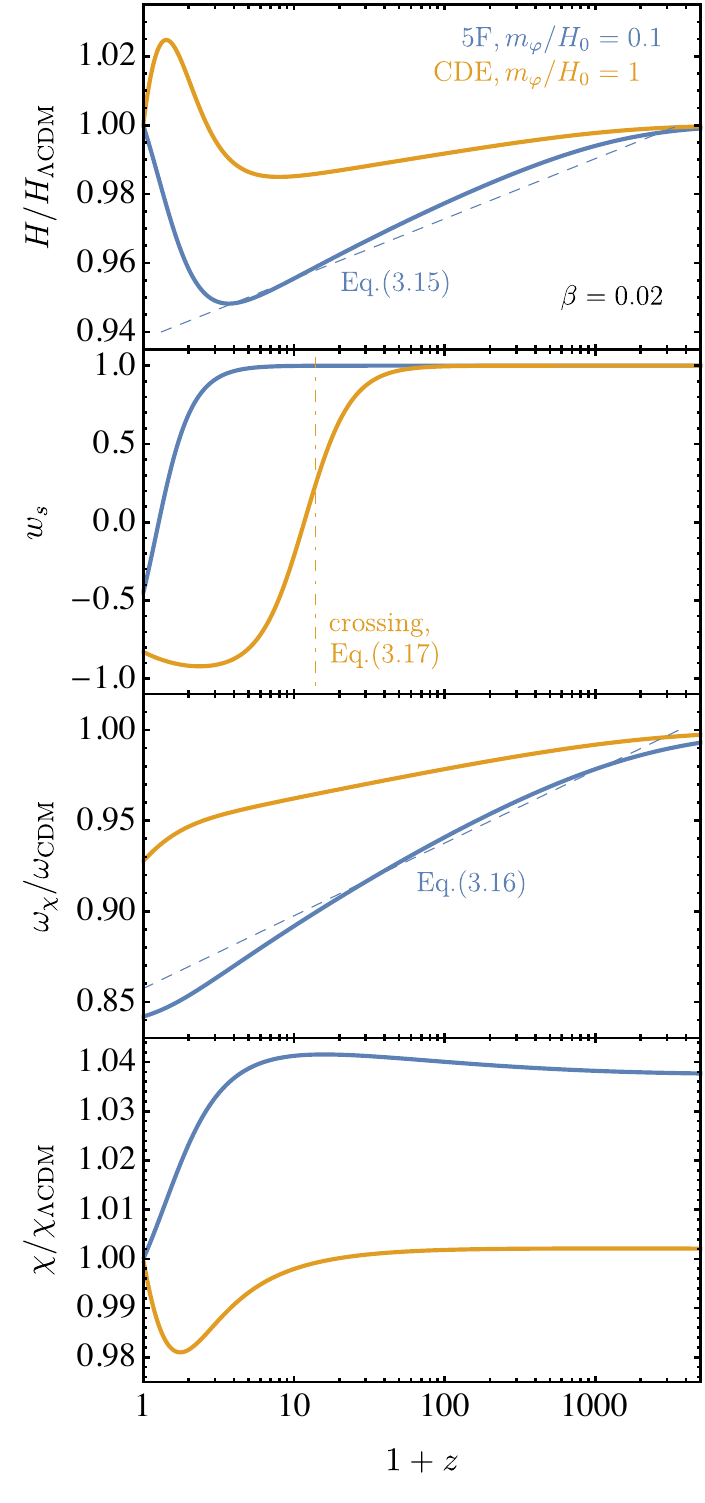}
    \caption{{\it (Left)} From top to bottom: Hubble normalized to $\Lambda$CDM, equation of state parameter for the 5th force field, DM density $\omega_\chi$ normalized to CDM, and comoving distance normalized to $\Lambda$CDM, for $\beta = 0.005$. We focus on the time evolution from matter-radiation equality onwards. Dashed and dot-dashed lines correspond to analytical approximations discussed in the text. {\it (Right)} Same as in the left panels, but for $\beta = 0.02$.}
    \label{fig:hubble}
\end{figure}

If $s$ is a coupled quintessence field the above discussion applies, but at some time during MD the equation of state eventually changes from $w_s = +1$ to $w_s \simeq -1$, crossing zero when the equality $(\bar{s}')^2 = 2 G_s a^2 V_s$ is satisfied. For the assumed quadratic form of $V_s$, this corresponds to
\begin{equation}\label{eq:crossing}
\log a_{\rm cross} \simeq \frac{2}{3}\log \left( \frac{\beta}{\bar{s}_{\rm ini}} \frac{\partial \log m_\chi (s)}{\partial s} \frac{\widetilde{\Omega}_d}{(\widetilde{\Omega}_d + \Omega_b^0)^{1/2}} \frac{H_0}{m_\varphi}\right). \quad (w_s\; \mathrm{crosses}\;0\,,\; \mathrm{CDE}\mathrm)
\end{equation}
Since $\bar{s}' \propto \beta$, increasing the 5th force coupling strength delays this crossing (i.e.~it decreases $|\log a_{\rm cross}|$) if all other parameters are kept fixed. On the other hand, in the limit $m_\varphi \ll H_0$ one would find the scaling $1 \ll \bar{s}_{\rm ini} \propto \sqrt{\beta} H_0 / m_\varphi$, leading to $a_{\rm cross} \propto (m_\varphi/H_0)^{2/3}$. Hence, a smaller $m_\varphi/H_0$ corresponds to an earlier crossing. Equation~\eqref{eq:crossing} is an excellent estimate of the time when the transition takes place, as demonstrated by the second row of panels in Fig.~\ref{fig:hubble}.
%Conversely, increasing $m_\varphi/H_0$ makes the crossing happen earlier, as the potential dominates sooner in Eq.~\eqref{eq:rho_p_s}. 

After the transition to $w_s \simeq - 1$, initially the potential term is still negligible in the KG equation and the approximate solution~\eqref{eq:s_MD} still applies, giving $\bar{\rho}_s \simeq m_\varphi^2 \bar{s}^2/ (2G_s) \simeq \mathrm{const}$, modulo a logarithmic evolution. Eventually, the potential energy dominates in the KG equation too and $s$ effectively becomes a decoupled quintessence field. This new regime, however, does not modify the scaling of the $s$ background energy density because the fluid equation~\eqref{eq:fluid_s} now has the decoupled solution $\bar{\rho}_s \propto a^{-3(1 + w_s)} \simeq \mathrm{const}\,$, since $w_s \simeq -1$.

Finally, once $\Omega_s$ becomes of $\mathcal{O}(1)$ the dynamics follows well-known solutions from (decoupled) quintessence models, see e.g. Ref.~\cite{Tsujikawa:2013fta} for a review.

\subsection{Numerical results}\label{eq:numerics_background}

Numerical solutions for background quantities are obtained through implementation in the Einstein-Boltzmann solver CLASS~\cite{Lesgourgues:2011re,Blas:2011rf}. In all cases the input parameters include $\{\beta,\, m_\varphi/H_0,\,\widetilde{\Omega}_d\}$ beyond the $\Lambda$CDM ones. Two distinct shooting procedures are applied depending on the scenario considered: for pure 5th force, $\bar{s}_{\rm ini}$ is set to $10^{-4}$ and $\omega_\Lambda$ is determined by imposing the closure condition today; for Coupled Dark Energy, $\omega_\Lambda$ is set to zero and $\bar{s}_{\rm ini}$ is determined from the closure requirement. The background solutions have also been cross-checked using a {\it Mathematica} code where the evolution equations were transformed into a system of first-order ODE, as done in Ref.~\cite{Baldi:2010vv} by generalizing previous methods~\cite{Copeland:1997et}. 

Figures from~\ref{fig:s_bkg_beta0p005} to~\ref{fig:hubble} show the evolution of background quantities, focusing on four benchmark cosmologies (beside $\Lambda$CDM): for the case of 5F we set $m_\varphi/H_0 = 0.1$, taking $\beta = 0.005$ or $0.02$; for CDE we take $m_\varphi/H_0 = 1$ and again $\beta = 0.005$ or $0.02$. Anticipating the results of Sec.~\ref{sec:constraints}, the smaller choice $\beta = 0.005$ corresponds approximately to the $95\%$ c.l. upper bound we have set in the 5F scenario using Planck and BAO data. On the other hand, the stronger coupling $\beta = 0.02$ is well within our bounds in the CDE scenario and is instructive for the 5F case as well, where it qualitatively represents the larger values $\beta \sim 0.01$ that may resolve the Hubble tension. For concreteness, in this section we set all standard $\Lambda$CDM parameters to their Planck best fit values \cite{Planck:2018vyg}. In addition, we take $\widetilde{\Omega}_d = 0.27$. The same benchmarks are also used in Sec.~\ref{sec:fluctuations} when presenting the dynamics of cosmological perturbations.

The second row of Fig.~\ref{fig:hubble} confirms that in the CDE scenario the transition from $w_s = +1$ to $w_s \approx -1$ takes place during late MD, as predicted by Eq.~\eqref{eq:crossing}. Furthermore, a departure from $w_s = + 1$ is observed in the 5F scenario as well, during the CC-dominated phase. This is a consequence of the choice $m_\varphi/H_0 = 0.1$; for smaller mass of the scalar, $w_s \approx +1$ would persist until today. Finally, the last row of Fig.~\ref{fig:hubble} shows the comoving distance $\chi(z) = \int_0^z dz'/H(z')$, normalized to $\Lambda$CDM. For 5F we observe the expected high-$z$ enhancement, while for CDE the comoving distance is close to its standard value.

The numerical implementation of the background equations also reveals the presence of an upper bound on $\beta$ in the 5F scenario, simply coming from the existence of a solution to Einstein's equations until $z=0$. For cosmological parameters close enough to the Planck $\Lambda$CDM best fit values, increasing $\beta$ to $\mathcal{O}(0.1)$ reduces so much the energy density in the $\chi$ fluid that the Universe never arrives at a CC dominated phase. The scalar field $s$ becomes dominant and quickly overcloses the Universe before any accelerated expansion can begin. This result highlights the importance of consistently including the effects of new degrees of freedom both at the level of the background and of the perturbations, even though the latter are naively considered to be most affected by the new dynamics. We will see another example of the importance of a consistent background evolution in Sec.~\ref{sec:sub_hor}.

An analogous bound does not exist in the CDE case. A larger $\beta$ is always compensated by a larger initial value of $\bar{s}$, yielding in all cases a smooth transition from a matter dominated to a CDE dominated Universe.

\section{Cosmological Perturbations}\label{sec:fluctuations}
Similarly to the cosmological background, the equations of motion at linear order in perturbation theory can be derived from the general equations for the DM fluid discussed in Sec.~\ref{sec:particles} and the ones for the scalar fifth force and the metric in Sec.~\ref{sec:fields}. Expanding to first order the continuity equation \eqref{eq:continuity} and Euler equation \eqref{eq:euler} for the DM fluid we obtain
\begin{align} \label{eq:delta_chi_conf}
\delta_{\chi}^{\prime} + \theta_\chi + 3 \Phi^\prime - \frac{\partial \log m_\chi (s)}{\partial s} \,\delta s^{\prime} - \frac{\partial^2 \log m_\chi (s)}{\partial s^2} \bar{s}^{\prime} \delta s  \;=&\; 0\,, \\
\label{eq:v_chi_conf} 
\theta_\chi^{\prime} + \Big( \mathcal{H} + \frac{\partial \log m_\chi (s)}{\partial s}\, \bar{s}^{\,\prime} \Big) \theta_\chi - k^2 \Big( \Psi + \frac{\partial \log m_\chi (s)}{\partial s}\, \delta s\Big) \;=&\; 0\,,
\end{align}
where $\nabla^i \to i k^i$ in Fourier space and we have made the definitions $\delta_x \equiv \delta \rho_x/\bar{\rho}_x$ and $\theta_x \equiv i v_x^i k_i\,$. The KG equation for the scalar fifth force in Eq.~\eqref{eq:KG_general} expanded at first order is\footnote{The term containing $\Psi$ in the second line appears to be missing from Eq.~(5.8) of Ref.~\cite{Saracco:2009df}.}
\begin{align} 
\label{eq:KG_first}
\delta s^{\prime\prime}+ 2 \mathcal{H} \delta s^\prime +k^2  \delta s +& \bar{s}^{\prime} (3 \Phi^{\prime} - \Psi^{\prime}) + \, a^2 G_s \big( V_{s,,s}\, \delta s + 2\, V_{s,s} \Psi \big)    \\   & \,+\, a^2 G_s \bar{\rho}_\chi \frac{\partial \log m_\chi (s)}{\partial s} (\delta_\chi + 2 \Psi)  + a^2 G_s \bar{\rho}_\chi \frac{\partial^2 \log m_\chi (s)}{\partial s^2} \delta s  = 0 \;.  \nonumber
\end{align} 
As we have seen in the previous section, at the background level the presence of the scalar fifth force impacts the Hubble flow in a highly non-trivial way, which also depends on whether the scalar is accounting or not for the DE in the Universe today. Now, from the equations for the DM perturbations one actually recognizes the imprint of the new force as a modification of the strength of the gravitational potential. In Eq.~\eqref{eq:v_chi_conf} we see how the velocity field is driven by the spatial derivatives of the gravitational potential $\Psi$ which is augmented by the presence of the 5th force potential $(\partial \log m_\chi (s)/\partial s)\, \delta s$. This effect dominates over time derivatives of the potential for modes whose wavelengths are much shorter than the horizon size, generating relative velocity and density perturbation between DM and the baryons that will grow over time. This is in striking contrast with the $\Lambda$CDM cosmology for which, given adiabatic initial conditions, relative velocity perturbations decay with time and relative density perturbations stay, at best, constant after recombination (ignoring reionization). 

Interestingly, the time dependence of the $s$ background affects the behavior of the fluctuations, in three different ways: i) in the continuity equation~\eqref{eq:delta_chi_conf} a term proportional to $\delta s$ is present and controlled by $\bar{s}'$; ii) similarly, in the Euler equation~\eqref{eq:v_chi_conf} a term proportional to $\bar{s}'$ enters the friction term; iii) moreover, the Hubble rate $\mathcal{H}$ depends non-trivially on $\bar{s}$, and this dependence cannot be dropped for a meaningful prediction of the CMB or matter power spectrum. For modes with wavelength much larger than the horizon, the terms proportional to $k^2$ in Eq.~\eqref{eq:v_chi_conf} can be dropped, yet the explicit dependence of the equations on $\bar{s}'$ remains together with the implicit one coming from $\mathcal{H}$. This implies that we expect differences with respect to a $\Lambda\mathrm{CDM}$ Universe even on super-horizon scales. 

In the KG equation~\eqref{eq:KG_first} the first line contains the standard terms for an uncoupled scalar, whilst the second line contains the new terms generated by the interaction of $s$ with the DM. Deep inside the horizon, the scalar field satisfies a Poisson equation of the form
\begin{align}\label{eq:new_poisson}
    k^2 \delta s = -\, a^2 4\pi G_N \beta  \frac{\partial \log m_\chi (s)}{\partial s} \bar{\rho}_\chi \delta_\chi\;, \qquad \text{(deep inside the horizon)}
\end{align}
to be compared with the corresponding equation for the Newtonian potential,
\begin{equation}
k^2 \Psi = -\, a^2 4\pi G_N  \sum_x \bar{\rho}_x \delta_x\;,
\end{equation}
as derived from the perturbed Einstein equations~\eqref{eq:EE_algebraic} and~\eqref{eq:EE_shear} given below. Thus, at the fluid level, the strength of the new interaction is dictated by $G_s (\partial \log m_\chi (s)/\partial s)$, rather than by $G_s$ alone. For example, for our representative model one has $G_s (\partial \log m_\chi (s)/\partial s) = G_s/(1+2 \bar{s})$. Since $\bar{s}$ does not evolve much over time, we could therefore fine-tune the initial condition of $\bar{s}$ to some very large value, effectively decoupling the DM from the scalar field. This is indeed what happens if one wants a very light and coupled scalar, $m_\varphi \ll H_0$, to play the role of DE at late times. In this regime the energy density in the scalar field behaves like $\sim m_{\varphi}^2 M_{\rm Pl}^2 \bar{s}_{\rm ini}^2/\beta$, and to reproduce the energy density in DE $\bar{s}_{\rm ini}^2 \gg \beta$ is needed, implying that the strength of the fifth force effectively goes to zero. For this reason we have chosen the largest possible value of the mediator mass, $m_\varphi = H_0$, as our CDE benchmark.

Finally, we write Einstein's equations in the form
\begin{align}
k^2 \Phi \;=&\; 4\pi G_N a^2 \sum_x \left[ \bar{\rho}_x \delta_x +  \frac{3\mathcal{H}}{k^2} (\bar{\rho}_x + \overline{\mathcal{P}}_x) \theta_x\right]\,, \label{eq:EE_algebraic}    \\
k^2 (\Phi + \Psi) \;=&\; - 12 \pi G_N a^2 \sum_x (\bar{\rho}_x + \overline{\mathcal{P}}_x) \sigma_x\,, \label{eq:EE_shear}
\end{align}
where $\sigma_x$ are the shear perturbations. To solve these equations we need to express the perturbed fluid variables for the scalar field in terms of the perturbed field variables~\cite{Hu:1998kj},
\begin{align}\label{eq:fluid_vars_NG}
    \delta \rho_s = \frac{\bar{s}' \delta s' - \bar{s}'^2 \Psi }{ G_s a^2} + V_{s,s} \delta s\,, \quad \delta \mathcal{P}_s = \delta \rho_s - 2 V_{s,s} \delta s\,,\quad (\bar{\rho}_s + \overline{\mathcal{P}}_s)v_{s}^i = -\frac{\bar{s}' \nabla^i \delta s}{G_s a^2}\,, 
\end{align}
while no shear perturbation is generated, since the scalar field is minimally coupled.

\subsection{Initial conditions}\label{sec:IC}
To set up the initial conditions (IC) for the perturbations in Newtonian gauge we follow closely Ref.~\cite{Ma:1995ey}. We assume radiation domination and solve for all the relevant variables by expanding in $k \tau \ll 1$. The IC for the metric perturbations and for all other species, except the DM and the scalar field, are unchanged and reported in Appendix~\ref{app:pert_SG}. In this section we describe how to consistently set up the IC for $\chi$ and $s$ at the lowest order in $k \tau$. Our implementation in CLASS includes the IC in synchronous gauge as well, which are discussed in Appendix~\ref{app:pert_SG}.

The simplest kind of IC are the so called adiabatic IC in the density perturbations, which can be obtained by requiring that the gauge invariant relative entropy perturbations between all species vanish:
\begin{align}
    \mathcal{S}_{ij}\equiv  3 \mathcal{H} \bigg(\frac{\delta \rho_j}{\bar{\rho}'_j}-\frac{\delta \rho_i}{\bar{\rho}'_i}\bigg) = 0\,.\label{eq:adiabaticdef}
\end{align}
This assumption for the IC can be realized for example in models of single field inflation. Imposing Eq.~\eqref{eq:adiabaticdef} to the photon and DM fluids at lowest order gives
\begin{align}
 \mathcal{S}_{\chi \gamma} =   -\frac{3}{4} \delta_\gamma + \cfrac{\delta_\chi}{1-\cfrac{\partial \log m_\chi (s)}{\partial s}\cfrac{\bar{s}'}{3\mathcal{H}}} = 0\,,
\end{align}
which very deep in radiation domination reduces to $\delta_\chi  = 3 \delta_\gamma/4$, since $\bar{s}'\tau \ll 1$ as shown by Eq.~\eqref{eq:s_RD}. 

The same procedure could be repeated for the fluid perturbations of $s$ with respect to the photon fluid, obtaining (recall that $\delta_s \equiv \delta \rho_s / \bar{\rho}_s$) 
\begin{align}
\label{eq:ad_s}
    \mathcal{S}_{ s \gamma} = -\frac{3}{4} \delta_\gamma + \frac{\delta_s/ (1 + w_s)}{1 + \cfrac{\bar{\rho}_\chi (\partial \log m_\chi (s)/ \partial s)\bar{s}'}{3\mathcal{H}\bar{\rho}_s (1 + w_s)}} = 0\,.
\end{align}
At early times we can safely assume $w_s = 1$ and the ratio appearing in the denominator of the second term, which was already found in Eq.~\eqref{eq:doom}, is equal to $-2/3$. Thus $\delta_s = \delta_\gamma/2$ to lowest order.\footnote{We notice in passing that the ratio in Eq.~\eqref{eq:doom} is precisely the ``doom'' factor defined in Ref.~\cite{Gavela:2009cy} which, if positive and large, could signal non-adiabatic instabilities in the cosmological fluctuations. As discussed more extensively in Appendix~\ref{app:fluid_eqs}, we find that in any model of scalar fifth force that admits a consistent microscopic description the doom factor is negative and does not lead to any instability.}

We can then use the definition of the perturbed fluid variables in Eq.~\eqref{eq:fluid_vars_NG}, together with the physical condition that the velocity divergence of $s$ vanish when $\tau \rightarrow 0\,$, to obtain the IC for the field, $\delta s = - \delta_\gamma \bar{s}' \tau  /4$, and for the velocity, $\theta_s =  -\, \delta_\gamma k (k \tau) /4 = \theta_\gamma$. Finally, we plug the solution for $\delta s$ into the equation for $\theta_\chi$ and obtain $\theta_\chi = \theta_\gamma$ to lowest order, once again because the long range interactions are strongly suppressed by $\bar{s}' \tau \ll 1$. 

We have seen in Sec.~\ref{sec:bkg_behavior} that the background dynamics of the fifth force field at early times is fully determined by its interaction with the DM. We can then ask whether the same is true at the level of the perturbations, and therefore if the IC for $\delta_\chi$ and $\theta_\chi$ fix the ones for the scalar field. We now show that this is the case. Indeed, for adiabatic IC, imposing Eq.~\eqref{eq:ad_s} is redundant and the adiabaticity of $s$ follows from the one of $\chi$. Deep in radiation domination and outside the horizon, the KG equation reads
\begin{align}\label{eq:KG_IC_NG}
    \delta s '' + \frac{2}{\tau} \delta s' -\frac{2\bar{s}'}{\tau} (\delta_\chi + 2 \Psi) - \frac{2 \bar{s}'}{\tau}  \frac{\partial^2 \log m_\chi (s)/\partial s^2}{\partial \log m_\chi (s)/\partial s }\, \delta s = 0\,,
\end{align}
where in the adiabatic case $\delta_\chi + 2 \Psi = -\delta_\gamma/4$ is constant in time. The general solution to this equation is
\begin{equation}\label{eq:deltas_sol_IC}
\delta s \simeq (\delta s)_{ \rm ad} + \alpha_{\rm na} \left( 1  +  \frac{\partial^2 \log m_\chi (s)/\partial s^2 }{\partial \log m_\chi (s)/\partial s} \bar{s}' \tau + \mathcal{O}(\bar{s}^{\prime\, 2} \tau^2)   \right) \,,
\end{equation}
where $(\delta s)_{ \rm ad}= - \,\delta_\gamma \bar{s}' \tau/4$ is the adiabatic piece and $\alpha_{\rm na}$ is a constant parametrizing the deviation from adiabaticity. The adiabatic term is found by inserting the solution into Eq.~\eqref{eq:fluid_vars_NG} and again imposing that the velocity divergence of $s$ goes to zero when $\tau \rightarrow 0$. It is then trivial to check that this implies $\delta_s = \delta_\gamma/2$ and therefore the adiabaticity of $s$. This is true only at leading order in the $k \tau$ expansion, with the subleading corrections easily calculated following the same steps outlined above, see Eq.~\eqref{eq:NLO_deltas}. 

It is also interesting to study non-adiabatic scalar field fluctuations. In this case the $\alpha_{\rm na}$ term in Eq.~\eqref{eq:deltas_sol_IC} generates an additional velocity contribution for $\chi$, whose IC now reads, neglecting terms suppressed by $\bar{s}^\prime \tau \ll 1$,
\begin{align}
\label{eq:theta_na}
    \theta_\chi \simeq (\theta_{\chi})_{\rm ad} + \frac{\alpha_{\rm na}}{2} \cfrac{\partial \log m_\chi (s)}{\partial s}\, k (k \tau)\,
\end{align}
with $(\theta_{\chi})_{\rm ad} = \theta_\gamma\,$. We see that the non-adiabatic piece is proportional to $\partial \log m_\chi(s)/\partial s$ and is therefore suppressed if $\bar{s}$ is large, as it happens in CDE scenarios with $m_\varphi \ll H_0$. In addition, the initial $s$ density perturbation is modified to
\begin{equation}
\delta_s = (\delta_s)_{\rm ad} + 2 \alpha_{\rm na} \frac{\partial^2 \log m_\chi (s)/\partial s^2 }{\partial \log m_\chi (s)/\partial s}\,,
\end{equation}
where $(\delta_s)_{\rm ad} = \delta_\gamma/2\,$. The IC for the DM density perturbation $\delta_\chi$ is unchanged. We note that $\alpha_{\rm na}$ is in general a function of $k$, which depends on inflationary dynamics. This is analogous to the initial curvature perturbation $\mathcal{R}(k) = 2\, C(k)$~\cite{Ma:1995ey}, which controls the spatial behavior of the adiabatic IC. %to the time-independent quantity $C$, which appears in the adiabatic part of the IC~\cite{Ma:1995ey} and is related to the initial curvature perturbation $\mathcal{R}$ by $C(k) = \mathcal{R}(k)/2$.}

In the presence of a dark long range 5th force we cannot therefore a priori assume that the velocity of DM and of all other species is the same even on super-horizon scales. A non-zero $\alpha_{\rm na}$ will alter dramatically the large scale evolution of cosmological perturbations, which in $\Lambda$CDM is based on the constancy of the comoving curvature perturbation outside the horizon, and as we will see in Sec.~\ref{sec:constraints} it is severely constrained by current data. 

\begin{figure}
    \centering
    \includegraphics[width = 0.45\textwidth]{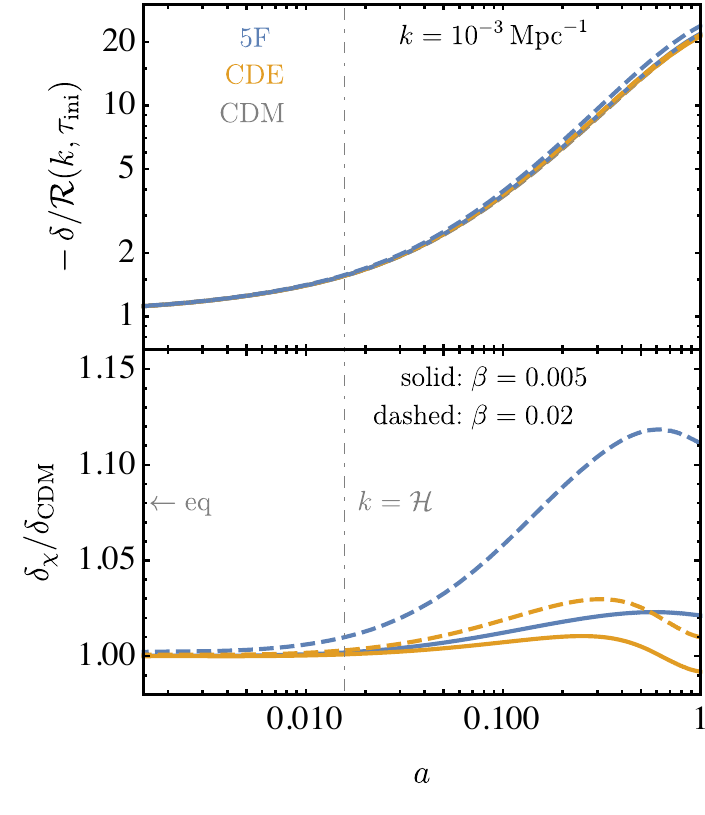}\hspace{2mm}
     \includegraphics[width = 0.45\textwidth]{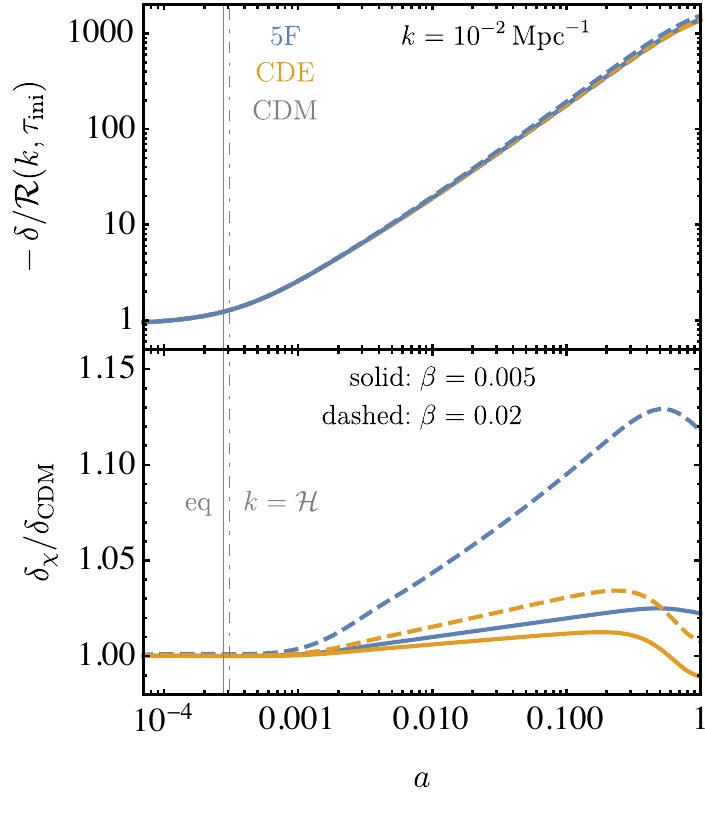}
     \includegraphics[width = 0.45\textwidth]{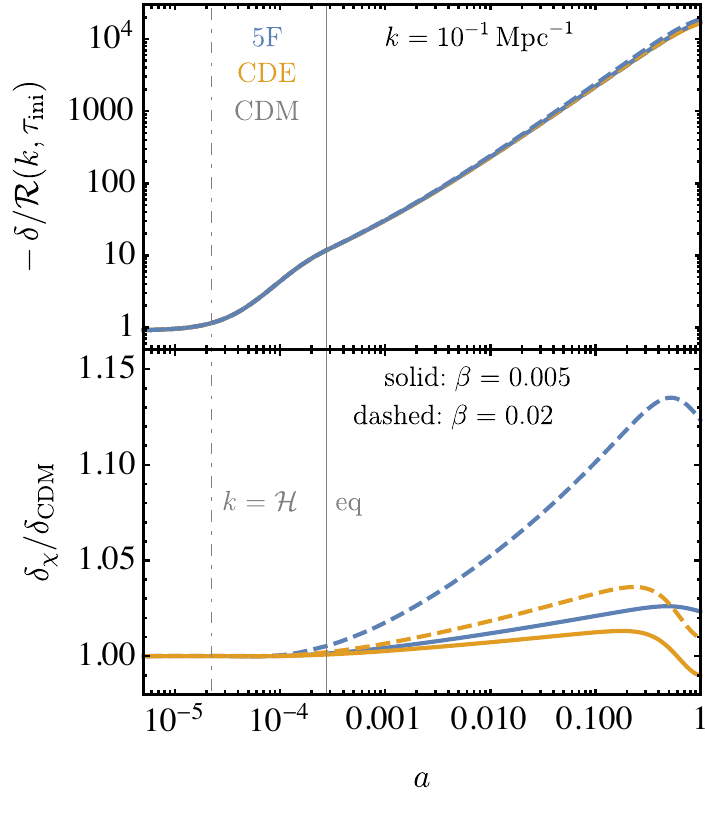}\hspace{2mm}
     \includegraphics[width = 0.45\textwidth]{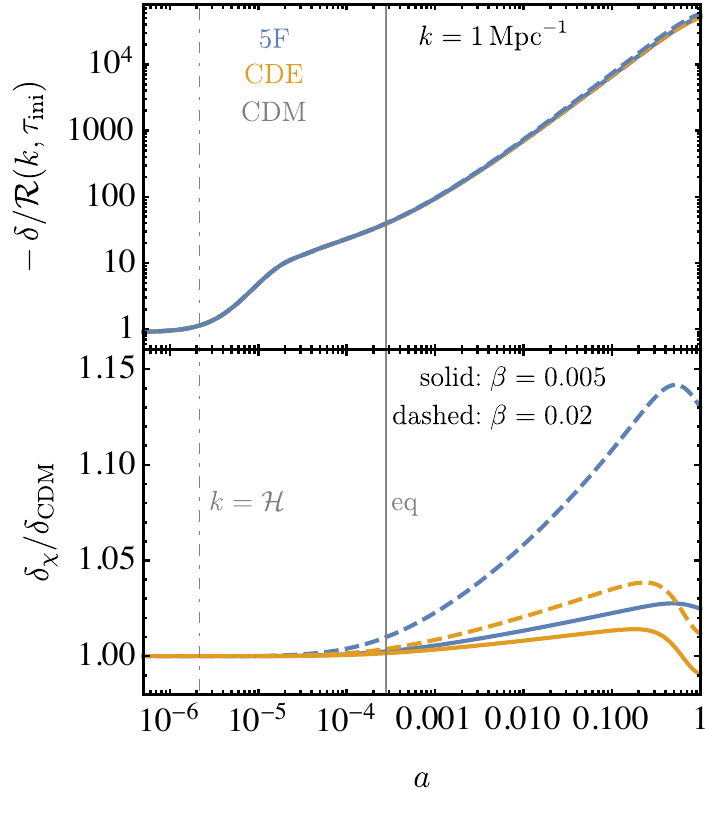}
    \caption{The time evolution of Dark Matter density perturbations in Newtonian gauge, for four different wavelengths: $k = 10^{-3} \text{ Mpc}^{-1}$ (upper left), $k = 10^{-2} \text{ Mpc}^{-1}$ (upper right), $k = 10^{-1} \text{ Mpc}^{-1}$ (lower left) and $k = 1 \text{ Mpc}^{-1}$ (lower right). The light blue, light orange, and gray lines show the 5F, CDE, and $\Lambda$CDM scenario, respectively. The solid lines correspond to $\beta = 0.005$, while the dashed ones to $\beta = 0.02$. The constant $C$ appearing in the IC and the initial curvature perturbation $\mathcal{R}$ are related by $C = \mathcal{R}/2$.}
    \label{fig:pert1}
\end{figure}

\subsection{The evolution of density fluctuations}
\label{sec:evolution}
As we did for the cosmological background, we have implemented to first order in perturbation theory all the relevant equations and their IC in CLASS. As a robustness test of our code we have checked that the Newtonian and synchronous gauges give identical results for physical observables, such as the CMB power spectrum. 

Figure~\ref{fig:pert1} shows the numerical evolution of the density perturbation of $\chi$ for four different modes, going from the largest observables scales ($k = 10^{-3} \text{ Mpc}^{-1}$, upper left panel), to small scales ($k = 1 \text{ Mpc}^{-1}$, lower right panel). To produce these plots we use the same benchmark cosmological parameters as in Sec.~\ref{sec:bkg}, setting in addition $\alpha_{\rm na} =0$ and the same IC for the perturbations in the baseline $\Lambda$CDM and in our models. We find that at early times, when modes are outside of the horizon, the evolution of DM density fluctuations is almost identical to $\Lambda$CDM, regardless of the wavelength. This indicates that the dynamics outside the horizon is still mostly set by the IC, at least in the adiabatic case. Moreover, since at the background level the evolution during radiation domination of DM and baryons is unchanged, we expect little to no difference, with respect to a standard scenario, on which scales enter the horizon before matter-radiation equality. In addition, modes that enter the horizon deep in the radiation era will evolve like in the $\Lambda$CDM case until equality, as we can see from the lower panels in Fig.~\ref{fig:pert1}. 
As soon as matter domination begins, perturbations can grow. Thanks to the presence of a dark fifth force, the fluctuations in DM grow more rapidly than the corresponding ones in $\Lambda$CDM.
In particular we notice that, at least in the 5F scenario where $\bar{s}\ll1$ always, the excess power is much larger than the naive $\mathcal{O}(\beta)$ expectation, similarly to what happened at the level of the background. In the CDE case the effect of long range interaction is less dramatic, due to the extra suppression of the interaction by $\partial \log m_\chi (s) / \partial s$, as further detailed in Sec.~\ref{sec:sub_hor}.

Finally, we note that at very late times, in either the CC or CDE dominated phase, the growth rate of DM perturbations is reduced compared to $\Lambda$CDM. At first, this might seem to clash with the intuition that a new attractive force should only increase the clustering of DM, but it can be understood by considering the effect of the new interaction on the cosmological background. In the 5F case it follows from the larger value of the CC required by the flatness constraint, which exponentially suppresses the growth of structure. In the CDE case it is a consequence of the faster expansion rate at very late times, see Fig.~\ref{fig:hubble}. In the CDE models, the $z=0$ amplitude of density fluctuations can even be suppressed with respect to a standard cosmological scenario.

\subsubsection{Sub-horizon solutions}
\label{sec:sub_hor}
While an analytical understanding of all the different stages of evolution of the DM perturbations shown in Fig.~\ref{fig:pert1} is not possible, some regimes can be investigated perturbatively for small $\beta$. This is the case of the sub-horizon evolution in the matter dominated era. In particular, we would like to quantitatively understand why the effect of the fifth force is much larger than the simple $\mathcal{O}(\beta)$ counting.

\begin{figure}
    \centering
    \includegraphics[width = 0.45\textwidth]{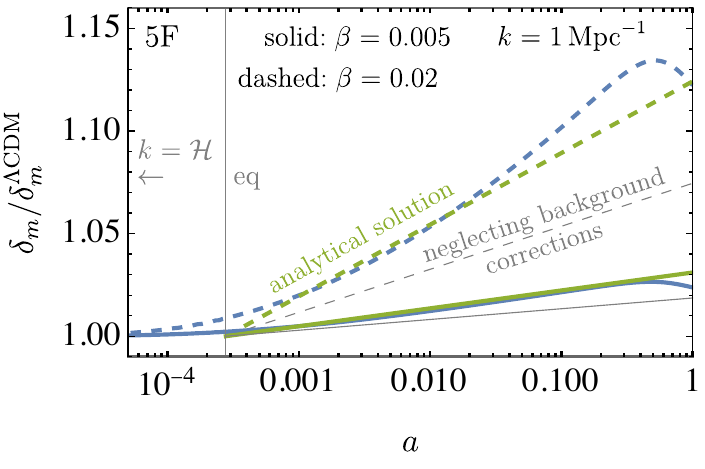}
    \caption{Ratio of total matter density perturbation to $\Lambda$CDM for the pure fifth force scenario (light blue), in Newtonian gauge and for the $k = 1$ Mpc$^{-1}$ mode, which enters the horizon much earlier than equality. Green lines show our perturbative analytical solution Eq.~\eqref{eq:deltam_anal}, whereas thin gray lines correspond to the solution obtained by neglecting corrections to the background cosmology, Eq.~\eqref{eq:deltam_literature}.}
    \label{fig:pert2}
\end{figure}

During matter domination and in the sub-horizon regime $k/\mathcal{H} \gg 1$, we can reduce the evolution of DM and baryons to a system of coupled second-order differential equations:
\begin{align}
  \delta_\chi '' \,+&\; \Big(\mathcal{H}+\cfrac{\partial \log m_\chi (s)}{\partial s}\, \bar{s}' \Big) \delta_\chi^\prime - \frac{3}{2}\Omega_m \mathcal{H}^2  \bigg[ f_\chi  \bigg(1+ \beta  \bigg(\cfrac{\partial \log m_\chi (s)}{\partial s}\bigg)^2\bigg) \delta_\chi + (1-f_\chi) \delta_b \bigg] = 0\,, \nonumber \\
  \delta_b '' \,+&\; \mathcal{H} \delta_b^\prime - \frac{3}{2} \Omega_m \mathcal{H}^2  \big( f_\chi \delta_\chi + (1-f_\chi) \delta_b \big) = 0\,, \label{eq:chi_b_subhor}
\end{align}
where $\bar{\rho}_m \equiv \bar{\rho}_\chi + \bar{\rho}_b$, $f_\chi \equiv \bar{\rho}_{\chi}/\bar{\rho}_m$, and $\Omega_m \equiv \bar{\rho}_m / \bar{\rho}_{\rm tot}$.\footnote{In the special case of exponential form of the field dependent mass, $m_\chi (s) = m_\chi e^{-\sqrt{2\beta} s}$, Eqs.~\eqref{eq:chi_b_subhor} recover the expressions given in Ref.~\cite{Gomez-Valent:2020mqn}.} Notice that all these quantities are time-dependent, with $\Omega_m \simeq 1 - \beta f_\chi^2 (\partial \log m_\chi(s)/\partial s)^2 /3$ deep in MD, see Eq.~\eqref{eq:rhos_MD}. The equation for $\delta_\chi$ is modified in two ways relative to $\Lambda$CDM: i) the friction term receives $\mathcal{O}(\beta)$ corrections due to background modifications; ii) the last term contains a new contribution to the potential, which can be traced back to Eq.~\eqref{eq:new_poisson}, as well as a suppression of $\Omega_m$ below unity due to the fraction of the total energy density that is made up by the non-clustering matter component $s$. Exact analytical solutions to Eqs.~\eqref{eq:chi_b_subhor} are difficult to obtain, due to the complicated time-dependence of the background. However, the system can be solved perturbatively in $\beta$. It is convenient to take as new variables 
\begin{equation}
\delta_m \equiv f_\chi \delta_\chi + (1-f_\chi) \delta_b\,,\qquad \delta_r \equiv \delta_\chi- \delta_b\, ,
\end{equation}
where the first is the total matter perturbation and the second the relative perturbation. 
Expanding as $\delta_i = \delta_i^{(0)} + \beta \delta_i^{(1)}$ ($i = m, r$), at leading order the total component grows with the scale factor, $\delta_{m}^{(0)}  = \mathcal{C} \tau^2$ where $\mathcal{C} \equiv (\widetilde{\Omega}_d + \Omega_b^0)H_0^2/4$, while we can set the relative density perturbation to zero,\footnote{This is not strictly correct because of the tight coupling of baryons with photons, see Ref.~\cite{Chen:2019cfu} for a discussion.} $\delta_r^{(0)}= 0$. Furthermore relative velocities, if present, simply decay. Up to first order in $\beta$, Eqs.~\eqref{eq:chi_b_subhor} can then be written in the form
\begin{align}
\delta_r^{\prime\prime} + \mathcal{H} \delta_r^\prime - \mathcal{H} f_\chi \beta \left(\frac{\partial \log m_\chi (s)}{\partial s}\right)^2 \Big( \delta_m^\prime + \frac{3}{2} \Omega_m \mathcal{H} \delta_m \Big) =\,& 0\,, \\
\delta_m^{\prime\prime} + \mathcal{H} \delta_m^\prime - \frac{3}{2}\Omega_m \mathcal{H}^2 \delta_m -  \mathcal{H} f_\chi^2 \beta \left(\frac{\partial \log m_\chi (s)}{\partial s}\right)^2 \Big( \delta_m^\prime + \frac{3}{2} \Omega_m \mathcal{H} \delta_m \Big) =\,& 0\,, \label{eq:deltam_subhor}
\end{align}
where $\mathcal{H}$ is expanded up to at most $\mathcal{O}(\beta)$ and reads
\begin{equation}
\mathcal{H} = \frac{a'}{a} = \frac{2}{\tau} \bigg( 1 - \beta f_\chi^2  \bigg( \frac{\partial \log m_\chi (s)}{\partial s} \bigg)^2 \, \bigg)\,,
\end{equation}
as derived from Eq.~\eqref{eq:a_MD_5F} for the 5F scenario. These equations are valid for modes that enter the horizon well before matter-radiation equality. To solve for the relative perturbation it is sufficient to retain the leading-order expressions of $\mathcal{H}$ and $\Omega_m$, yielding the solution
\begin{equation}
\label{eq:delta_r_eds}
\delta_r (\tau) = \frac{5}{3} \beta f_\chi  \left(\frac{\partial \log m_\chi (s)}{\partial s}\right)^2 \big[\delta_{m}^{(0)}(\tau) - \delta_{m}^{(0)} (\tau_{\rm eq})\big]\,,
\end{equation}
which grows linearly with the scale factor. The relative velocity between Dark Matter and baryons also grows accordingly. On the other hand, to solve for the total matter density up to first order we need to include the first corrections to $\mathcal{H}$ and $\Omega_m$. Plugging these into Eq.~\eqref{eq:deltam_subhor}, we arrive at the differential equation
%and expanding as $\delta_m = \delta_m^{(0)} + \beta \delta_m^{(1)}$,
\begin{align}
 \delta_m^{\prime\prime} + \frac{2}{\tau} \bigg( 1 - 2 \beta f_\chi^2 \left(\frac{\partial \log m_\chi (s)}{\partial s}\right)^2 \hspace{-0.25mm} \bigg) \delta_m^{\prime} -\frac{6}{\tau^2} \bigg( 1 - \frac{4}{3}\beta f_\chi^2 \left(\frac{\partial \log m_\chi (s)}{\partial s}\right)^2 \hspace{-0.25mm}   \bigg) \delta_m = 0\,, 
\end{align}
whose solution is simply
\begin{equation}
  \delta_m (\tau) = \delta_m^{(0)} (\tau)\,,  %\left[ 1 + \frac{2}{5} \beta f_\chi^2 \bigg( \frac{\partial \log m_\chi (s)}{\partial s} \bigg)^2 \log\frac{\tau}{\tau_{\rm eq}} \right]\,,
\end{equation}
namely $\delta_m^{(1)} = 0$. This result can be expressed in terms of the scale factor as
\begin{equation}
\delta_m (a) = \delta_m^{(0)}(a) \left[ 1 + \beta f_\chi^2 \bigg( \frac{\partial \log m_\chi (s)}{\partial s} \bigg)^2 \log \frac{a}{a_{\rm eq}} \right]\,,  \quad (\mathrm{our}\;\mathrm{result}) \label{eq:deltam_anal} 
\end{equation}
where $\delta_m^{(0)}(a) = a\,$. We thus find that the two new physics effects in Eq.~\eqref{eq:chi_b_subhor} conspire to produce a growth stronger by a factor of $5/3$ compared to the solution where background corrections are not taken into account,
\begin{equation} \label{eq:deltam_literature}
\delta_m (a) = \delta_m^{(0)}(a) \left[ 1 + \frac{3}{5} \beta f_\chi^2 \bigg(\frac{\partial \log m_\chi (s)}{\partial s}\bigg)^2 \log \frac{a}{a_{\rm eq}} \right]\,.\qquad (\mathrm{neglecting}\;\mathrm{background})
\end{equation}
The strong impact of background corrections is further highlighted by Fig.~\ref{fig:pert2}, demonstrating the good agreement of our analytical solution Eq.~\eqref{eq:deltam_anal} with the numerical result.

For completeness we also report the individual solutions for $\chi$ and the baryons during MD,
\begin{align}
\delta_\chi (\tau) =&\; \delta_m^{(0)} (\tau) - \frac{5}{3} \beta f_\chi (f_\chi - 1) \Big( \frac{\partial \log m_\chi (s)}{\partial s} \Big)^2 \big[ \delta_m^{(0)} (\tau) - \delta_m^{(0)} (\tau_{\rm eq}) \big]\,,  \\
\delta_b (\tau) =&\; \delta_m^{(0)} (\tau) - \frac{5}{3} \beta f_\chi^2 \Big( \frac{\partial \log m_\chi (s)}{\partial s} \Big)^2 \big[ \delta_m^{(0)} (\tau) - \delta_m^{(0)} (\tau_{\rm eq}) \big]\,.
\end{align}
Thus far, in the discussion of sub-horizon solutions we have focused on the 5F scenario. For the CDE case one can derive analogous analytical solutions, but the increase of power is far less pronounced.
\begin{figure}
    \centering
    \includegraphics[width = 0.47\textwidth]{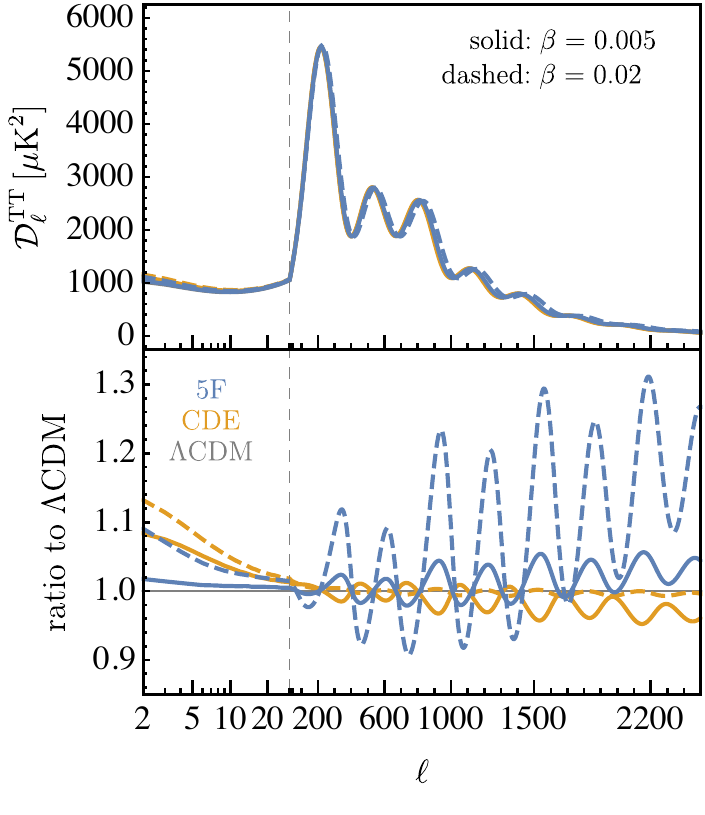}\hspace{5mm}
     \includegraphics[width = 0.47\textwidth]{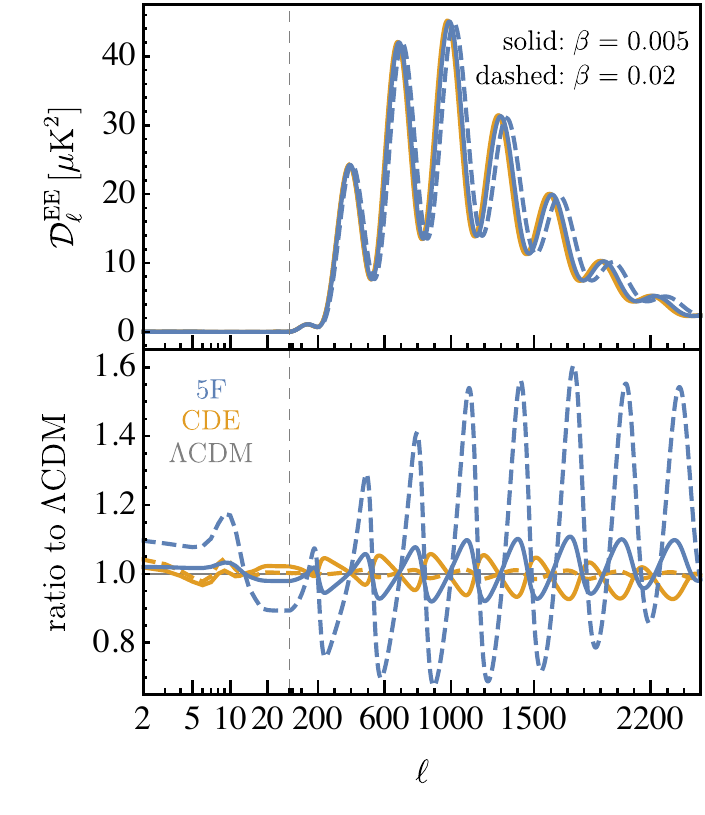}\\\vspace{2mm}
     \includegraphics[width = 0.47\textwidth]{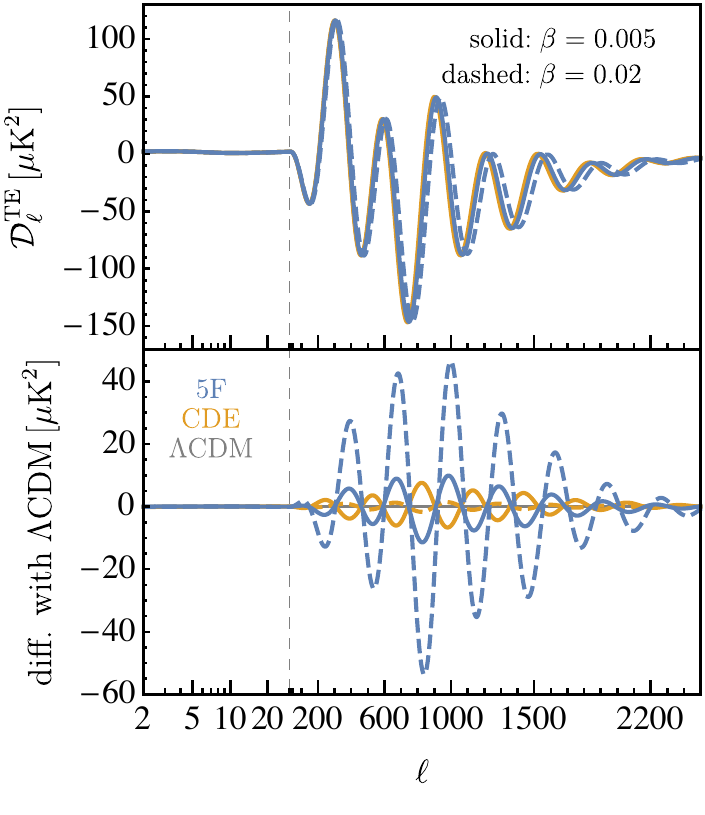}\hspace{5mm}\includegraphics[width = 0.47\textwidth]{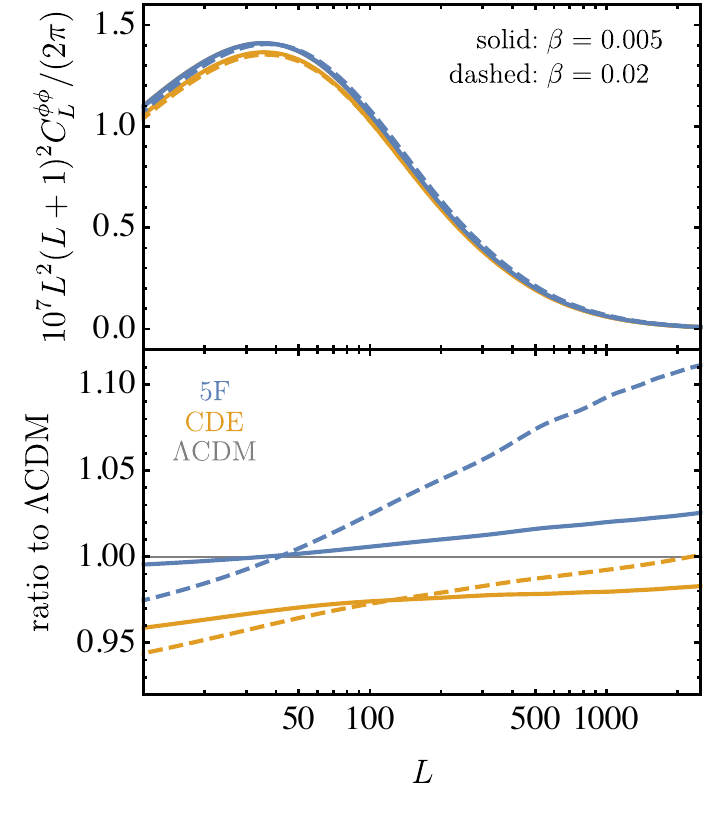}
    \caption{CMB power spectra: TT (top left), EE (top right), TE (bottom left) and lensing $\phi\phi$ (bottom right). The light blue, light orange, and gray lines show the 5F, CDE, and $\Lambda$CDM scenario, respectively. The solid lines correspond to $\beta = 0.005$, while the dashed ones to $\beta = 0.02$. In the first three panels, at $\ell = 30$ (marked by the gray dashed line) the scale on the horizontal axis switches from logarithmic to linear. Note that for the TE power spectrum we show the difference, not the ratio, between the models.}
    \label{fig:C_ell}
\end{figure}

\subsection{The CMB and the total matter power spectrum}

The evolution of density perturbations over time and at fixed wavelength, discussed in the previous section and shown in Fig.~\ref{fig:pert1}, is useful to understand the physical effects generated by the fifth force, but it is not directly observable. 
In particular, in this work we are interested in constraining new long range forces with CMB and BAO data.

Figure~\ref{fig:C_ell} shows the CMB temperature (upper left panel), $E$-mode polarization (upper right panel), cross temperature-polarization (lower left panel), and lensing (lower right panel) power spectra. As in previous plots, solid lines correspond to $\beta = 0.005$, while dashed lines to $\beta = 0.02$. The color coding is also the same as before, with the prediction for the 5F scenario in light blue, and the one for CDE in light orange. For reference the corresponding $\Lambda$CDM prediction is shown in gray. The shape of the CMB primary power spectra is mainly determined by the physical scales relevant for the tight coupling between baryons and photons, and by their projection onto observed angles on the sky. As we have seen in Sections~\ref{sec:bkg} and~\ref{sec:evolution}, before matter-radiation equality a Universe containing a fifth force with small $\beta$ is for all practical purposes indistinguishable from a $\Lambda$CDM one. Since last scattering happens close enough to the onset of the new interaction, we expect small differences in the physical scales at recombination, which are governed by the values of $\omega_b$, $\omega_\chi$, and $\omega_\gamma$. For example, the sound horizon changes only by $\mathcal{O}(\beta)$. On the other hand the projection on the sky depends on the angular diameter distance between $z=0$ and last scattering, and it is severely affected by the new force, see Fig.~\ref{fig:hubble}. To first approximation we thus expect, compared to the $\Lambda$CDM case, a shift in the location of the peaks and troughs of the CMB power spectrum. This is indeed what we see in the bottom panels of Fig.~\ref{fig:C_ell}: the residuals around the standard model oscillate around unity. At sufficiently large multipoles we also start seeing the difference in the angular projection of diffusion damping, and therefore an increase~(decrease) of power in the 5F~(CDE) scenario. 

The difference at large angular scales in the CMB power spectra is again due to projection, more specifically to the Integrated Sachs-Wolfe effect (ISW). While the Bardeen potentials in our model are never constant in time, even during matter domination, it is still the case that the largest difference with respect to a $\Lambda$CDM Universe happens at low redshift, which maps the late ISW anisotropies to large angular separation.

Finally, the lower right panel in Fig.~\ref{fig:C_ell} shows the CMB lensing power spectrum. The amplitude of the lensing power spectrum depends on the total amount of matter in the Universe, which is reduced compared to a standard cosmological model by the long range force, as shown in Fig.~\ref{fig:hubble}. This will act towards decreasing the overall amplitude of the lensing power spectrum. On the other hand, the CMB lensing kernel peaks at relatively high redshift, at which the clustering of DM is enhanced by the new force. This effect partially cancels the lower matter density, and it results in a relatively small difference in the lensing power spectrum when compared to a $\Lambda$CDM one. It is also worth pointing out that different multipoles $L$ receive contributions from different physical scales $k$, e.g. larger multipoles are mostly sourced by smaller scales, hence the scale dependence we see in the lensing power spectrum in Fig.~\ref{fig:C_ell} is compatible with the one we observed in Fig.~\ref{fig:pert1}. It should be kept in mind that the numerical evaluation of the lensing power spectrum requires a prescription for the fully nonlinear matter power spectrum. This is not available for the models described in this work, hence the precise value of the high-$L$ lensing power spectrum should be taken with a grain of salt. For CDE models some results have appeared in Refs.~\cite{Keselman:2009nx,Baldi:2010vv,Baldi:2008ay}, but they are not yet at the same maturity level of the prediction in the standard model.
\begin{figure}
    \centering
    \includegraphics[width = 0.47\textwidth]{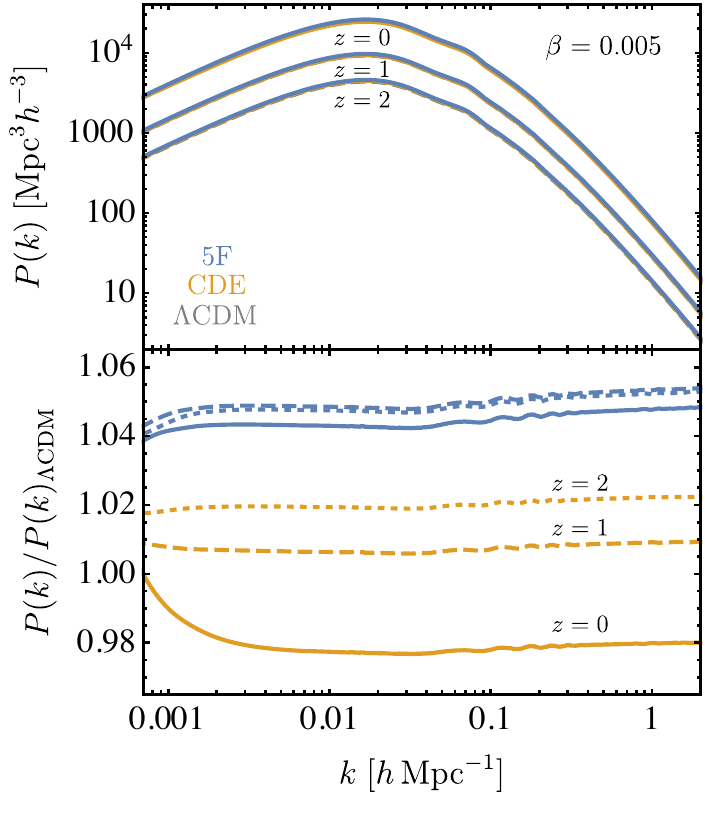}\hspace{2mm}
     \includegraphics[width = 0.47\textwidth]{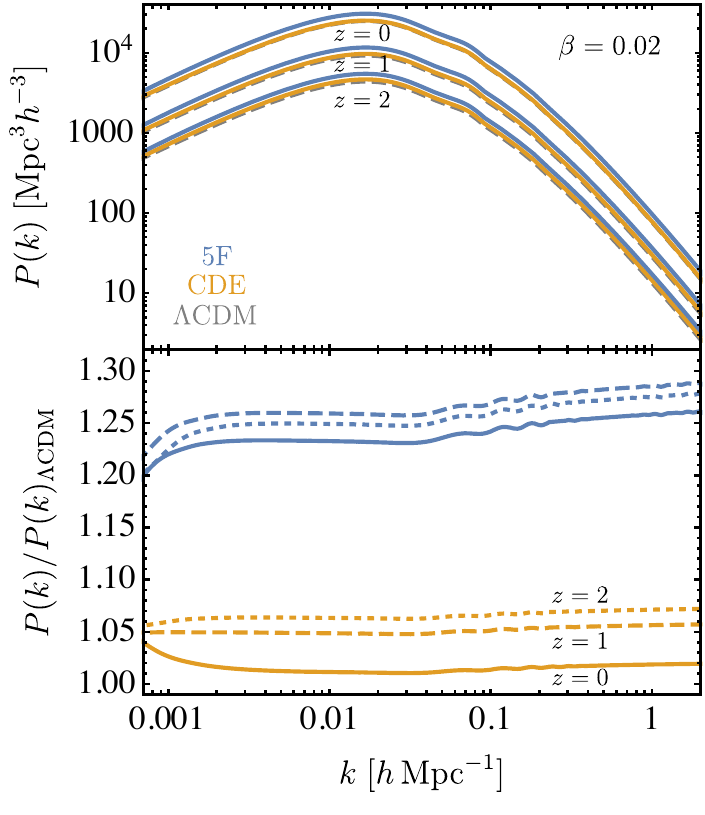}
    \caption{{\it (Left)} Total matter power spectrum for $\beta = 0.005$, at three different redshifts. {\it (Right)} Same as in the left panels, but for $\beta = 0.02$. The light blue, light orange, and gray lines show the 5F, CDE, and $\Lambda$CDM scenario, respectively. Solid lines show the results at $z=0$, dashed lines at $z=1$, and dotted ones at $z=2$. }
    \label{fig:Pk}
\end{figure}

Figure~\ref{fig:Pk} shows the linear matter power spectrum for different redshifts. While not directly observable, the total matter power spectrum is closely related to the galaxy power spectrum measured in redshift surveys and to the cosmic shear and galaxy-galaxy lensing measurements of imaging surveys. Roughly speaking, we expect the difference in the matter power spectrum between a $\Lambda$CDM model and a cosmology with dark long range force to be twice the effect we saw in Fig.~\ref{fig:pert1} and Sec.~\ref{sec:evolution}. For this reason we will not repeat here the same discussion, but rather highlight a few key features of the models described in this work. First of all, as presented in Sec.~\ref{sec:sub_hor}, the effect of the new force is much larger than the naive $\mathcal{O}(\beta)$ counting. On the other hand, the new long range interaction primarily changes the amplitude of the power spectrum, with only a mild scale dependence, roughly of $\mathcal{O}(\beta)$. In comparison with a $\Lambda$CDM model, and at fixed $H_0$, the excess power is larger at higher redshift, and then decreases towards $z=0$. As discussed in Sec.~\ref{sec:evolution}, this is due to the difference in the expansion rate at very low redshift between a Universe with a fifth force in the dark sector and a standard one.   
\begin{figure}
    \centering
\includegraphics[width = 0.5 \textwidth]{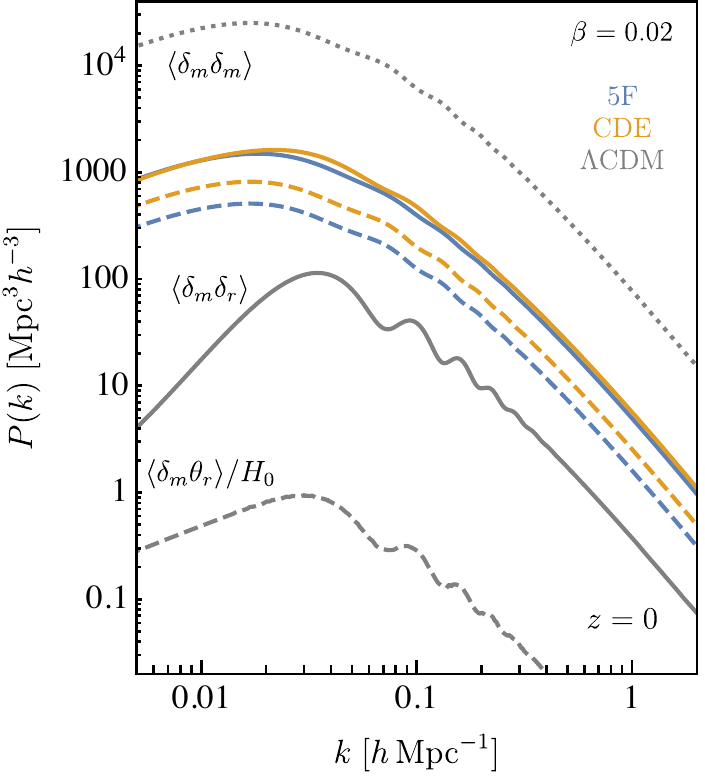}
    \caption{Cross power spectra at $z = 0$ between the total matter and relative density perturbations $\delta_r$ (solid lines) and between the total matter and relative velocity divergence perturbations $\theta_r$ (dashed lines). The total matter power spectrum in $\Lambda$CDM is also shown for reference (dotted line). The light blue, light orange, and gray lines show the 5F, CDE, and $\Lambda$CDM scenario, respectively.}
    \label{fig:rel_pt}
\end{figure}

Finally, in light of the discussion in the next section on the use of BAO data, it is useful to look at the amplitude of the relative density and velocity power spectra. Figure~\ref{fig:rel_pt} shows, at $z=0$ and in the $\Lambda$CDM case, the matter power spectrum (gray dotted line), the cross power spectrum between the total matter and the relative density perturbations, $\langle \delta_m \delta_r \rangle$ (gray solid line), and the cross power spectrum between the total matter and the relative velocity divergence perturbations, $\langle \delta_m \theta_r \rangle$ (gray dashed line).\footnote{For dimensional reasons, the relative velocity divergences are multiplied by $1/H_0$.} As it is well known, in a $\Lambda$CDM model relative density perturbations are at most $1\%$ of the total matter and relative velocities are negligible. In light blue and light orange colors we show the same cross power spectra for the 5F and CDE scenarios with $\beta = 0.02$. The cross power spectrum between the total matter and the relative density perturbations in cosmologies with a dark fifth force, solid lines, is approximately an order of magnitude larger than the corresponding quantity in the $\Lambda$CDM case, and it can reach $10\%$ of the total power spectrum. The BAO oscillations are also out of phase with those in the matter power spectrum. The cross power spectrum between the total matter and the relative velocity divergence perturbations, dashed lines, is $\mathcal{O}(100\,\text{-}1000)$ times larger than in the standard cosmological model. 
For comparison, at $z=10$ the relative density perturbations in cosmologies with dark fifth forces are still more than two times larger than the corresponding quantity in a $\Lambda$CDM scenario, which becomes a factor of 30 for the relative velocities.

\section{Constraints and Discussion}\label{sec:constraints}
In this section we present our constraints on the cosmological models discussed in this work. We employ a Markov Chain Monte Carlo approach, and in particular the Metropolis-Hastings algorithm, to scan the parameter space of interest until the standard criteria for the convergence of the chains are reached. Our implementation relies on the MontePython code~\cite{Audren:2012wb,Brinckmann:2018cvx}.
Datasets used in this work include the most recent Planck temperature and polarization data of the CMB~\cite{Planck:2018vyg,Planck:2019nip}, and measurements of the Baryon Acoustic Oscillations scale in spectroscopic galaxy surveys~\cite{Kazin:2014qga,Beutler:2011hx,BOSS:2016wmc,Ross:2014qpa}, to which we collectively refer as BAO. We will also briefly comment on the possibility that long range forces in the dark sector could alleviate the well known tension between the CMB and local determinations of $H_0$ (see for example Refs.~\cite{Knox:2019rjx,DiValentino:2021izs,Schoneberg:2021qvd} for extensive reviews of the problem and of the proposed solutions to it). We will therefore show additional constraints that include a Gaussian prior on the Hubble constant, $H_0 = 74.03 \pm 1.42$ km/s/Mpc at 68\% c.l., from Ref.~\cite{Riess:2019cxk}.\footnote{We note that more recent measurements of $H_0$ from the SH$_0$ES + Pantheon teams have been presented  in \cite{Riess:2021jrx,Brout:2022vxf}, but they would not change quantitatively our results. As emphasized by the SH$_0$ES team and by others \cite{Camarena:2021jlr,Benevento:2020fev,Efstathiou:2021ocp}, a more robust way to implement the local distance ladder would be through a prior on the Supernovae absolute magnitude $M_b$, and then to refit the Pantheon data assuming a given background cosmological model. We will return to these issues in a future publication.}
We vary 5 $\Lambda$CDM parameters $\{ \omega_b,\, H_0,\,n_s,\,A_s,\,
\tau\}$, where $\tau$ is the reionization optical depth, to which we add the dark sector parameters $\{\widetilde{\Omega}_{d},\,\beta\}$. Regardless of the combination of datasets, we find that most $\Lambda$CDM parameters are unchanged, with the exception of the Hubble constant. We will therefore only show the constraints for $\widetilde{\Omega}_{d},\,\beta,$ and $H_0$.

Due to the marked differences in the time evolution of the background and perturbations between a 5F cosmology and a CDE one, we present the parameter constraints separately, in Sections~\ref{sec:Hubbletension} and~\ref{sec:CDE_data} respectively. We refer the reader to Sections~\ref{sec:bkg} and~\ref{sec:fluctuations} for a detailed comparison of the two scenarios and recall here only the key distinctions. 

In the 5F scenario the cosmological energy density in the scalar field is negligible at all times. This is obtained for initial field background values $\bar{s}_{\rm ini}\ll 1$, which in turn implies that $\partial \log m_\chi (s) / \partial s \simeq 1$. The dependence on the scalar mass $m_\varphi$ is very weak provided it is smaller than $H_0$. It is therefore the fifth force coupling $\beta$ that primarily controls the size of the differences with respect to the $\Lambda$CDM model.  

In the CDE scenario the scalar field is required to provide the accelerated expansion at very late times. This implies a non-trivial interplay between $\bar{s}_{\rm ini}$, $m_\varphi$ and $\beta$, as picking a value for two of them fixes the third one. At fixed $\beta\simeq 10^{-2}$ and $m_\varphi = H_0$, we find $\bar{s}_{\rm ini} \sim \mathcal{O}(1)$ and therefore $\partial \log m_\chi (s) / \partial s < 1$, as seen in Sec.~\ref{sec:bkg}. If $m_\varphi$ is reduced below $H_0$, then $\bar{s}_{\rm ini}$ increases to keep the potential energy in the scalar field the same, which in turn yields a stronger suppression of $\partial \log m_\chi(s) / \partial s \ll 1$. Thus, in a CDE Universe the effective strength of the fifth force is generically reduced compared to the 5F case, hence weaker constraints on $\beta$ are expected. 

In Sec.~\ref{sec:alpha_data} we present the constraints on the parameter $\alpha_{\rm na}$, defined in Sec.~\ref{sec:IC}, which quantifies departures from the adiabatic initial conditions for DM velocity perturbations.

\subsection{Pure Fifth Force and the Hubble tension}\label{sec:Hubbletension}

\begin{figure}
    \centering
    \includegraphics{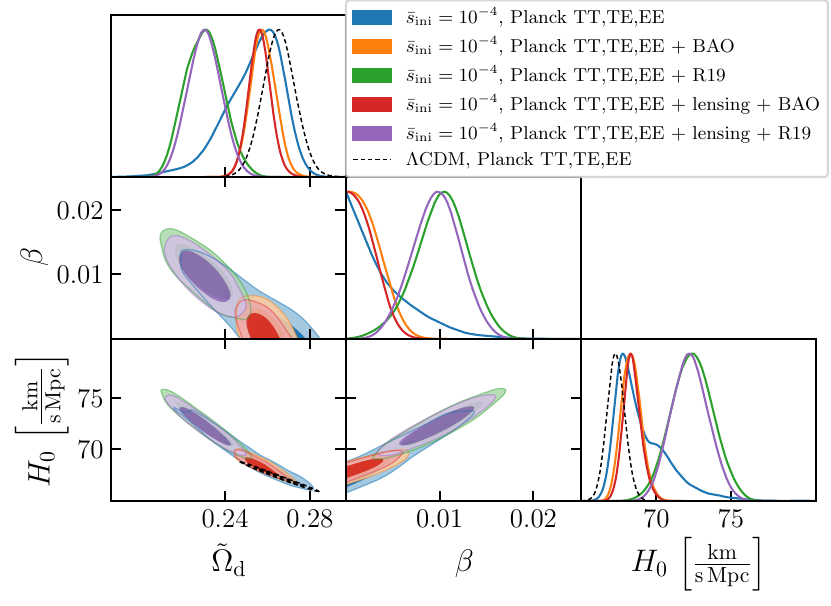}
    \caption{Bounds on the pure 5th force scenario with $m_\varphi/H_0 = 0.1$ and $\bar{s}_{\text{ini}}=10^{-4}$. The blue shaded region is favored by Planck CMB data~\cite{Planck:2018vyg,Planck:2019nip} at 1$\sigma$ and 2$\sigma$. The orange shaded region also includes BAO measurements~\cite{Kazin:2014qga,Beutler:2011hx,BOSS:2016wmc,Ross:2014qpa}, while the red shaded region adds CMB lensing measurements from Planck~\cite{Planck:2018lbu}. The green shaded contours combine Planck CMB data with a $H_0$ prior from local measurements~\cite{Riess:2019cxk}, whereas the purple region also includes lensing. The dashed contour shows the 2$\sigma$ region favored by Planck CMB data in $\Lambda$CDM.}
    \label{fig:5th_bounds}
\end{figure}

Figure~\ref{fig:5th_bounds} shows the constraints on our model parameters in a 5F scenario with $m_\varphi/H_0 = 0.1$. Different colors display different datasets, and for comparison we plot as dashed lines and contours the corresponding bounds in $\Lambda$CDM from CMB-only data. Notice that in the standard model $\widetilde{\Omega}_d$ simply reduces to the present day value of the CDM energy density, $\Omega_{\rm CDM}$. 
Using Planck data alone (blue lines and contours in Fig.~\ref{fig:5th_bounds}) the bound on the strength of the fifth force normalized to gravity is $\beta < 0.011$ at 95\% credible levels (c.l.). This is quite a strong constraint considering that the cosmological model at recombination is mostly unchanged with respect to the standard scenario (see Sections~\ref{sec:bkg} and \ref{sec:fluctuations}), and thus most of the constraining power comes from the geometric projection of physical scales at last scattering onto observed angles on the sky. 

Degeneracies between $\beta$ and both $H_0$ and $\widetilde{\Omega}_d$ are clearly visible, and can be understood in the following way. In a flat $\Lambda$CDM model, keeping the physical density of photons, baryons and CDM fixed, a change in the value of the Hubble constant can be compensated, to ensure flatness, by a different value of the CC energy density $\Omega_{\Lambda}$, which in turn implies a different value of the present day matter density and hence a strong degeneracy in the $\Omega_m\,$-$\,H_0$ plane. This geometric degeneracy is not exact in flat models, and it is internally broken by CMB data, as shown by the dashed contour in Fig.~\ref{fig:5th_bounds}. In particular one cannot take arbitrarily low values of $\Omega_m$ in a $\Lambda$CDM model and still fit the data. 
In the presence of a dark fifth force, and for fixed values of the energy densities at recombination, we are now allowed to increase the Hubble constant to keep the distance to last scattering the same as in $\Lambda$CDM, see the light blue lines in Fig.~\ref{fig:hubble}. This explains the positive correlation between $H_0$ and $\beta$ we observe in Fig.~\ref{fig:5th_bounds}. The other degeneracy lines are now just consequences of this fact, because to ensure flatness with a larger value of $H_0$, a larger $\Omega_\Lambda$, i.e.~a lower $\widetilde{\Omega}_{d}$, is needed. Percent level strengths for $\beta$ correspond to very low values for $\widetilde{\Omega}_d$, and the resulting
degeneracy between the present day DM density (approximately given by $\widetilde{\Omega}_d)$ and the Hubble parameter is much larger than in the standard cosmological model. 
Using only CMB data, the posterior distribution for all three parameters is quite non Gaussian, which, combined with the larger degeneracies discussed above, results in large errorbars for the parameters of interest. In particular we find $66.40 < H_0 / (\text{km/s/Mpc}) < 72.90$ and $0.226 < \widetilde{\Omega}_d < 0.277$ at 95\% c.l.. 

There are several ways to break geometric degeneracies in primary CMB data. The most constraining one is to include BAO data. BAO measures, in units of the sound horizon at the baryon drag epoch $r_d$, the Hubble parameter and the angular diameter distance to the redshift of a given galaxy sample, therefore allowing to constrain the matter density $\Omega_m$ independently from the CMB. BAO measurements are robust to changes in the background and are pretty insensitive to the broadband shape of the galaxy power spectrum and correlation function \cite{Aubourg:2014yra,Sherwin:2018wbu,Carter:2019ulk}. 

However, all BAO measurements assume that the way $r_d$ is imprinted in the distribution of galaxies follows the distribution of the total matter overdensity $\delta_m$. Equivalently, relative density and velocity perturbations between DM and baryons are typically neglected in BAO analyses. It was long ago realized that this approximation might lead to systematic biases in BAO analyses \cite{Tseliakhovich:2010bj,Dalal:2010yt}, since in $\Lambda$CDM the oscillatory features in $\delta_r$ and $\vec{v}_r$ are out of phase with the ones in $\delta_m$. 
Of particular concern is the fact that relative perturbations enter the prediction of the galaxy power spectrum multiplied by unknown free parameters, which can easily be of order one or larger \cite{Schmidt:2016coo,Barreira:2019qdl}. 
As said earlier in Sec.~\ref{sec:fluctuations}, in the $\Lambda$CDM model relative density perturbations do not grow and relative velocities decay, and recent work in Refs.~\cite{Blazek:2015ula,Chen:2019cfu,Givans:2020sez} indicates that, for current surveys and in the standard model, biases to the distances inferred via the BAO method are smaller than the typical measurement error. However, in the presence of long range interactions acting only on one species, like the ones we consider here, relative perturbations actually grow with time, as discussed in Sec.~\ref{sec:fluctuations} and shown in Fig.~\ref{fig:rel_pt}. As a benchmark, for $\beta = 0.005$, which is roughly the 68\% c.l.~bound from Planck data, the relative densities are approximately a factor of $3$ larger, on BAO scales and at $z=0$, than the corresponding fluctuations in the standard model (a factor of 10 at the turnaround of the power spectrum). The relative velocity divergence, $\theta_r$, is now a factor of 100 larger than in $\Lambda$CDM, again at $z=0$. 
%In addition, the free multiplicative bias parameters discussed above could be larger in models with 5th forces, since they measure the response of the small scales galaxy number density to the presence of long wavelength modes. 
At $k = 0.1 \,\text{Mpc/}h$, we find $\delta_r \sim 0.02\, \delta_m$ for $\beta = 0.005$, and the ratio increases linearly with $\beta$. For reference, the current best BAO measurements have few percent accuracy~\cite{BOSS:2016wmc}.

On the other hand, these new relative fluctuations are, to first order, proportional to $\delta_m$, see Eq.~\eqref{eq:delta_r_eds}, and are thus not expected to produce additional phase shifts of the BAO. They add to the off-phase piece produced at recombination, and eventually dominate over the latter after redshift $z\sim 10\,$-$15$. Making these statements more quantitative would require a dedicated BAO analysis framework that includes the effects of relative perturbations in the $\Lambda$CDM model and beyond. Such pipeline is not currently available, and preparing one certainly goes beyond the scope of this work.
Given the numbers in Ref.~\cite{Chen:2019cfu}, we roughly expect that $\beta \sim 0.01$ is the ballpark value at which relative perturbations effects could become significant at BAO scales. This corresponds to the 95\% c.l.~upper bound on $\beta$ from CMB data alone in the 5F scenario.

It is therefore reasonable, albeit with the caveats mentioned above, to combine Planck data, which do not favor a percent value of $\beta$, with BAO data to further improve the bounds. The constraints from Planck plus BAO are shown in orange in Fig.~\ref{fig:5th_bounds}. As expected, BAO break the degeneracy between $H_0$ and $\widetilde{\Omega}_d\,$, both parameters move closer to their $\Lambda$CDM counterparts and are much more Gaussian distributed. The allowed range for the coupling strength also shrinks, with now $\beta < 0.0054$ at 95\% c.l..~To date, this is the strongest bound on dark fifth forces from cosmological data, corresponding to $g_D < 2 \times 10^{-32}\, m_\chi/(10^{-3}\,\mathrm{eV})$ when expressed in terms of the coupling constant of the underlying field theory description.

Given the tail at large values of the Hubble constant in the $H_0$ posterior from Planck data alone, we can also ask whether a dark fifth force can provide a solution to the Hubble tension. We therefore run the CMB likelihood including a prior on $H_0$ from Ref.~\cite{Riess:2019cxk}. The results are shown in green in Fig.~\ref{fig:5th_bounds}. In the $\widetilde{\Omega}_d\,$-$\,H_0$ plane the green contours occupy the leftmost region that was allowed by the CMB-only fits, i.e., a large $H_0$ requires small $\widetilde{\Omega}_d$. The best fit value for the Hubble constant is $H_0 = 72.41 \pm 1.40$ km/s/Mpc at 68\% c.l., clearly compatible with the local measurements. 
Interestingly, there is now evidence, at more than $3.5\,\sigma$ level, for a non-vanishing 5th force strength in the dark sector, with $\beta = 0.0102\, \pm\, 0.0028$ at 68\% c.l.. Clearly, such large values of $\beta$ results in very different distances to the mean redshift of galaxy surveys and in a different matter power spectrum at late times compared to a $\Lambda$CDM model, see for example Fig.~\ref{fig:Pk}. However, as discussed above, a rigorous analysis of late time data, such as the BAO and the Full Shape of the galaxy power spectrum, in presence of nonzero long range forces requires a number of new tools that are yet to be developed.
For these reasons we decided  not to combine a local $H_0$ prior with BAO data. We intend to return to these issues in a forthcoming publication, hoping that our results will further motivate the community to develop more general analysis pipelines of Large Scale Structure data.

Another way to partially lift parameter degeneracies is to include CMB lensing information. This comes with the caveats mentioned in Sec.~\ref{sec:fluctuations} about matter non linearities in the presence of a fifth force. However, given the current uncertainties of the Planck lensing power spectrum and the relatively tight constraints on $\beta$ from primary CMB, we do not expect large biases in the inferred parameters by adding Planck lensing data. This might not be the case for ground based CMB experiments, targeting the lensing power spectrum at much smaller scales \cite{SPT:2019fqo}.
In a $\Lambda$CDM model, CMB lensing is primarily sensitive to a combination of $H_0$, $\Omega_m$ and $\sigma_8$, the amplitude of matter fluctuations at 8 Mpc$/h$. A measurement of the angular size of the sound horizon at last scattering with primary CMB data then breaks this degeneracy, which in turn helps to break the geometric degeneracy in the primary CMB.
The inclusion of the Planck CMB lensing power spectrum data on our constraints is shown in Fig.~\ref{fig:5th_bounds} by the red lines and contours, for the primary CMB plus BAO combination, and in purple for the primary CMB plus local $H_0$ prior case. Unfortunately, in the extended parameter space of our models, Planck lensing information does not yet provide meaningful improvements of the uncertainties over other datasets, as one can see from the minor difference between the orange and red lines, or between the green and purple ones.

Finally, since we discussed a possible explanation of the $H_0$ tension by a dark fifth force, a comment is warranted about the impact on $\sigma_8$. As is well known, this is affected by a moderate but persistent discrepancy between CMB and Large Scale Structure measurements, with the latter giving smaller results~\cite{Heymans:2020gsg,Chen:2022jzq}. At face value (in particular, for fixed bias parameters), in the 5F scenario studied here the $\sigma_8$ tension is expected to worsen.

\subsection{Coupled Dark Energy}
\label{sec:CDE_data}
\begin{figure}
    \centering
    \includegraphics{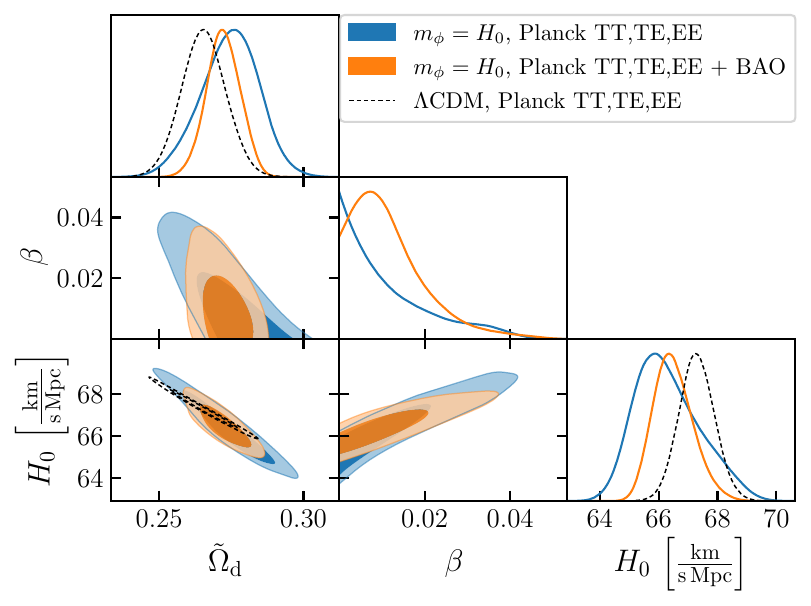}
    \caption{Bounds on the CDE scenario with $m_\varphi/H_0 = 1$. The blue shaded region shows the 1$\sigma$ and 2$\sigma$ contours favored by CMB data from Planck~\cite{Planck:2018vyg,Planck:2019nip}. The orange shaded region also includes BAO measurements~\cite{Kazin:2014qga,Beutler:2011hx,BOSS:2016wmc,Ross:2014qpa}. The dashed contour shows the 2$\sigma$ region favored by Planck CMB data in $\Lambda$CDM.}
    \label{fig:CDE}
\end{figure}

Similar considerations about parameter degeneracies apply to the bounds in the CDE scenario. In this case the effective strength of the fifth force, which contains powers of $\partial \log m_\chi (s)/\partial s$, is suppressed by the fact that $\bar{s}\gtrsim \mathcal{O}(1)$ in order for the new mediator to play the role of the Dark Energy at late times. We thus expect the bound on $\beta$ to be weaker than in the 5F scenario. Figure~\ref{fig:CDE} shows the constraints on our three reference parameters in a CDE Universe with $m_\varphi / H_0 = 1$. 
From the discussion in Sections~\ref{sec:bkg_behavior} and \ref{sec:fluctuations} it is clear that the bound on $\beta$ weakens for smaller mass of the scalar field, hence Fig.~\ref{fig:CDE} corresponds to the most constrained CDE scenario. The blue lines and contours show results using only CMB data. We find that the contours are closer to their $\Lambda$CDM counterparts than the ones for 5F. In particular, the $H_0$ posterior does not have a tail at larger values, implying we should not combine CMB data with a local prior on the Hubble parameter. The bound on $\beta$ is significantly weaker than in the 5F case, with $\beta < 0.034$ at 95\% c.l..~Adding BAO data does not improve the constraints as much as in the 5F case, again because the tuning of the initial conditions to reproduce the CC at late times renders the effect of the fifth force vanishing small, see Figs.~\ref{fig:C_ell} and \ref{fig:Pk}. The upper bound on the coupling parameter becomes $\beta < 0.029$ at 95\% c.l.. We do not show results including lensing information, because this does not add significant constraining power. A summary of the main parameter constraints, for both the 5F and CDE scenarios, is given in Table~\ref{tab:constraints}.\vspace{-4mm}

\renewcommand{\tabcolsep}{4.5pt}
\begin{table}[t]
\centering
\begin{tabular}{cc|c|c|cl}
%\begin{tabular}{p{0.8cm}|p{1.3cm}|p{1.7cm}|p{4.5cm}|p{4.7cm}}
\multicolumn{2}{c|}{95\% c.l.} &   $\widetilde{\Omega}_d$ & $\beta$ & $H_0$ {\small [km/s/Mpc]}   \\\hline
\multirow{3}{*}{\bf{5F}, $m_\varphi/H_0 = 0.1$} & Planck TT, TE, EE   & [0.226, 0.277] & $< 0.0110$ & [66.40, 72.90] \\ 
 & TT, TE, EE$\,+\,$BAO   &  [0.246, 0.270]  & $< 0.0054$  & [67.10, 69.70] \\ %\hline
 & TT,TE,EE+lens.+BAO   &  [0.246, 0.267]  & $<0.0048$  & [67.24, 69.61] \\ \hline
\multirow{2}{*}{\bf{CDE} , $m_\varphi/H_0 = 1$} & Planck TT, TE, EE                  &   [0.253, 0.294] & $< 0.0345$ & [64.40, 68.60] \\ 
      &       TT, TE, EE$\,+\,$BAO        &  [0.260, 0.283] &  $< 0.0287$ & [65.25, 67.85]  \\
\end{tabular}
\caption{Summary of the main constraints obtained in this work on the pure fifth force (5F) and Coupled Dark Energy (CDE) scenarios.}
\label{tab:constraints}
\end{table}

It is worthwhile to comment on how our CDE scenario differs from previous studies of DM-DE interactions. Formally, our setup can be described within the parametrization developed in Refs.~\cite{Gleyzes:2015pma,Gleyzes:2015rua},\footnote{In the notation of Refs.~\cite{Gleyzes:2015pma,Gleyzes:2015rua}, our CDE model corresponds to
\begin{equation*}
C_c = m_\chi^2 (s)\,,\qquad D_c = 0\,,\qquad c_s^2 \alpha =  \frac{2\hspace{0.2mm}\dot{\bar{s}}^2}{\beta H^2}\,, \qquad  \beta_\gamma = - \sqrt{\beta}\, \frac{\partial \log m_\chi (s)}{\partial s}\,,
\end{equation*}
as well as $ \alpha_M = \alpha_B = \alpha_T = 0$, reflecting the fact that gravity per se is not modified. Furthermore, we have the identifications $\pi = \delta s/\dot{\bar{s}}$ and $v_x = - a\, \theta_x/k^2$, and $\Phi_{\rm there} = \Psi$ and $\Psi_{\rm there} = - \Phi$.} but it is not straightforward to compare the numerical results of those works to ours, because of the assumptions made there about the time evolution of the cosmological parameters~\cite{Gleyzes:2015rua}. In particular, Hubble was taken to evolve as in $w$CDM (with constant $w$) and the coupling $\beta_\gamma$ was taken to be time-independent, neither of which applies in our framework. In general, we emphasize that the simplest microscopic model discussed here -- just an ultralight scalar with a trilinear interaction with the DM field -- leads to a complex cosmological evolution, where the dynamics of both the background and the fluctuations are modified.   

General forms for the DM-DE interaction were also introduced in Ref.~\cite{Pourtsidou:2013nha},\footnote{In the notation of Ref.~\cite{Pourtsidou:2013nha}, our CDE model corresponds to a Type-1 theory with $\alpha = \log m_\chi (\varphi)$ and $F = Y + V(\varphi)$. Note that their perturbed KG and continuity equations (91) and (92) are incomplete: each of them is missing a term $\sim \alpha_{\phi \phi}\,$, given in our Eqs.~\eqref{eq:KG_first} and~\eqref{eq:delta_chi_conf} by the pieces proportional to $\partial^2 \log m_\chi (s)/ \partial s^2$.} which however limited the discussion to a qualitative illustration of the expected impact on the CMB and matter power spectra. Notice also that the results in Ref.~\cite{Pourtsidou:2013nha} were presented assuming an exponential form of the field dependent mass $m_\chi (s)$, which is not considered in this work and will be discussed elsewhere.

\subsection{Non-adiabatic initial conditions}
\label{sec:alpha_data}
\begin{figure}
    \centering
    \includegraphics{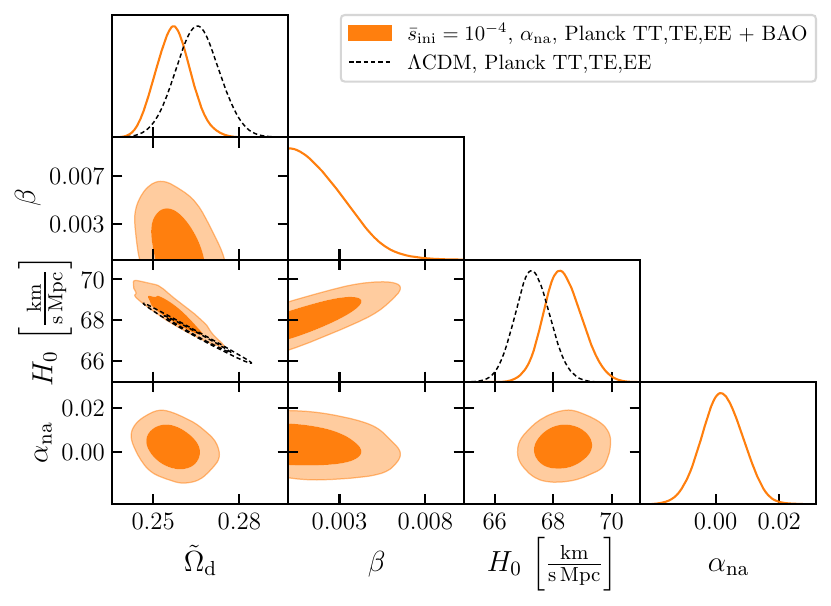}
    \caption{Bounds on the pure 5th force scenario from CMB measurements and BAO. We fix $m_\varphi/H_0 = 0.1$ and include non-adiabatic initial conditions for the perturbations, as parametrized by $\alpha_{\rm na}$ defined in Eq.~\eqref{eq:deltas_sol_IC}.}
    \label{fig:alpha_na}
\end{figure}

Finally, we can use cosmological data to constrain $\alpha_{\rm na}$, which parametrizes possible deviations from adiabatic initial conditions. As discussed in Sec.~\ref{sec:IC}, a non-zero $\alpha_{\rm na}$ sources a non zero relative velocity between the DM and the other species on super-horizon scales. We assume $\alpha_{\rm na}$ to be scale-independent and, for simplicity, we only show in Fig.~\ref{fig:alpha_na} the bounds in the 5F scenario, combining CMB and BAO data. As expected, the constraint in CDE models is weaker because the effective non-adiabaticity turns out to be proportional to $\alpha_{\rm na}(\partial \log m_\chi (s)/\partial s)$, see Eq.~\eqref{eq:theta_na}. For 5F we find $\alpha_{\rm na} = 0.0021 \pm 0.0065$ at 68\% c.l., which is clearly compatible with zero. For such small values, the constraints on the other model parameters are basically unchanged. We can understand this by noticing that $\beta$ does not modify much the evolution of super-horizon modes, as illustrated by Fig.~\ref{fig:pert1}. The constraint on $\alpha_{\rm na}$ is quantitatively of the same order of the ones on other kinds of non-adiabatic initial conditions~\cite{Planck:2018vyg}, e.g.~isocurvature perturbations.

\section{Interplay with Visible Fifth Forces}\label{sec:implications_particle}
So far we assumed the long range fifth force to be entirely confined to the dark sector. However, it is interesting to consider the possibility that the same fifth force couples to the visible matter. In this situation, depending on the specific realization, one can study the interplay between the new bounds on dark fifth forces derived in Sec.~\ref{sec:constraints} and the existing constraints on visible long range forces. Since we focus on scenarios where $m_\varphi \lesssim H_0$, the mediator can be taken as effectively massless on all the scales relevant for experimental tests of gravity in the visible sector. We assume the DM mass to satisfy the naturalness bound in Eq.~\eqref{eq:naturalness_bound}, which requires it to be smaller than approximately $0.01$~eV. In this regime the DM is an ultralight boson, possibly produced in the early Universe via the misalignment mechanism~\cite{Preskill:1982cy,Abbott:1982af,Dine:1982ah}.

For $m_\varphi \lesssim H_0$ the effect of the visible fifth force on the interaction between two bodies with masses $m_A$ and $m_B$ at distance $r$ can be parametrized as 
\begin{equation}
V_{\text{visible}}=-G_N \frac{m_A m_B}{r}\left[1+\alpha_A\alpha_B\right]\ ,
\end{equation}
where the Yukawa factor $e^{-m_\varphi r}$ can be dropped for such a light mediator, so that no  deviation from the Newtonian potential would arise.\footnote{The experimental tests of Newton's inverse square law range from length scales of $10^{-6}\,\text{m}$ to few au and are sensitive to mediator masses down to $10^{-18}\,\text{eV}$~\cite{Adelberger:2003zx}. Below this mass, deviations from the inverse square law decouple like $m_\varphi L_{\text{exp}}$ where $L_{\text{exp}}$ is the size of the experimental apparatus. This would result in tiny deviations at the $10^{-15}$ level or below for the range of mediator masses considered in this work.} The coupling $\alpha_{A}$ can be related to the field dependence of the macroscopic masses, $\alpha_A = (4\pi G_N)^{-1/2} (\partial \log m_A(\varphi)/\partial \varphi)$. In general this coupling contains a universal part, which results in an unobservable shift of the Newton constant, and a non-universal part which depends non-trivially on the composition of the material and results in effective violation of the EP on macroscopic scales. The latter effect can be measured as the difference in the acceleration of two bodies with different compositions and has been extensively tested experimentally (see e.g. Ref.~\cite{Tino:2020nla} and references therein). 

The macroscopic coupling $\alpha_A$ can be expressed in terms of the microscopic couplings of the light scalar with the light Standard Model (SM) fields, which can be written in general as
\begin{equation}\label{eq:dilaton_couplings}
\mathcal{L} = \sqrt{4\pi G_N}\, \varphi \bigg( \frac{d_e}{4e^2}F_{\mu\nu}F^{\mu\nu} - \frac{d_g \beta_3}{2g_3}G_{\mu\nu}^a G^{\mu\nu\,a} - d_{m_e} m_e \bar{e}e - \sum_{q\, =\, u,d} (d_{m_q} +  \gamma_{m_q}d_g) m_q \bar{q} q  \bigg)\,,
\end{equation}
where $e$ and $g_3$ are the electromagnetic and QCD couplings respectively, $\beta_3 = -b_3 g_3^3/16\pi^2$ with $b_3 = (11-2 N_f/3)$ is the beta function encoding the evolution of the QCD coupling constant with energy, and $\gamma_{m_q}$ are the anomalous dimensions of the quark masses. Notice that a non-canonical normalization has been adopted for the electromagnetic gauge field. The analysis of Refs.~\cite{Damour:2010rm,Damour:2010rp} calculated the leading effects of the microscopic coefficients $d_i$ on the macroscopic parameter $\alpha_A$,
\begin{equation}
\alpha_A \simeq d_{g}^{\ast} + (d_{\hat{m}} - d_g) [Q^\prime_{\hat{m}}]_A + d_e [Q'_e]_A\,,\label{eq:alphaeff}
\end{equation}
where $d_g^\ast= d_g + 0.093(d_{\hat{m}} - d_g) + 2.7 \times 10^{-4}\, d_e$ corresponds to the composition independent part and we defined $d_{\hat{m}} = (d_{m_u} m_u + d_{m_d}m_d)/(m_u + m_d)$. The coupling of the scalar to the electromagnetic field strength, $d_e\,$, is numerically suppressed in the matching to $\alpha_A$ compared to $d_g$ and $d_{\hat{m}}$, due to the weak dependence of nuclear binding energies on electromagnetism. $Q^\prime_{\hat{m}}\simeq-0.036\, A^{-1/3}- 1.4\times 10^{-4} Z(Z-1) A^{-4/3}$ and $Q'_e\simeq 7.7\times 10^{-4} Z(Z-1) A^{-4/3}$ in Eq.~\eqref{eq:alphaeff} are material dependent ``dilaton charges'' ($A$ is the atomic mass number and $Z$ the atomic number) as approximately derived in Refs.~\cite{Damour:2010rm,Damour:2010rp}. 
Experimental tests of the EP place important constraints on new long range interactions in the visible sector. The best Earth-based limit comes from the E\"ot-Wash experiment~\cite{Schlamminger:2007ht}, which set a bound on the parameter combination $|d_g^\ast (d_{\hat{m}} - d_g + 0.22\,d_e)| < 5.1\times 10^{-11}$ at $2\sigma$ level. An even stronger limit has been obtained by the MICROSCOPE space mission~\cite{Touboul:2017grn,MICROSCOPE:2022doy}, which found at $2\sigma$
\begin{equation}
\vert  d_g^\ast\left( d_{\hat{m}} - d_g + 0.62\, d_e\right)\vert< 1.7\times 10^{-12}\, .\label{eq:EPmicroscope}
\end{equation}
This corresponds to the constraints $|d_g|< 1.3\times 10^{-6}$, $|d_{\hat{m}}|< 4.2\times 10^{-6} $ and $|d_e| < 9.9\times 10^{-5}$ if one coupling is switched on at a time in Eq.~\eqref{eq:dilaton_couplings}. Again we see that the electromagnetic coupling is subject to a sizably weaker bound. One should also keep in mind that, as pointed out in Ref.~\cite{Oswald:2021vtc}, the constraints on EP violation can be significantly relaxed if the microscopic couplings conspire to give macroscopic effects that resemble very much the universality of Newtonian gravity (at least for the class of elements that have been tested experimentally so far).

The direct couplings of $\varphi$ to the SM in Eq.~\eqref{eq:dilaton_couplings} induce temporal variations of the SM parameters. These can be tested by comparing the transition frequencies of different atomic clocks, as explored in Refs.~\cite{Flambaum:2002de,Arvanitaki:2014faa,Safronova:2017xyt}. Since the scalar masses considered here are much smaller than the inverse of the typical run time of the atomic clock experiments ($\sim \text{years}$), the only observable effect induced by the variation of $\varphi$ is a steady drift in the frequency ratio between the two clocks. For two different optical transitions, the drift rate is controlled by the change of the fine structure constant $\alpha$, while for one hyperfine microwave transition and one optical transition there is also sensitivity to the change of $\mu_A/\mu_{\rm B}$, the ratio between the nuclear magnetic moment and the Bohr magneton, which is linearly proportional to the inverse of $\mu = m_p/m_e$~\cite{Flambaum:2004tm}. 

The drifts in the fundamental constants can be related through Eq.~\eqref{eq:dilaton_couplings} to the time evolution of the scalar field,
\begin{equation} \label{eq:drift_rate_A}
\frac{\dot{\alpha}}{\alpha} = \sqrt{4\pi G_N}\, d_e \dot{\varphi}\ ,\qquad
\frac{\dot{\mu}}{\mu} = \sqrt{4\pi G_N}\, (d_g - d_{m_e} - M_A d_{\hat{m}} ) \dot{\varphi}\ ,
\end{equation}
where $d_{\hat{m}}$ is defined below Eq.~\eqref{eq:alphaeff} and $M_A$ was estimated in Ref.~\cite{Flambaum:2004tm} for several nuclei. Next, recalling Eq.~\eqref{eq:rho_p_s} we write $\dot{\varphi}=-\sqrt{\bar{\rho}_{s}(1+w_s)}$, where the negative sign is the correct one in our setup. This allows us to finally express the present day values of the drifts as (see also Ref.~\cite{Euclid:2021cfn})
\begin{align} \label{eq:drift_rate_B}
&\left( \frac{\dot{\alpha}}{\alpha} \right)_0 = -\,d_e H_0\sqrt{3\Omega_s^0(1+w_s^0)/2}\ ,\\
&\left( \frac{\dot{\mu}}{\mu} \right)_0=-\,(d_g - d_{m_e} - M_A d_{\hat{m}} )H_0\sqrt{3\Omega_s^0(1+w_s^0)/2}\ .
\end{align}
These equations make it apparent that the atomic clock constraints can be weakened if the present day energy density of the scalar field is small ($\Omega_s^0 \ll 1$), or its equation of state is near the CC one ($w_s^0 \approx -1$).

The best experimental constraints on the drifts of the fine structure constant and proton-to-electron mass ratio come from measurements with ytterbium ion clocks and caesium clocks~\cite{Lange:2020cul,Filzinger:2023zrs}. The resulting $2\sigma$ bounds $\left(\dot{\alpha}/\alpha \right)_0 = (1.8\, \pm\, 5.0) \times 10^{-19}\; \mathrm{yr}^{-1}$ and $\left( \dot{\mu}/\mu\right)_0 = (- 8  \pm 72) \times 10^{-18}\; \mathrm{yr}^{-1}$ constitute improvements by a factor of 90 and 2, respectively, with respect to previous measurements~\cite{Rosenband,McGrew:19}. In terms of the couplings in Eq.~\eqref{eq:dilaton_couplings}, the $2\sigma$ bounds read
\begin{equation}\label{eq:clocksmeasure}
|d_e| < \frac{6.1\times 10^{-9}}{\sqrt{\Omega_s^0(1+w_s^0)}}\,, \qquad
|d_g - d_{m_e} - M_A d_{\hat{m}}| < \frac{8.8\times 10^{-7}}{\sqrt{\Omega_s^0(1+w_s^0)}}\; ,
\end{equation}
where the central values of the measurements were neglected for simplicity. 

In the 5F scenario, $w_s^0$ is far from $-1$ (it is $\approx +1$ if $m_\varphi \ll H_0$) and the present day fraction of energy density in the scalar field is approximately given by $\Omega_s^0 \simeq \beta f_\chi^2/3\,$, where $\beta$ is the coupling of the scalar to DM, see Eq.~\eqref{eq:rhos_MD}. In light of the constraints on $\beta$ from CMB+BAO data presented in Sec.~\ref{sec:Hubbletension}, we thus expect $\Omega_s^0 \lesssim 10^{-3}$. Accounting for this suppression, we find by comparison with the MICROSCOPE bounds reported below Eq.~\eqref{eq:EPmicroscope} that the sensitivity of atomic clocks to the scalar-photon coupling is stronger than EP tests by at least two orders of magnitude. On the other hand, for the scalar-gluon coupling MICROSCOPE remains the most sensitive probe. In the CDE scenario, $\Omega_s^0 \sim \mathcal{O}(1)$ and $w_s^0$ can deviate at the $10\%$ level from the CC equation of state, as shown in Fig.~\ref{fig:hubble}, only leading to a mild weakening of the atomic clock constraints. In this case atomic clocks always dominate over EP tests.  

In the remainder of this section we discuss in explicit scenarios the interplay of the MICROSCOPE and atomic clock constraints with the dark fifth force bounds presented in Sec.~\ref{sec:constraints}. This depends on how the couplings in Eq.~\eqref{eq:dilaton_couplings} scale with $\beta$. In Sec.~\ref{sec:loops} we discuss the most generic scenario where the fifth force coupling to the SM is only induced through DM loops, while in Sec.~\ref{sec:visible} we consider a model where the fifth force couples directly to both DM and the SM.

\subsection{Visible fifth force from Dark Matter loops}\label{sec:loops}
\begin{figure}
    \centering
    \includegraphics[width = 0.75\textwidth]{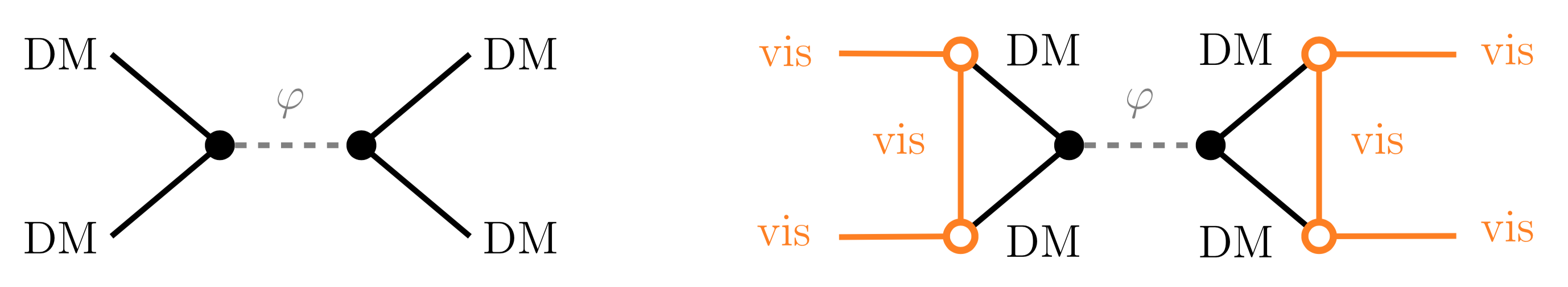}
    \caption{Generic interplay between dark and visible fifth forces. According to Eq.~\eqref{eq:axion_model}, the light force mediator is coupled to DM at tree level, while the coupling to the SM only arises via DM loops.}
    \label{fig:Feynman_loop}
\end{figure}

We consider a scenario where the fifth force is sequestered within the dark sector and interacts only with DM at tree level. On the other hand, we assume that DM couples to the SM directly and we want to estimate the size of the fifth force coupling to the visible sector induced by DM loops, see Fig.~\ref{fig:Feynman_loop}. Specifically, having in mind QCD axion models, we consider the simple case of pseudoscalar DM $a$ coupled to photons and gluons at one loop and to the dark fifth force field $\varphi$ at tree level,
\begin{equation}\label{eq:axion_model}
\mathcal{L} \supset \frac{\alpha}{8\pi}\frac{E}{N}\frac{a}{f_a} F_{\mu\nu}\widetilde{F}^{\mu\nu} + \frac{\alpha_3}{8\pi}\frac{a}{f_a} G_{\mu\nu}^a\widetilde{G}^{\mu\nu\,a} - g_D m_a \varphi a^2\,.
\end{equation}
In this simple scenario, taking $E/N\sim \mathcal{O}(1)$ a DM loop would induce a coupling between the fifth force and the SM photons and gluons with the approximate size 
\begin{align}
&d_e\simeq\; \sqrt{\beta} \left( \frac{m_a}{4\pi f_a}\right)^2 \frac{\alpha^2}{16\pi^2} \simeq 2\times 10^{-10}\;\sqrt{\frac{\beta}{0.01}} \left(\frac{m_a}{f_a}\right)^2 \, ,\label{eq:loop_induced_5Fe}\\
&d_g\simeq  \sqrt{\beta} \left( \frac{m_a}{4\pi f_a}\right)^2 \frac{\alpha_3}{8\pi b_3}\simeq 3\times 10^{-6}  \;\sqrt{\frac{\beta}{0.01}} \left(\frac{m_a}{f_a}\right)^2\; , \label{eq:loop_induced_5Fg}
\end{align}
where the couplings to the gauge field strengths were defined in Eq.~\eqref{eq:dilaton_couplings}. If a nonzero value of $\beta$ were to be found by cosmological measurements, for instance $\beta\simeq 0.01$ as motivated by the present Hubble tension (see Sec.~\ref{sec:Hubbletension}), the constraint on EP violation in the visible sector given in Eq.~\eqref{eq:EPmicroscope} would translate into bounds on the axion  parameter space. The existence of a fifth force could then have implications for the detectability of axion DM. For the axion-photon coupling, the suppression of $d_e$ given by the smallness of $\alpha$ and the suppressed experimental sensitivity of fifth force constraints do not lead to any interesting bounds on $m_a/f_a$. For the axion-gluon coupling, the strength of the visible long range force in Eq.~\eqref{eq:loop_induced_5Fg} is at the level of the current experimental sensitivity in Eq.~\eqref{eq:EPmicroscope} and Eq.~\eqref{eq:clocksmeasure} only for $m_a/f_a\sim \mathcal{O}(1)$, which is already abundantly excluded by laboratory and astrophysical probes of light axions. Thus, the sequestering of the fifth force within the dark sector is enough to keep the axion DM detectable in future direct searches (see for example Ref.~\cite{Graham:2015ouw} for a review).    

As a final remark on the parametrics of Eqs.~\eqref{eq:loop_induced_5Fe} and~\eqref{eq:loop_induced_5Fg}, we notice that the weakness of the loop-induced visible fifth force is essentially related to the smallness of $m_a/f_a$, which is constrained to be less than about $10^{-18}$ if the mass of the axion is fixed on the QCD line $m_a=5.7 \,\mu \mathrm{eV}\,(10^{12}\;\mathrm{GeV}/f_a)$~\cite{GrillidiCortona:2015jxo}. This suggests that models with larger $m_a/f_a$ could realize parametrically larger visible fifth forces. As discussed in Sec.~\ref{sec:naturalness}, however, heavier DM would require either a symmetry mechanism in the dark sector (such as supersymmetry) to preserve the naturalness of the fifth force mass, or to accept a large amount of fine-tuning. Under this non-minimal assumption, the implications of the existence of a dark fifth force on the direct detection prospects for WIMP DM were previously investigated in Refs.~\cite{Bovy:2008gh,Carroll:2008ub}.

\subsection{Visible and dark fifth forces from the radial mode}\label{sec:visible}
\begin{figure}
    \centering
    \includegraphics[width = 0.65\textwidth]{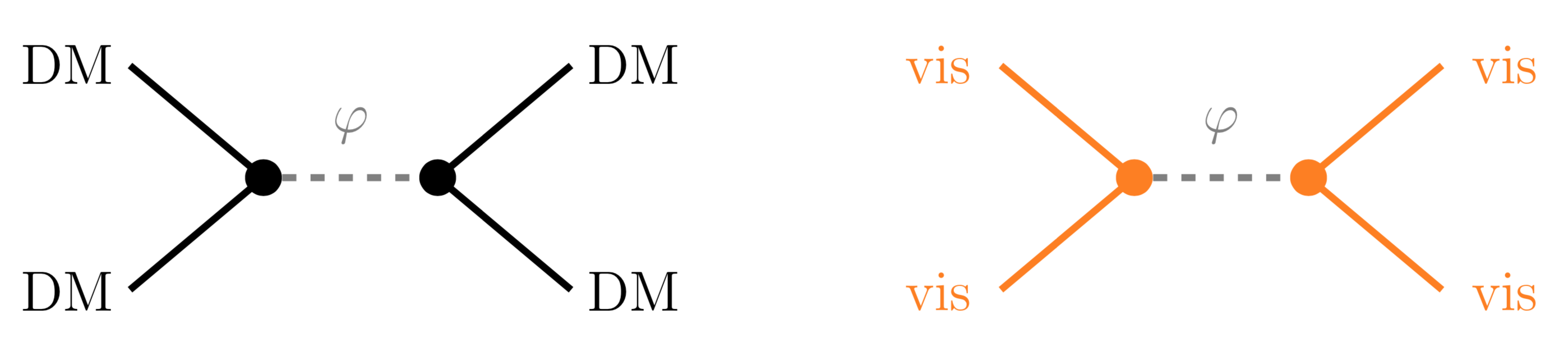}
    \caption{Second scenario illustrating the interplay between dark and visible fifth forces. The light force mediator is coupled to DM at tree level, Eq.~\eqref{eq:trilinearcoupling}, while the coupling to the SM arises from the effective interaction in Eq.~\eqref{eq:photoncoupling}.}
    \label{fig:Feynman_tree}
\end{figure}

We now consider a simple model where the fifth force couples directly to both DM and the SM, see Fig.~\ref{fig:Feynman_tree}. The scalar DM $\chi$ is an axion, while the fifth force mediator $\varphi$ is the radial mode of a complex scalar field $\Phi$ with Lagrangian
\begin{equation}\label{eq:L_Phi}
\mathcal{L}_{\Phi} = -\, \partial_\mu \Phi^\ast \partial^\mu \Phi +m^2 \Phi^\ast \Phi - \lambda (\Phi^\ast \Phi)^2 + \frac{1}{4}\delta m^2 (\Phi - \Phi^\ast)^2 \,,  
\end{equation}
where the $U(1)$ symmetry acting on the complex field ($\Phi\to e^{i\alpha}\Phi$) is both spontaneously broken by the Mexican-hat potential and explicitly broken by the last term controlled by $\delta m^2$. The latter preserves the $Z_2$ symmetry $\Phi \to \Phi^\ast$ that stabilizes the DM $\chi$. Parametrizing the complex field as $\Phi = (f + \varphi)e^{i \chi/f}/\sqrt{2}$ with the vacuum expectation value given by $f = \sqrt{m^2/\lambda}\,$, we obtain
\begin{align}
\mathcal{L}_{\Phi} =\,& - \frac{1}{2}\partial_\mu \varphi \partial^\mu \varphi - \frac{1}{2} \Big( 1 + \frac{\varphi}{f} \Big)^2 \partial_\mu \chi \partial^\mu \chi - \frac{1}{2}(2\lambda f^2) \varphi^2 + \mathcal{O}(\varphi^3, \varphi^4) \nonumber\\
-\,&\frac{1}{2}\delta m^2 \Big( \chi^2 + 2\frac{\varphi\chi^2}{f} + \frac{\varphi^2 \chi^2}{f^2} + \mathcal{O}(\chi^4) + \ldots \Big)\,.\label{eq:trilinearcoupling}
\end{align}
Matching to Eq.~\eqref{eq:Lag_scalar} leads to the following parameter identifications,
\begin{equation}
\delta m^2 = m_\chi^2 (1 + 2 \bar{s}_0)\,,\quad f = \sqrt{\frac{2}{\beta}}\,  (1 + 2 \bar{s}_0) M_{\rm Pl} \,,\quad \lambda =  \frac{\beta}{4(1 + 2 \bar{s}_0)^2} \frac{m_\varphi^2}{M_{\rm Pl}^2}\,, \quad m^2 = \frac{m_\varphi^2}{2}\, ,
\end{equation}
where from the equations above we can see that an extremely small quartic coupling and a super Planckian decay constant are required to realize the dark fifth force mediator as the radial mode of a vanilla axion DM model. For $\beta \sim 10^{-2}$ and $m_\varphi \sim H_0$ we need $\lambda \sim 10^{-123}$. The trilinear coupling between the axion DM and the fifth force mediator is given by $\delta m^2/f= \sqrt{\beta/2}\, m_\chi^2/M_{\text{Pl}}$. Furthermore, several interactions are generated beyond those we have considered in our cosmological analysis, including a derivative $(\varphi/f)(\partial \chi)^2$ coupling as well as a $\varphi^2 \chi^2$ term. However, these are unlikely to substantially change the results obtained in Sec.~\ref{sec:constraints}.

We note that, since $\delta m^2 \gg \lambda f^2$, we cannot view the last term in Eq.~\eqref{eq:L_Phi} as a small perturbation to the $U(1)$ invariant Lagrangian for $\Phi$. Therefore, the choice to include only the specific explicit breaking structure $(\Phi - \Phi^\ast)^2$ appears to be rather ad-hoc. In particular, there is no symmetry argument justifying the suppression of the explicit breaking operator $(\Phi + \Phi^\ast)^2$, which would induce a mass term for $\varphi$ of order $m_\chi^2$. In this respect, the theory discussed here should be considered as a  toy model, of somewhat limited theoretical interest but still useful to compare the parametric sensitivity of different fifth force probes.

Now assuming the presence of some electrically charged chiral fermions with Yukawa couplings to $\Phi$, the effective operator
\begin{equation}
\delta \mathcal{L}_\Phi = c_\psi \frac{\alpha}{4\pi} \frac{\Phi^\ast \Phi}{f^2} F_{\mu\nu}F^{\mu\nu} \quad \to \quad c_\psi \frac{\alpha}{4\pi} \frac{\varphi}{f} F_{\mu\nu}F^{\mu\nu}   \label{eq:photoncoupling}
\end{equation}
is generated with $c_\psi \sim \mathcal{O}(1)\,$, together with the usual anomalous coupling of the axion DM to photons. The corresponding value of the effective coupling in Eq.~\eqref{eq:dilaton_couplings} is $d_e =c_\psi \alpha \sqrt{\beta}/[(1 + 2 \bar{s}_0)\pi]$ and the resulting $2\sigma$ bound from MICROSCOPE is $\beta <  0.0018\,(1 + 2 \bar{s}_0)^2/c_\psi^2$. Interestingly, this is comparable to our new bound from dark fifth forces. However, atomic clocks give much more stringent constraints both in the 5F and CDE scenarios, requiring 
\begin{align}
&\beta < 4.5\times 10^{-6}\,\frac{1}{c_\psi f_\chi\sqrt{1+w_s^0}}\;\,,\qquad\;\; \text{(5F)} \\
&\beta < 6.8\times 10^{-12}\,\frac{(1+2 \bar{s}_0)^2}{c_\psi^2 \Omega^0_s(1+w_s^0)}\;. \qquad (\text{CDE})
\end{align}
To summarize, in this minimal model the current sensitivity of atomic clock tests in the visible sector surpasses the bounds on dark fifth forces we have derived here. A similar result would hold if we considered a coupling of $\Phi$ to gluons. In passing, we note that the super Planckian decay constants obtained in this simple model leave no hope of testing the axion directly in ultralight DM searches. 

Finally, we briefly go back to the issue of fine-tuning discussed in Sec.~\ref{sec:naturalness}. The introduction of new couplings of the ultralight scalar field to the SM raises again the concern of how the tiny $m_\varphi$ can be stabilized against quantum corrections. In the simple setup presented here this issue can be explicitly quantified by estimating the one-loop contribution to the quartic coupling $\lambda$ induced by the non-renormalizable interaction in Eq.~\eqref{eq:photoncoupling}, $\Delta\lambda\simeq c_\psi^2\alpha^2\Lambda_{
\text{UV}}^4/(4\pi f)^4$. The ultraviolet cutoff $\Lambda_{\text{UV}}$ corresponds to the masses of the new electrically charged chiral fermions, which must be heavier than the electroweak scale for $\mathcal{O}(1)$ charges, thereby generating a severe fine-tuning problem for the scalar field mass. The issue of introducing testable fifth forces coupled to the SM without fine-tuning is a longstanding one, as reviewed e.g.~in Ref.~\cite{Arvanitaki:2014faa}. At present, a convincing dynamical solution does not seem to exist, though interesting attempts have been made~\cite{Damour:1994zq,Damour:1994ya,Damour:2002mi,Damour:2002nv}. Remarkably, for a fifth force that is sequestered within the dark sector and whose couplings to the SM arise only from DM loops, as discussed in Sec.~\ref{sec:loops}, this fine-tuning issue is solved by taking the DM mass to be sufficiently light, as quantified in Eq.~\eqref{eq:naturalness_bound}.

\section{Outlook}\label{sec:conclusions}

This work starts a systematic investigation of the phenomenology of dark long range interactions in cosmology. Here we briefly summarize our findings and present a number of future directions. 

In this first exploration we focused on scalar dark fifth forces, and derived the corresponding equations for background and perturbations. We focused on the cosmology of dark fifth forces that can be mapped to minimal  microscopic theories with natural parameter choices. The analytical results we derived in Sections~\ref{sec:bkg} and~\ref{sec:fluctuations} show the build-up of the effects of dark fifth forces over time, and clarify why cosmology is such a powerful probe of long range dynamics in the dark sector: even a parametrically small effect can become relevant after more than 13 billion years.

A very light scalar mediator could also account for the Dark Energy, with a very different phenomenology compared to the case where the new degree of freedom purely mediates a new interaction. We dub the former the Coupled Dark Energy (CDE) scenario, and the latter the pure 5th force (5F) scenario. The fundamental parameter of the theory is $\beta$, the ratio between the strength of the new interaction and the Newton constant. In the 5F case, we find $\beta<0.0053$ at 95\% c.l.~by combining CMB and BAO data. This bound weakens by approximately a factor of 8 in the CDE case with $m_\varphi = H_0$, due to the fine-tuning required to match the value of the Cosmological Constant. For smaller masses of the CDE field the bound weakens further, scaling approximately at least as $m_\varphi/H_0$. These are the strongest constraints on Equivalence Principle violation in the dark sector.

Our work opens up two sets of different but complementary questions: i) on the phenomenological side, it would be important to explore the nonlinear cosmology of dark fifth forces; ii) on the theoretical side, it would be interesting to explore non-minimal scalar theories, vector mediators and the interplay with inflationary models. We now briefly present these future directions in turn.  

The rich phenomenology beyond linear theory of our models is yet to be investigated. For example, we made the compelling case for a careful study of the mildly nonlinear regime of structure formation, since we have shown that dark fifth forces could alleviate the well-known $H_0$ tension between CMB and local measurements by simply invalidating the use of BAO information. This was possible because the traditional BAO analyses cannot be applied to cosmologies with large density or velocity perturbations between the different species, as it is the case if DM violates the Equivalence Principle. The development of a dedicated BAO analysis pipeline is therefore of the highest priority. More generally, the shape of the power spectrum is much more affected by the presence of new dark long range interactions than the CMB anisotropies. Analyzing galaxy clustering data however requires extending next-to-leading-order perturbative calculations of Large Scale Structure beyond the current state of the art. For reference, a value of $\beta = 0.005$, close to the current 95\% c.l.~upper bound from CMB plus BAO data, yields approximately 5\% differences in the shape of the power spectrum compared to the $\Lambda$CDM case.  

Moving to smaller scales, many astrophysical systems could be affected by the presence of a new dark long range interaction, and we list here just a few examples. 

At very high redshifts, it is well known that relative velocities between DM and baryons have an important effect on the formation of the first stars and galaxies, and can vastly modify the shape of the 21~cm power spectrum at cosmic dawn. In our models, the relative velocities are one order of magnitude larger than in a $\Lambda$CDM scenario, and it would be therefore interesting to study their consequences for the high redshift Universe. Other examples concern the collapse of DM halos and their internal structure.  We expect more halos in cosmology with a fifth force, simply because matter can accrete faster. Their internal structure could also be different than in $\Lambda$CDM, and this may be tested for instance with strong lensing data. The abundance and the profile of small mass halos are also related to the number of satellite galaxies, which could be studied assuming a model for the baryonic physics in the presence of a 5th force.
Finally, the dynamics of stars and of stellar streams is sensitive to the DM distribution inside galaxies, and could be tested with kinematics data of JWST (the former) and of Gaia (the latter). 
Our modified version of CLASS provides the baseline for all the above investigations. 

The other set of questions are related to the theoretical foundations of the framework. While we discussed initial conditions from the perspective of initializing the linear theory equations, a period of cosmological inflation can provide a dynamical origin for the initial conditions of the Universe. It is therefore interesting to ask how to embed a new very light and interacting degree of freedom into inflation. Actually, a general prediction of inflation models with light scalar spectator fields is the generation of non-adiabaticity in the initial conditions and of Primordial non-Gaussianities, with important observational consequences.

One could also consider more general interactions with scalar fields than the ones studied in this work. One example are dilaton-like couplings, which are interesting because they would evade the suppression of the effective strength of the fifth force in CDE scenarios that we discussed here. 

Finally, it is tempting to look into light Abelian vector mediators, whose very small mass can be technically natural. Abelian gauge theories are screened at large distances if there is no net dark charge in the Universe, therefore one would expect a less rich phenomenology than in the scalar case. However, at the level of the perturbations, light interacting vector bosons will generate vector perturbations, which are severely constrained by data. Clearly, if the total dark charge in the Universe is not zero we can anticipate major differences with $\Lambda$CDM at all scales.

We hope to hit the road in all these various directions in future work.

\section*{Acknowledgments}
We thank Marko Simonovi\'c for collaboration at the initial stage of this project. We thank Sergey Sibiryakov for useful discussions, and Christophe Grojean, Filippo Vernizzi and Tomer Volansky for stimulating comments. We are grateful to Salvatore Bottaro and Marco Costa for pointing out some mistakes in an earlier version of the paper (v2). We thank Marco Costa, Cyril Creque-Sarbinowski, Olivier Simon and Zachary Weiner for pointing out that in the previous v3 some coefficients in the analytical discussion of sub-horizon solutions (Sec.~\ref{sec:sub_hor}) were incorrect. ES acknowledges partial support from the EU’s Horizon 2020 programme under the MSCA grant agreement 860881-HIDDeN.

\appendix

\section{Particle vs Field Description}\label{sec:appendixFields}
Here we show an alternative method to derive the equations governing the evolution of $\chi$, making use of the field-theoretical viewpoint. For ease of exposition we focus on scalar DM. The starting point is the microscopic Lagrangian, which can be written as
\begin{equation}
\mathcal{L} = - \frac{1}{2}\partial_\mu \chi \partial^\mu \chi - V_\chi - \frac{1}{2G_s} \partial_\mu s \partial^\mu s - V_s - V_{\rm int}\,.
\end{equation}
We define the $\chi$ fluid as ``containing the interaction'', 
\begin{align}
(T_{\chi})_{\mu\nu} =&\, \partial_\mu \chi \partial_\nu \chi + g_{\mu\nu} \Big( - \frac{1}{2} \partial_\alpha \chi \partial^\alpha \chi - V_\chi - V_{\rm int} \Big)\,, \\
(T_s)_{\mu\nu} =&\, \frac{1}{G_s} \partial_\mu s \partial_\nu s + g_{\mu\nu} \Big( - \frac{1}{2G_s} \partial_\alpha s \partial^\alpha s - V_s \Big)\,,
\end{align}
so that the equations expressing the non-conservation of the energy-momentum tensor for each species read
\begin{equation}\label{eq:Tmunu_non_cons}
\nabla_\mu (T_\chi)^{\mu}_{\;\;\;\nu} = - \frac{\partial V_{\rm int}}{\partial s}\, \partial_\nu s \,, \qquad \nabla_\mu (T_s)^{\mu}_{\;\;\;\nu} = + \frac{\partial V_{\rm int}}{\partial s}\, \partial_\nu s \,.
\end{equation}
Notice that the transfer four-vector appearing on the right-hand sides is proportional to the scalar field four-velocity, $\partial^\nu s \propto u_{s}^\nu$. Writing the DM stress tensor as~\cite{Langlois:2006iq}
\begin{equation} \label{eq:Tmunu_newframe}
T^{\mu\nu}_\chi = (\rho_\chi^{\rm RF} + \mathcal{P}_\chi^{\rm RF}) u^\mu u^\nu + \mathcal{P}_\chi^{\rm RF} \,g^{\mu\nu}\,,\qquad u^\mu = \gamma (1, \vec{v}/a)\,,
\end{equation}
with $\gamma \equiv 1/ \sqrt{1 - v^2}$ (satisfying $u^\mu u^\nu g_{\mu \nu} = -1$) and RF denoting rest-frame quantities, the energy density and pressure are given by
\begin{align}
-\, (T_\chi)^{0}_{\;\;0} =\,&\,  \rho_\chi =  \gamma^2 \rho^{\rm RF}_\chi + (\gamma^2 - 1) \mathcal{P}^{\rm RF}_\chi\,,  \\
\frac{1}{3} \delta^i_{\;\;j} (T_\chi)^j_{\;\;i} =\,&\, \mathcal{P}_\chi = \mathcal{P}^{\rm RF}_\chi + \frac{1}{3} \gamma^2 v^2 (\rho_\chi^{\rm RF} + \mathcal{P}^{\rm RF}_\chi)\,.
\end{align}
In the rest frame we also have
\begin{equation}
\rho_\chi^{\rm RF} = -\frac{1}{2} \partial^0 \chi \partial_0 \chi + V_\chi + V_{\rm int}\,, \qquad \mathcal{P}_\chi^{\rm RF} = -\frac{1}{2} \partial^0 \chi \partial_0 \chi  - V_\chi - V_{\rm int}\,. \end{equation}
Assuming $\chi$ to be pressureless in its rest frame, $\mathcal{P}_\chi^{\rm RF} = 0\,$, we thus find $\rho_\chi^{\rm RF} = 2 (V_\chi + V_{\rm int})$ and finally
\begin{equation}
 \rho_\chi - 3 \mathcal{P}_\chi = \rho_\chi^{\rm RF} = \frac{\partial V_{\mathrm{int}}/\partial s} {\partial \log m_\chi(s)/ \partial s}\,,
\end{equation}
where the second equality can be easily verified for the class of potentials $V_{\chi}\,, V_{\rm int}$ considered in this work, which are both quadratic in $\chi$. In the pressureless limit, the above equation reduces to Eq.~\eqref{eq:micro_macro}.

On the background, the time component of the first equation in~\eqref{eq:Tmunu_non_cons} takes the form
\begin{equation}\label{eq:DM_Tmunu_interm}
\nabla_\mu (T_\chi)^{\mu}_{\;\;\;0} = (\dot{T}_\chi)^0_{\;\;0} + 3 H (T_\chi)^0_{\;\;0} - H \delta^i_{\;\;j} (T_\chi)^j_{\;\;i} = -\, \frac{\partial V_{\rm int}}{\partial s}\dot{\bar{s}}\,,
\end{equation}
which making use of the above results we rewrite as
\begin{equation}\label{eq:bkg_chi_rel}
\dot{ \bar{\rho} }_\chi + 3 H (\bar{\rho}_\chi + \overline{\mathcal{P}}_\chi) = (\bar{\rho}_\chi - 3 \overline{\mathcal{P}}_\chi) \frac{\partial \log m_\chi(s)}{\partial s}\, \dot{\bar{s} }\,.
\end{equation}
This reduces to Eq.~\eqref{eq:chi_bkg} in the pressureless limit.

Equation~\eqref{eq:bkg_chi_rel} can also be derived from the particle viewpoint, by integrating the Vlasov equation~\eqref{eq:BE} in $d^3p\, E/(2\pi)^3$, performing integration by parts, and recalling the definitions $\rho_\chi = \int d^3p E f_\chi/(2\pi)^3$ and $\mathcal{P}_\chi = \int d^3 p\, p^2 f_\chi / [(2\pi)^3 3E]\,$. We note that there is yet another way to obtain the above equation from the particle perspective~\cite{Farrar:2003uw}, by regarding the DM as a collection of point particles with number density $n_\chi$ related to energy density and pressure by $\rho_\chi =  \gamma m_\chi(s) n_\chi $ and $\mathcal{P}_\chi = \gamma v^2  m_\chi(s) n_\chi /3$. Therefore $\rho_\chi - 3\mathcal{P}_\chi = \sqrt{1 - v^2}\, m_\chi(s) n_\chi $, which manifests the expected decoupling in the ultra-relativistic limit $v\to 1$. The connection with the field perspective is made through the relations $n_\chi   = \gamma m_\chi(s) \chi^2$ for real scalar DM, $n_\chi   = 2 \gamma m_\chi(s) \chi^\ast \chi$ for complex scalar DM, and $n_\chi = \gamma \overline{\chi}\chi$ for Dirac fermion DM.

The field approach can also be extended to first order, by making use of the relation between the perturbed fluid and field variables. In Newtonian gauge,
\begin{equation}
\delta \rho_\chi = -\, (\delta T_\chi)^0_{\;\;0} = \dot{\bar{\chi}}\dot{\delta \chi} - \dot{\bar{\chi}}^2 \Psi + \frac{\partial V_\chi}{\partial \chi} \delta \chi +  \frac{\partial V_{\rm int}}{\partial s} \delta s +  \frac{\partial V_{\rm int}}{\partial \chi} \delta \chi \,,
\end{equation}
which should be contrasted with the first of Eqs.~\eqref{eq:fluid_vars_NG}. In addition the fluid velocity is given by $(\bar{\rho}_\chi + \overline{\mathcal{P}}_\chi) v_{\chi}^i = - \dot{\bar{\chi}} \nabla^i \delta \chi /a$. By employing these relations and the field equations of motion at the background and first-order levels, in the pressureless limit the continuity and Euler equations~\eqref{eq:delta_chi_conf} and~\eqref{eq:v_chi_conf} are straightforwardly obtained.

\section{Perturbations in Synchronous Gauge}\label{app:pert_SG}

Following Ref.~\cite{Ma:1995ey} we call $h,\eta$ the two synchronous gauge (SG) potentials, defined in momentum space by
\begin{equation}
ds^2 =  -\, \dd t^2 + a^2 (\delta_{ij} + h_{ij}) \dd x^i \dd x^j , \quad h_{ij} = \int d^3 k\, e^{i \vec{k} \cdot \vec{x}} \Big[ \frac{k_i k_j}{k^2} h + \Big(\frac{k_i k_j}{k^2} - \frac{\delta_{ij}}{3} \Big) 6 \eta \Big].
\end{equation}
The value of the SG scalar field perturbation is related to the one in Newtonian gauge (NG) by
\begin{equation}
\delta s^{\rm NG} = \delta s^{\rm SG} + \bar{s}' \alpha\,, \qquad \alpha \equiv \frac{1}{2k^2} (h' + 6 \eta').
\end{equation}
The continuity equation reads
\begin{equation}
\delta'_\chi + \theta_\chi + \frac{h'}{2} - \frac{\partial \log m_\chi (s)}{\partial s} \,\delta s' - \frac{\partial^2 \log m_\chi(s)}{\partial s^2} \bar{s}' \,\delta s  = 0\,,
\end{equation}
the Euler equation
\begin{equation}\label{eq:Euler_sync}
\theta_\chi^\prime + \Big( \mathcal{H} + \frac{\partial \log m_\chi (s)}{\partial s} \,\bar{s}' \Big) \theta_\chi - k^2 \frac{\partial \log m_\chi (s)}{\partial s} \,\delta s = 0\,,
\end{equation}
and the perturbed KG equation is
\begin{equation}
\delta s^{\prime\prime}+ 2 \mathcal{H}  \delta s^\prime + k^2 \delta s + \frac{\bar{s}^\prime h^\prime}{2} + a^2 G_s V_{s,,s} \delta s  + a^2 G_s \bar{\rho}_\chi \frac{\partial \log m_\chi (s)}{\partial s} \delta_\chi   + a^2 G_s \bar{\rho}_\chi \frac{\partial^2 \log m_\chi (s)}{\partial s^2} \delta s  = 0 \,. \label{eq:KG_pert_sync_conf}
\end{equation}
As is well known, the above equations are obtained from those in NG by setting
\begin{equation}\label{eq:quick_gauge_transform}
\Psi \to 0\,,\qquad\quad \Phi \to h/6\,.
\end{equation}
The Einstein's equations take the standard form~\cite{Ma:1995ey}. It may be useful to recall the relation between our definitions of the NG potentials and those in Ref.~\cite{Ma:1995ey}: $\Psi^{\rm here} = \psi^{\rm MB}$ and $\Phi^{\rm here} = -\, \phi^{\rm MB}$.

\subsection{Initial conditions in synchronous gauge}
The adiabatic initial conditions in SG for all fluids except $\chi, s$, as well as for the gravitational potentials, are found using standard methods~\cite{Ma:1995ey}. By definition we have $\theta_{c} = 0$ and, retaining only the leading terms in the $k \tau \ll 1$ expansion,
\begin{align} \label{eq:ICs_SG}
\frac{4}{3} \delta_{c} &\,= \frac{4}{3} \delta_b = \delta_\nu = \delta_\gamma = -\frac{2}{3} C (k \tau)^2 , \qquad \theta_b = \frac{15 + 4 R_\nu}{23 + 4 R_\nu} \theta_\nu =  \theta_\gamma = - \frac{C}{18} k (k \tau)^3 \,,\nonumber \\
&\sigma_\nu = \frac{4C}{3} \frac{(k \tau)^2}{15 + 4 R_\nu}\, , \qquad h = C (k \tau)^2\,, \,\quad \eta = 2 C - \frac{C}{6} \frac{5 + 4 R_\nu}{15 + 4 R_\nu}  (k \tau)^2 \,,
\end{align}
with $C$ a dimensionless normalization constant and $R_\nu \equiv \bar{\rho}_\nu /    (\bar{\rho}_\gamma + \bar{\rho}_\nu)$. The initial condition for $\delta_\chi$ is found to satisfy adiabaticity, namely $\delta_\chi \simeq \frac{3}{4}\delta_\gamma$ up to corrections strongly suppressed by $\bar{s}'\tau \ll 1$, as obtained from the RD solution in Eq.~\eqref{eq:s_RD}.
%obtained by requiring that the gauge-invariant entropy perturbation relative to photons vanish,
%\begin{equation}
%0 = \mathcal{S}_{\chi \gamma} = 3 \mathcal{H} \bigg( \frac{\delta \rho_\gamma}{\bar{\rho}_\gamma^{\,\prime}} - \frac{\delta \rho_\chi}{\bar{\rho}_\chi^{\,\prime}}  \bigg) = - \frac{3}{4}\delta_\gamma + \frac{\delta_\chi}{1 - \frac{\partial \log m_\chi (s)}{\partial s} \frac{\bar{s}'}{3\mathcal{H}}} \approx - \frac{3}{4}\delta_\gamma + \delta_\chi\,,
%\end{equation}
%where the last equality holds to high accuracy since $\bar{s}'\tau \ll 1$, as obtained from the RD solution in Eq.~\eqref{eq:s_RD}. Hence $\delta_\chi  =3\delta_\gamma/4\,$.
Then, assuming RD when $V_s$ is negligible and taking the super-horizon limit, the KG equation~\eqref{eq:KG_pert_sync_conf} admits the solution
\begin{equation} \label{eq:s_field_SG_IC}
\delta s \simeq - \frac{C}{6}\bar{s}^\prime k^2 \tau^3 + \alpha_{\rm na} \left( 1 + \frac{\partial^2 \log m_\chi (s)/\partial s^2 }{\partial \log m_\chi (s)/\partial s}\bar{s}'\tau + \mathcal{O}(\bar{s}^{\prime\,2}\tau^2) \right) 
\end{equation}
where $\alpha_{\rm na}$ parametrizes the deviation from adiabaticity. This solution holds provided $\partial^2 \log m_\chi (s)/\partial s^2 \neq 0$, namely excluding the case $m_\chi (s) \propto e^{-s}$. Plugging Eq.~\eqref{eq:s_field_SG_IC} into the definition of the $s$ energy density perturbation (in SG, the expression of the $s$ fluid variables is given by Eq.~\eqref{eq:fluid_vars_NG} setting $\Psi = 0$) we arrive at
\begin{equation}
\delta_s = (1 + w_s) \frac{\delta s'}{\bar{s}'}\simeq \frac{3\delta_\gamma}{2} + 2\hspace{0.2mm} \alpha_{\rm na}\, \frac{\partial^2 \log m_\chi (s)/\partial s^2}{\partial \log m_\chi (s)/\partial s}\,.
\end{equation}
Thus for $\alpha_{\rm na} = 0$ we find $\delta_s = 3\delta_\gamma/2$ in SG, which differs from $\delta_s = \delta_\gamma/2$ found at leading order in NG. There is, however, no inconsistency, as we now show by going to the next order. The SG potentials read
\begin{equation}
h = C(k\tau)^2 + a_h (k \tau)^4\,,\qquad \eta = 2 C - \frac{C}{6} \frac{5 + 4 R_\nu}{15 + 4 R_\nu}  (k \tau)^2 + a_\eta (k\tau)^4\,, 
\end{equation}
where $a_h, a_\eta$ are constants whose explicit expressions are not needed for our purposes. The NG density perturbations are found to be
\begin{align}
\delta_\gamma^{\rm NG} =& - \frac{40C}{15 + 4 R_\nu} - \Big( \frac{2C}{3} + 8 (a_h + 6 a_\eta)\Big)(k \tau)^2\,, \nonumber \\ \delta_s^{\rm NG} =& - \frac{20C}{15 + 4 R_\nu} - \Big( C + 4 (a_h + 6 a_\eta)\Big)(k \tau)^2\,. \label{eq:NLO_deltas} 
\end{align}
The relative entropy perturbation between the scalar field and the photons in the two gauges is then, upon application of the background relation Eq.~\eqref{eq:doom},
\begin{equation}
\mathcal{S}_{s \gamma} = - \frac{3}{4}\delta_\gamma^{\rm NG} + \frac{3}{2}\delta_s^{\rm NG} = - C (k\tau)^2 = - \frac{3}{4}\delta_\gamma^{\rm SG} + \frac{3}{2}\delta_s^{\rm SG}\,,    
\end{equation}
neglecting $\mathcal{O}(k^4\tau^4)$ terms. Thus $\mathcal{S}_{s\gamma}$ vanishes only at leading order, whereas it is nonzero (and gauge invariant as it must be) at next-to-leading order. A similar observation was made in Ref.~\cite{Ballesteros:2010ks}.

Turning to the SG velocity potentials, for $s$ we find
\begin{equation}
\theta_s = \frac{k^2}{\bar{s}'} \delta s \simeq 3\theta_\gamma +  \frac{\alpha_{\rm na} k^2}{\bar{s}'} \,.    
\end{equation}
Finally, the initial condition for the $\chi$ velocity potential is obtained from the Euler equation~\eqref{eq:Euler_sync},
\begin{equation}
\theta_\chi \simeq \frac{\alpha_{\rm na}}{2} \frac{\partial \log m_\chi (s)}{\partial s}\, k (k \tau)\,.  
\end{equation}
Thus, for $\alpha_{\rm na} \neq 0$ we find $\theta_\chi \sim \mathcal{O}(k^2 \tau)$, potentially much larger than for the photons and baryons which have $\theta \sim \mathcal{O}(k^4 \tau^3)$.

\subsection{Initial conditions in Newtonian gauge, summary}

The adiabatic initial conditions in NG for all fluids except $\chi, s$, and for the gravitational potentials, are at leading order in the $k \tau \ll 1$ expansion~\cite{Ma:1995ey}
\begin{align} \label{eq:ICs_NG}
\frac{4}{3} \delta_{c} =\,& \frac{4}{3} \delta_b = \delta_\nu = \delta_\gamma = - \frac{40 \,C}{15 + 4 R_\nu}\,,  \qquad \theta_{c} = \theta_b = \theta_\nu =  \theta_\gamma = \frac{10 \,C}{15 + 4 R_\nu} k (k \tau)\,, \nonumber \\
& \sigma_\nu = \frac{4C}{3} \frac{(k \tau)^2}{15 + 4 R_\nu} \,, \qquad \Psi = \frac{20\,C}{15 + 4 R_\nu}\,,\quad \Phi = - \Psi \Big( 1 + \frac{2}{5} R_\nu \Big) . 
\end{align}
For $\chi$ and $s$ we summarize the results of Sec.~\ref{sec:IC}: for the field perturbation we have found
\begin{equation}
\delta s \simeq  \frac{10C}{15 + 4 R_\nu}\bar{s}' \tau  + \alpha_{\rm na} \left( 1 + \frac{\partial^2 \log m_\chi (s)/\partial s^2 }{\partial \log m_\chi (s)/\partial s}\bar{s}'\tau + \mathcal{O}(\bar{s}^{\prime\,2}\tau^2) \right)
\end{equation}
and for the fluids, recalling Eq.~\eqref{eq:fluid_vars_NG},
\begin{align}
\delta_\chi =&\, \frac{3}{4}\delta_\gamma\,,\qquad \delta_s = \frac{\delta_\gamma}{2} + 2 \alpha_{\rm na} \frac{\partial^2 \log m_\chi (s)/\partial s^2}{\partial \log m_\chi (s)/\partial s} \,,\\
\theta_\chi =&\; \theta_\gamma + \frac{\alpha_{\rm na}}{2} \frac{\partial \log m_\chi (s)}{\partial s}\, k (k \tau) \,,\qquad \theta_s = \theta_\gamma + \frac{\alpha_{\rm na} k^2}{\bar{s}'} \,.
\end{align}

\section{Perturbations in the Fluid Description}\label{app:fluid_eqs}
For completeness, in this section we provide the equations of motion for the mediator written in fluid form, in NG. By applying the relations between the field perturbations and fluid perturbations given in Eq.~\eqref{eq:fluid_vars_NG}, we express $\delta s, \delta s^\prime$ in terms of $\delta_s, \theta_s$, thus obtaining two first-order equations in place of the perturbed KG equation~\eqref{eq:KG_first}. 

For $\delta_s$ we find
\begin{align}
\delta_s^\prime  +&\,  3 \mathcal{H} (1 - w_s) \delta_s  + \frac{ \bar{\rho}_\chi }{\bar{\rho}_s} \frac{\partial \log m_\chi (s)}{\partial s} \bar{s}^\prime  \Big( \Psi + \delta_\chi - \frac{w_s}{1 + w_s} \delta_s \Big) + 3 (1 + w_s) \Phi' \nonumber \\
\;\,+&\, \bigg[ \frac{1 - c_\varphi^2}{2(1+w_s)} \Big(6 \mathcal{H} (1 + w_s) +  \frac{ \bar{\rho}_\chi }{\bar{\rho}_s} \frac{\partial \log m_\chi (s)}{\partial s} \bar{s}^\prime \Big)  \Big(3 \mathcal{H} ( 1 + w_s) +  \frac{ \bar{\rho}_\chi }{\bar{\rho}_s} \frac{\partial \log m_\chi (s)}{\partial s} \bar{s}^\prime \Big) \nonumber \\ 
\;\,&\,\qquad\qquad  +\, \frac{ \bar{\rho}_\chi }{\bar{\rho}_s}  \frac{\partial^2 \log m_\chi (s)}{\partial s^2} (\bar{s}^\prime)^2  + (1 + w_s) k^2  \bigg]  \frac{\theta_s}{k^2} = 0   \;, \label{eq:deltas}
\end{align}
where the adiabatic sound speed squared is defined as
\begin{equation}\label{eq:ad_soundspeed}
c_\varphi^2 \equiv \frac{\overline{\mathcal{P}}^{\,\prime}_s}{\bar{\rho}_s^{\,\prime}} = 1 + \frac{2 a^2 G_s V_{s,s}}{3\mathcal{H} \bar{s}^\prime + a^2 G_s \frac{\partial \log m_\chi (s)}{\partial s} \bar{\rho}_\chi} \,.
\end{equation}
This should not be confused with the sound speed squared $c_{s\varphi}^2 \equiv (\delta \mathcal{P}_s/\delta \rho_s)_{\rm RF}\,$, which always equals $1$ for a scalar field. Instead, the adiabatic sound speed satisfies the gauge-invariant relation
\begin{equation} \label{eq:deltaP}
\delta \mathcal{P}_s = \delta \rho_s + ( 1 - c_\varphi^2) \Big(3 \mathcal{H} \bar{\rho}_s ( 1 + w_s) + \bar{\rho}_\chi \frac{\partial \log m_\chi (s)}{\partial s}  \bar{s}' \Big) \frac{\theta_s}{k^2}\,.
\end{equation}
The equation for the $s$ velocity perturbation is found to be
\begin{equation} \label{eq:thetas}
\theta_s^\prime -2 \mathcal{H} \Big( 1 +\frac{3}{2} \frac{ \bar{\rho}_\chi }{3   \mathcal{H} \bar{\rho}_s (1 + w_s)} \frac{\partial \log m_\chi (s)}{\partial s} \bar{s}^\prime \Big)  \,\theta_s - k^2\Psi  - \frac{k^2 \delta_s}{1 + w_s}  = 0\,.
\end{equation}
The second term in the parenthesis of the friction term contains the doom factor defined in Eq.~\eqref{eq:doom}, and we now see the reason for this terminology. 
If the doom factor is large and positive, there exists, even in the absence of sources, a solution to the Euler equation which goes as $\tau^n$, where $n$ is proportional to the doom factor. In this scenario, perturbations become unstable at very early times \cite{Valiviita:2008iv,Gavela:2009cy}. For the case of Yukawa interactions discussed in this work, which rests on a microscopic Lagrangian description, the doom factor is negative and therefore does not lead to any runaway instabilities of the perturbations (in fact, the $\theta_s$ term in Eq.~\eqref{eq:thetas} vanishes during early RD). We believe the same conclusion applies to any physical model of long range forces in the dark sector.

We also rewrite the equations for $\chi$ in the same variables: Eq.~\eqref{eq:delta_chi_conf} as
\begin{align} 
\delta_\chi^\prime \,+&\,  \theta_\chi + 3 \Phi^\prime - \frac{\partial \log m_\chi (s)}{\partial s} \bar{s}^\prime \Big( \Psi + \frac{\delta_s}{1 + w_s}  \Big)    \\
\,-&\,  \bigg[   \frac{1 - c_\varphi^2}{2(1 + w_s)} \frac{\partial \log m_\chi (s)}{\partial s} \bar{s}^\prime \Big( 3 \mathcal{H}(1 + w_s) + \frac{\bar{\rho}_\chi}{\bar{\rho}_s} \frac{\partial \log m_\chi (s)}{\partial s} \bar{s}^\prime \Big) + \frac{\partial^2 \log m_\chi (s)}{\partial s^2}(\bar{s}^\prime)^2 \bigg] \frac{\theta_s}{k^2} = 0\,,  \nonumber
\end{align} 
and Eq.~\eqref{eq:v_chi_conf} as
\begin{equation}
\theta_\chi^\prime + \Big( \mathcal{H} + \frac{\partial \log m_\chi (s)}{\partial s} \bar{s}^\prime \Big) \theta_\chi - k^2 \Psi  - \frac{\partial \log m_\chi (s)}{\partial s} \bar{s}^\prime \theta_s  = 0\,.
\end{equation}
It is tedious but straightforward to check that these equations agree with Refs.~\cite{Valiviita:2008iv,Beyer:2014djk}, once the appropriate identifications are made. The SG equations are simply obtained by performing the replacements~\eqref{eq:quick_gauge_transform} in the above expressions.

\bibliography{5thforce}
\end{document}